\documentclass[12pt]{article}
\pdfoutput=1

\input{epsf}
\usepackage{epsfig}
\usepackage{amssymb}

\usepackage{amsfonts}
\usepackage{amsbsy}
\usepackage[cmtip,all]{xy}
\usepackage{amsmath}
\usepackage{esint}
\usepackage{collectbox}
\usepackage{amscd}
\usepackage{mathrsfs}
\usepackage{amsmath,amsthm}
\usepackage{enumitem}
\usepackage{dsfont}
\usepackage{soul}
\usepackage{comment}
\usepackage{cite} 
\usepackage[most]{tcolorbox}
\usepackage{stmaryrd}
\usepackage{xcolor}
\definecolor{burgundy}{rgb}{0.5, 0.0, 0.13}
\definecolor{olive}{rgb}{0.50, 0.50, 0.0}
\usepackage[linktocpage=true,colorlinks=true,linkcolor=burgundy,citecolor=black!20!blue,urlcolor=violet]{hyperref}
\usepackage{accents}

\usepackage{tikz}
\usepackage{tikz-cd}
\usetikzlibrary{arrows}
\usetikzlibrary{arrows.meta}
\usetikzlibrary{positioning}
\usetikzlibrary{shapes,snakes}
\usetikzlibrary{fit}
\usetikzlibrary{decorations.pathmorphing,decorations.pathreplacing,decorations.markings}
\usetikzlibrary{calc}

\theoremstyle{definition}
\newtheorem{assumption}{Assumption}

\DeclareMathAlphabet{\mathpzc}{OT1}{pzc}{m}{it}

\def\exp{{\rm exp}}

\def\I{{\rm i}}

\def\log{{\rm log}}

\def\Tr{{\rm Tr}}

\def\p{\partial}

\def\bDelta{\boldsymbol{\Delta}}

\def\CC {{\cal C}}
\def\CD {{\cal D}}

\def\CF {{\cal F}}

\def\CI {{\cal I}}
\def\CJ {{\cal J}}

\def\CL {{\cal L}}

\def\CN {{\cal N}}
\def\CO {{\cal O}}
\def\CP {{\cal P}}

\def\CO {{\cal O}}
\def\CZ {{\cal Z}}

\def\CI {{{\cal I}}}

\def\CS {{\cal S}}

\def\CZ{{\cal Z}}

\def\IC{\mathbb{C}}

\def\IN{\mathbb{N}}

\def\IP{\mathbb{P}}
\def\IR{{\mathbb{R}}}

\def\IZ{{\mathbb{Z}}}

\def\ff{\mathfrak{f}}

\def\fg{\mathfrak{g}}

\def\fl{\mathfrak{l}}

\def\fo{\mathfrak{o}}

\def\fq{\mathfrak{q}}

\def\fs{\mathfrak{s}}

\def\fq{\mathfrak{q}}

\def\fs{\mathfrak{s}}

\def\fB{\mathfrak{B}}

\def\bSigma{{\boldsymbol{\Sigma}}}

\def\Dslash{\,{\raise.15ex\hbox{/}\mkern-12mu \CD}}

\def\lm{\limits}

\numberwithin{equation}{section}

\newcommand\Kappa{\mathrm{K}}
\def\myY{\mathsf{Y}}
\def\myW{\mathcal{W}}
\def\BAE{{\bf BAE}}
\def\myequiv{:= }
\def\mya{a}

\def\mye{e}
\def\myf{f}
\def\myt{t}
\def\myk{\psi}

\def\mys{\texttt{s}}

\def\myblue{white!40!blue}
\newcommand{\longsquiggly}{\xymatrix{{}\ar@{~>}[r]&{}}}

\def\coL{\boldsymbol{\CL}}
\def\coe{{\bf e}}
\def\cof{{\bf f}}

\def\copsi{\boldsymbol{\psi}}

\tcbset{boxrule=0.6pt,colback=white!97!blue}

\newcommand\sqbox[1]{{
	\setbox0=\hbox{\mbox{$\Box$}}
	\setbox1=\hbox{\mbox{\raisebox{0.35ex}{\tiny #1}}}
	\mbox{\raisebox{-0.2ex}{\rlap{\hbox to \wd0{\hss{\box1}\hss}}\box0}}
}}
\newcommand{\mytimes}{ \tikz[baseline=-.55ex] \node [inner sep=0pt,cross out,draw,line width=1pt,minimum size=1ex] (a) {};}

\usepackage{accents}
\renewcommand{\ddot}[1]{{\accentset{\mbox{\large\bfseries .}}{#1}}}

\def\myitem{\item[\textcolor{\myblue}{\textbullet}]}

\begin{document}
	
\pagenumbering{Alph} 
\begin{titlepage}
		
\begin{center}
		
{\bf\Large{Gauge/Bethe correspondence \\ \bigskip from quiver BPS algebras}}
\vskip 0.2in
\renewcommand{\thefootnote}{\fnsymbol{footnote}}
{Dmitry Galakhov$^{1,2,}$\footnote[2]{e-mail: dmitrii.galakhov@ipmu.jp; galakhov@itep.ru},  Wei Li$^{3}$\footnote[3]{e-mail: weili@mail.itp.ac.cn} and Masahito Yamazaki$^{1,}$\footnote[4]{e-mail: masahito.yamazaki@ipmu.jp}} 
\vskip 0.2in 
\renewcommand{\thefootnote}{\roman{footnote}}
{\small{ 
                \textit{$^1$Kavli Institute for the Physics and Mathematics of the Universe (WPI), }\vskip -.4cm
                \textit{University of Tokyo, Kashiwa, Chiba 277-8583, Japan}
                \vskip 0 cm 
                \textit{$^2$Institute for Information Transmission Problems,}\vskip -.4cm
                \textit{ Moscow, 127994, Russia}
                 \vskip 0 cm 
                \textit{$^3$Institute for Theoretical Physics, Chinese Academy of Sciences}\vskip -.4cm
                \textit{ Beijing, 100190, China}
}}
\end{center}

\vskip 0.2in
\baselineskip 16pt

\begin{abstract}
We study the Gauge/Bethe correspondence for two-dimensional $\mathcal{N}=(2,2)$ supersymmetric quiver gauge theories associated with toric Calabi-Yau three-folds, whose BPS algebras have recently been identified as the quiver Yangians. 
We start with the crystal representations of the quiver Yangian, which are placed at each site of the spin chain. 
We then construct integrable models by combining the single-site crystals into crystal chains by a coproduct of the algebra, which we determine by a combination of representation-theoretical and gauge-theoretical arguments. 
For non-chiral quivers, we find that the Bethe ansatz equations for the crystal chain coincide with the vacuum equation of the quiver gauge theory, thus confirming the corresponding Gauge/Bethe correspondence. 
For more general chiral quivers, however, we find obstructions to the $R$-matrices satisfying the Yang-Baxter equations and the unitarity conditions, and hence to their corresponding Gauge/Bethe correspondence. 
We also discuss trigonometric (quantum toroidal) versions of the quiver BPS algebras, which correspond to three-dimensional $\mathcal{N}=2$ gauge theories and arrive at similar conclusions. 
Our findings demonstrate that there are important subtleties in the Gauge/Bethe correspondence, often overlooked in the literature.
\end{abstract}

\date{June, 2022}
\end{titlepage}
\pagenumbering{arabic} 
	
\newpage
\tableofcontents
\newpage
	

\section{Introduction}

The Gauge/Bethe correspondence \cite{Nekrasov:2009uh,Nekrasov:2009ui} claims interesting relations between integrable models and 2D/3D/4D 
supersymmetric gauge theories with four supercharges. 
The vacua of the latter theory (compactified onto 2D for the 3D/4D cases) are described by the extremization of the effective twisted superpotential $\myW(\sigma)$,
\begin{align}
  \exp \left( \frac{\partial \myW(\sigma)}{\partial \sigma}\right)=1 \,
\label{eq.BAE}
\end{align}
where $\sigma$ collectively denotes scalars in the twisted vector multiplet.
The non-trivial statement is that the resulting gauge-theory vacuum equation can be identified with the Bethe ansatz equations (BAE) for
the associated integrable model. Many examples of such developments have been worked out, and have led to fascinating interactions between gauge theories, geometries and representation theories, among others.\footnote{The literature on this topic is huge and it is unfortunately impossible to list all the relevant references.
As of this writing both of the foundational papers \cite{Nekrasov:2009uh,Nekrasov:2009ui} have more than 200 citations.}

Despite years of research on this subject, however, there remains a fundamental question regarding this correspondence---how general
can the correspondence \eqref{eq.BAE} be? 

In the literature, it is often assumed that 
the correspondence \eqref{eq.BAE} works in full generality, for 
{\it any} 2D $\mathcal{N}=(2,2)$ supersymmetric gauge theory as well as their 3D/4D cousins.
Indeed, the vacuum equation \eqref{eq.BAE}, as derived from supersymmetric gauge theory, is often automatically called the BAE,
even when the integrable model in question is unknown.
We emphasize, however, that it is actually rather non-trivial to verify this assumption explicitly---we need a general algorithm to identify the associated integrable model from a given supersymmetric gauge theory.

In fact, since there exists a zoo of 2D $\mathcal{N}=(2,2)$ theories (and their 3D/4D counterparts) with various matter contents, it is natural to imagine that we will inevitably need new types of integrable models not discussed in the literature, beyond those associated e.g.\ to representations of the standard Yangians $\myY(\mathfrak{g})$ associated with (affine) Lie (super)algebras $\mathfrak{g}$ and their trigonometric/elliptic deformations.  This means that any complete understanding of the Gauge/Bethe correspondence requires a systematic study of such new algebras and 
associated integrable models.

In this paper, we tackle this question when the supersymmetric gauge theories in question are quiver gauge theories whose quivers and superpotentials are associated with toric Calabi-Yau three-folds. 
The advantage of working with this class of theories is that the associated BPS algebras have already been identified as the
quiver Yangians \cite{Li:2020rij,Galakhov:2020vyb,Galakhov:2021xum} and their trigonometric/elliptic counterparts \cite{Galakhov:2021vbo,Noshita:2021dgj,Noshita:2021ldl}, 
whose generators and relations have been worked out very explicitly.
Moreover, a very general class of representations of the algebras have been constructed already \cite{Li:2020rij,Galakhov:2021xum} in terms of the statistical-mechanics model of crystal melting \cite{Ooguri:2008yb,Yamazaki:2010fz}. One can then construct spin chains from these representations, and work out the BAE explicitly.
If the resulting equations matches the vacuum equations of the corresponding gauge theories, we have 
verified the Gauge/Bethe correspondence for our theories.

In this paper we present two main results: one is a ``yes-go'' result and another is a  ``no-go'' result.
Which case we end up on depends both on whether the quiver is chiral or non-chiral (corresponding to whether the toric CY$_3$ has compact four-cycle or not, respectively) and on the ``shift" of the quiver Yangians (as defined in  \cite{Galakhov:2021xum}).

The  ``yes-go'' result applies to un-shifted 
quiver Yangians for non-chiral quivers, which are 
associated with toric CY$_3$ without compact four-cycles and include 
unshifted affine Yangian of $\mathfrak{g}$ where $\mathfrak{g}=\mathfrak{gl}_{m|n}$ or $D(2,1;\alpha)$. 
The allowed representations are given by certain 2D crystals, which can be easily constructed (see \cite{Galakhov:2021xum}) based on the original 3D crystal representations of \cite{Li:2020rij}.
Within this class, we can derive the Bethe ansatz equations and verify that the equation coincides with the gauge-theory vacuum equations, as expected.

A surprise comes when we discuss shifted quiver Yangians for non-chiral quivers and all quiver Yangians for chiral quivers.
From the viewpoint of BPS state counting, we do not encounter any problem
and we still have well-defined representations from crystal melting.
We find, however, that we will run into inconsistencies in the Yang-Baxter equations if we follow the standard procedures often assumed in the literature.
Under certain assumptions, we find several no-go arguments (from both the algebraic viewpoint and the gauge-theoretical one) that rule out both shifted quiver Yangians for non-chiral quivers and all quiver Yangians for chiral quivers.
What these two problematic classes have in common is that in the mode expansions of their Cartan generators, there exist negative modes (namely modes that multiplying positive powers of the spectral parameter) --- as long as the representations are not too trivial.
While our argument relies on some assumptions, we will clarify our assumptions and 
also present representation-theoretical and gauge-theoretical motivations for the assumptions.

Finally, our discussion is not limited to the rational algebras (i.e.\ the quiver Yangians), which correspond to 2D $\mathcal{N}=(2,2)$ supersymmetric gauge theories --- we have also included the
trigonometric versions of the algebra (namely the quiver toroidal algebras), which correspond to 3D $\mathcal{N}=2$ supersymmetric quiver gauge theories, and reached the same conclusion. 

We believe that the results of our paper point out important subtleties
in the discussion of the Gauge/Bethe correspondence, which are often overlooked in the literature.

The rest of this paper is organized as follows.
In Section~\ref{sec:review} we collect some review material needed for understanding the rest of this paper.
In Section~\ref{sec:coproduct} we discuss coproduct structures of the BPS algebras,
which are needed for the construction of the spin chain.
In Section~\ref{sec:gauge_deriv} we outline the gauge-theoretical arguments for the Gauge/Bethe correspondence,
which support some of the assumptions made in Section~\ref{sec:coproduct}.
In Section~\ref{sec:no-go} we present arguments against the Gauge/Bethe correspondence
for shifted quiver algebras and those for chiral quivers.
In Section~\ref{sec:BAE} we derive the BAE for non-chiral quivers and 
derive the Gauge/Bethe correspondence for these examples.
While we focus on the rational cases of the quiver Yangians in most of this paper,
we discuss in Section~\ref{sec:trig} the trigonometric cases of quantum toroidal quiver algebras,
where we find similar ``no-go'' results despite some important differences.
We end in Section~\ref{sec:summary} with comments on future directions.
We have also included several appendices containing  technical computations.

\section{Reviews}\label{sec:review}

In this section, we will first review the quiver BPS algebras (Section~\ref{sec:algebra}) and their crystal representations (Section~\ref{sec:cry_rep}). 
We then review some basic aspects of integrable models, such as coproducts, $R$-matrices, and the Yang-Baxter equations, for the reader's convenience and to fix our notations (Section~\ref{sec:integrable_review}). 
Finally, we briefly summarize the Gauge/Bethe correspondence for quiver gauge theories in Section~\ref{sec:Bethe-gauge},
which will be the motivation for the rest of this paper. Readers familiar with the  quiver Yangians (and/or integrable models) are encouraged to skip Sections \ref{sec:algebra} and \ref{sec:cry_rep} (and/or Section~\ref{sec:integrable_review}).

\subsection{Quiver BPS algebras}\label{sec:algebra}

Quiver BPS algebras refer to the ``algebras of BPS states" \cite{Harvey:1996gc} of string/M/F theory compactified on toric Calabi-Yau threefolds (toric CY${}_3$).
The resulting 4D/5D/6D theory is a supersymmetric gauge theory, with the low energy effective description of the BPS sectors given by 1D $\mathcal{N}=4$, 2D $\mathcal{N}=(2,2)$, and 3D $\mathcal{N}=2$ quiver gauge theories, respectively. 
The BPS algebras are directly defined in terms of the corresponding quiver data (together with the superpotential), hence the name ``quiver BPS algebras".
The trichotomy of string/M/F correspond to the rational/trigonometric/elliptic versions of the quiver BPS algebras, which are called quiver Yangian \cite{Li:2020rij,Galakhov:2020vyb,Galakhov:2021xum,Yamazaki:2022cdg}, toroidal quiver algebras \cite{Galakhov:2021xum,Noshita:2021dgj,Noshita:2021ldl}, and  elliptic quiver algebras \cite{Galakhov:2021xum}, respectively. 

In this paper, we will mostly focus on the rational versions of quiver BPS algebras, namely the quiver Yangians;
the trigonometric cases, the toroidal quiver algebras, will be discussed later in Section~\ref{sec:trig}.
In this section we list some definitions and relations associated with quiver BPS algebras, which we will use throughout the paper.
We will not provide original motivations and proofs for this construction; the interested reader is referred to the original references \cite{Li:2020rij,Galakhov:2020vyb,Galakhov:2021xum}.

Starting from a given toric Calabi-Yau three-fold (toric CY${}_3$), its corresponding quiver--superpotential pair $(Q,W)$ can be obtained by e.g.\ the procedure of ``brane tiling" \cite{Hanany:2005ve, Franco:2005rj,Franco:2005sm,Kennaway:2007tq, Yamazaki:2008bt} and this is a one-to-many map in general. 
We will always start with the quiver data $(Q,W)$.
We will use lowercase letters $a$, $b$, $\cdots$, to label the quiver nodes and denote the set of quiver nodes as $Q_0$; and we will use uppercase letters $I$, $J$, $\cdots$, to label the arrows of the quiver and denote their set as $Q_1$.
If we need to stress that an arrow $I$ flows from node $a$ to node $b$ we will mark this arrow as
$$
I:\,a\to b\,.
$$
We also define 
\begin{align}
\begin{array}{cl}
     \{a\to b\}: &\mbox{  the set of all arrows pointing  from node }a\mbox{ to node }b \;,\\
     |a\to b|: &\mbox{  the total number of arrows pointing from node }a\mbox{ to node }b \;.
\end{array}
\end{align}
For a pair of nodes $a,b\in Q_0$, we define their chirality as:
\begin{align}\label{chirality}
\chi_{ab}\myequiv|a\to b|-|b\to a|\,.
\end{align}
A quiver is called \textit{non-chiral} when $\chi_{ab}=0$ for any pair of nodes $a,b\in Q_0$; 
otherwise the quiver is called \textit{chiral}. 
A quiver that corresponds to a toric CY${}_3$ with (resp.\ without) compact four-cycles is chiral (resp.\ non-chiral).

To define the quiver BPS algebra, we need to enhance the quiver data $(Q,W)$ with an additional structure.
To each arrow $I\in Q_1$, we associate a complex-valued equivariant parameter $h_I$.
In the case of a toric CY${}_3$, the set of parameters $h_I$ are subjected to the loop and vertex constraints \cite{Li:2020rij}: 
\begin{align}
\begin{split}
    \textrm{loop constraints:}\qquad&\sum\lm_{I\in L}h_I=0,\quad \forall L\,,\\
    \textrm{vertex constraints:}\qquad&\sum\lm_{I:\bullet \to a}h_I-\sum\lm_{J:a \to \bullet}h_J=0,\quad \forall a\in Q_0\,,
\end{split}
\end{align}
where $L$ are all the loops in the quiver lattice (see Section~\ref{sec:cry_rep} and \eqref{lattice}).
These constraints leave among all $h_I$ only two independent parameters \cite{Li:2020rij}, which we denote by $\texttt{h}_1$ and $\texttt{h}_2$, so that $h_I$ take values in an integral lattice parameterized by $\texttt{h}_{1,2}$:
\begin{align}
    h_I=\alpha_I \texttt{h}_1+\beta_I\texttt{h}_2,\quad \alpha_I,\beta_I\in\IZ\,.
\end{align}
These two parameters $\texttt{h}_{1,2}$ correspond to the two equivariant parameters of the torus action that preserves the holomorphic Calabi-Yau three-form.

A Chevalley basis for a quiver BPS algebra $\myY(Q,W)$ consists of a triplet of generators $(e^{(a)}(z),\psi^{(a)}(z), f^{(a)}(z))$ for each $a\in Q_0$, where $z$ is the spectral parameter.
The generators $\psi^{(a)}(z)$ are Cartan generators, whereas $e^{(a)}(z)$ (resp.\ $f^{(a)}(z)$) are the raising (resp.\ lowering) operators.\footnote{In the crystal representation, all crystal states are eigenstates of all the Cartan generators $\psi^{(a)}(z)$, and $e^{(a)}(z)$ (resp.\ $f^{(a)}(z)$) adds (resp.\ removes) an atom of color $a$ to the existing crystal state.
To each node $a\in Q_0$, we can assign a non-negative integer $N_a$ as its quiver dimension. The generator $e^{(a)}(z)$ (resp.\ $f^{(a)}(z)$) increases (resp.\ decreases) $N_a$ by $1$, whereas  $\psi^{(a)}(z)$ leaves all $N_a$ invariant.}
There are trigonometric and elliptic versions of these algebras having non-trivial central elements such that $\psi^{(a)}(z)$ are non-commutative among themselves. We will, however, not consider those cases in this paper.

The algebra $\myY$ is a superalgebra in general.
The $\IZ_2$-grading (Bose/Fermi statistics) of the generators $e^{(a)}(z)$ and $f^{(a)}(z)$ is defined as
\begin{align}
    |a|=\left(|a\to a|+1\right)\;{\rm mod}\;2\,,
\end{align}
while the generators $\psi^{(a)}(z)$ are always even (Bose statistics).

The quiver Yangian is defined by the commutation relations \cite{Li:2020rij}:
\begin{align}
		\begin{aligned}\label{QiuvYangian}
			\myk^{(a)}(z)\mye^{(b)}(w) &\simeq \varphi^{a\Leftarrow b}\left(z-w\right) \mye^{(b)}(w) \myk^{(a)}(z)\,,\\
			\myk^{(a)}(z)\myf^{(b)}(w) &\simeq \varphi^{a\Leftarrow b}\left(z-w\right)^{-1} \myf^{(b)}(w) \myk^{(a)}(z)\,,\\
			\mye^{(a)}(z)\mye^{(b)}(w) &\sim (-1)^{|a||b|}\,\varphi^{a\Leftarrow b}(z-w) \mye^{(b)}(w) \mye^{(a)}(z)\,,\\
			\myf^{(a)}(z)\myf^{(b)}(w) &\sim (-1)^{|a||b|}\, \varphi^{a\Leftarrow b}(z-w)^{-1} \myf^{(b)}(w) \myf^{(a)}(z)\,,\\
			\left[\mye^{(a)}(z)\,,\myf^{(b)}(w)\right\}&\sim -\delta_{a,b} \frac{\myk^{(a)}\left(z\right)- \myk^{(a)}\left(w\right)}{z-w}\,,
		\end{aligned}
\end{align}
where the equality sign $\simeq$ ($\sim$) equates Taylor series on both sides up to terms of the form $z^nw^{m\geq 0}$ ($z^nw^{m\geq 0}$ and $z^{n\geq 0}w^m$), $[\star,\star\}$ is a super-commutator:
\begin{align}
    \left[X,Y\right\}=XY-(-1)^{|X||Y|}YX\,,
\end{align}
and the bond factor $\varphi$ is defined as:
\begin{align}
    \label{bond_factor}
		\varphi^{a\Leftarrow b} (u)\myequiv (-1)^{|b\rightarrow a| \chi_{ab}} \frac{\prod_{I\in \{a\rightarrow b\}} \left(u+h_{I}\right) }{\prod_{J\in \{b\rightarrow a\}} \left( u-h_{J} \right)} \;.
\end{align}

\subsection{Crystal representations} \label{sec:cry_rep}

The BPS algebras have crystal representations, where vectors are labeled by (generically 3D) molten crystals.
In this paper, we will need non-trivial tensor products of the \emph{2D subcrystal} representations.
For this, we will review the molten crystals as descriptions of BPS states \cite{Ooguri:2008yb,Yamazaki:2010fz}, the 3D canonical crystal representations of BPS algebras \cite{Li:2020rij}, the 2D subcrystal representations \cite{Galakhov:2021xum}, and finally the naive tensor representations \cite{Galakhov:2020vyb, Galakhov:2021vbo}.
We will not attempt to give a thorough review here, but refer the interested reader to these papers.

\subsubsection{Molten crystals as BPS states}

The molten crystal construction appears in the counting problem for Donaldson-Thomas invariants \cite{MR1634503} for toric CY${}_3$'s \cite{Okounkov:2003sp,Iqbal:2003ds,Szendroi,MR2836398,MR2592501,Jafferis:2008uf,Ooguri:2008yb}.
The effective low energy dynamics of D-branes wrapping a toric CY${}_3$ $\mathscr{X}$ is described by a 4D $\CN=1$ quiver gauge theory with gauge-matter content encoded in a pair of quiver $Q$ and superpotential $W$.
The initial counting problem becomes a counting problem for BPS states in this system,
which can be identified with the equivariant cohomologies of the quiver moduli spaces.
Due to the localizing properties of the equivariant cohomologies, the BPS wave-function can be approximated by a semi-classical expression localized to the classical vacua.
The classical vacua of the quiver gauge theory are given by the field configurations that satisfy a set of constraints coming from the $D$-terms and the $F$-terms in the Lagrangian, as well as fixed-point constraints coming from the equivariant action.

Let us denote the complex scalar fields constructed from the compactified gauge holonomies associated to quiver nodes as $X_a$, $a\in Q_0$, and complex scalars in the chiral multiplets associated to quiver arrows as $\phi_I$, $I\in Q_1$.
Then the $D$-term and $F$-term can be summarized in the form of real and complex moment maps. 
The $D$-term constraint, the $F$-term constraint, and the fixed point constraint are given by
\begin{align}\label{fp}
	\begin{split}
		&\mu_{\IR}^{(a)}:=r_a-\sum\lm_{I:\,\bullet\to a}\phi_I\phi_I^{\dagger}+\sum\lm_{J:\,a\to\bullet}\phi_J^{\dagger}\phi_J=0\,,\\
		&\mu_{\IC}^{(I)}:=\p_{\phi_I}W=0\,,\\
		&\hat G(X)\cdot\phi_{(I:\,a\to b)}:=X_b\phi_{(I:\,a\to b)}-\phi_{(I:\,a\to b)}X_a=0\,,
	\end{split}
\end{align}
where $r_a\in \IR$ are Fayet-Iliopoulos (FI) parameters of the model.

Classical vacua---solutions to \eqref{fp}---in the cyclic BPS chamber (all $r_a>0$) for framed quivers can be enumerated by molten crystals.
The construction of a molten crystal starts with a lattice construction.
First we notice that the toric CY${}_3$ pair of a quiver and a superpotential corresponds to a periodic quiver on a torus or a quiver lattice
$\mathscr{L}$. For the simplest example of $\IC^3$ this is a simple triangular lattice: 
\begin{align}\label{lattice}
	\IC^3:\,\begin{array}{c}
		\begin{tikzpicture}
			\tikzset{arr/.style={ 
					postaction={decorate},
					decoration={markings, mark= at position 0.6 with {\arrow{stealth}}},
					thick}}
			\draw[thick, black!30!red, postaction={decorate},decoration={markings, 
				mark= at position 0.7 with {\arrow{stealth}}}] (0,0) to[out=60,in=0] (0,1) to[out=180,in=120] (0,0);
			\node[above,black!30!red] at (0,1) {$\scriptstyle (X,\texttt{h}_1)$};
			\begin{scope}[rotate=120]
				\draw[thick, \myblue, postaction={decorate},decoration={markings, 
					mark= at position 0.7 with {\arrow{stealth}}}] (0,0) to[out=60,in=0] (0,1) to[out=180,in=120] (0,0);
				\node[below left,\myblue] at (0,1) {$\scriptstyle (Y,\texttt{h}_2)$};
			\end{scope}
			\begin{scope}[rotate=240]
				\draw[thick, black!60!green, postaction={decorate},decoration={markings, 
					mark= at position 0.7 with {\arrow{stealth}}}] (0,0) to[out=60,in=0] (0,1) to[out=180,in=120] (0,0);
				\node[below right,black!60!green] at (0,1) {$\scriptstyle (Z,\texttt{h}_3)$};
			\end{scope}
			\draw[fill=white] (0,0) circle (0.08);
			\node[below] at (0,-1) {$W=\Tr \,{\color{black!30!red} X}\left[{\color{\myblue} Y},{\color{black!60!green} Z}\right]$};
		\end{tikzpicture}
	\end{array}\longleftrightarrow
	\begin{array}{c}
		\begin{tikzpicture}[scale=0.7]
			\tikzset{arr1/.style={ 
					postaction={decorate},
					decoration={markings, mark= at position 0.6 with {\arrow{stealth}}},
					thick,black!30!red}}
			\tikzset{arr2/.style={ 
					postaction={decorate},
					decoration={markings, mark= at position 0.6 with {\arrow{stealth}}},
					thick,\myblue}}
			\tikzset{arr3/.style={ 
					postaction={decorate},
					decoration={markings, mark= at position 0.6 with {\arrow{stealth}}},
					thick,black!60!green}}
			\foreach \x in {0,1,...,5}
			\foreach \y in {0,1,...,3}
			{
				\draw[arr1] (\x + 0.5*\y, 0.866025*\y) -- (\x + 0.5*\y + 0.5, 0.866025*\y + 0.866025);
				\draw[arr2] (\x + 1 + 0.5*\y, 0.866025*\y) -- (\x + 0.5*\y, 0.866025*\y);
				\draw[arr3] (\x + 0.5*\y + 0.5, 0.866025*\y + 0.866025) -- (\x + 1 + 0.5*\y, 0.866025*\y);
			}
			\foreach \x in {0,1,...,5}
			\foreach \y in {4}
			{
				\draw[arr2] (\x + 1 + 0.5*\y, 0.866025*\y) -- (\x + 0.5*\y, 0.866025*\y);
			}
			\foreach \x in {6}
			\foreach \y in {0,1,...,3}
			{
				\draw[arr1] (\x + 0.5*\y, 0.866025*\y) -- (\x + 0.5*\y + 0.5, 0.866025*\y + 0.866025);
			}
			\foreach \x in {0,1,...,6}
			\foreach \y in {0,1,...,4}
			{
				\draw[fill=white] (\x + 0.5*\y, 0.866025*\y) circle (0.1);
			}
			\draw[-stealth] (-1.5, -0.866025) -- (2.5, -0.866025);
			\draw[-stealth] (-1.5, -0.866025) -- (0., 1.73205);
			\node[above] at (0., 1.73205) {$\scriptstyle\texttt{h}_1$};
			\node[right] at (2.5, -0.866025) {$\scriptstyle -\texttt{h}_2$};
		\end{tikzpicture}
	\end{array}
\end{align}
The nodes of the quiver lattice correspond to coherent sheaves on $\mathscr{X}$, whereas the arrows are the corresponding Ext-functors, and the superpotential is generated by disk amplitudes \cite{Yamazaki:2008bt,Eager:2016yxd} (see also \cite{Banerjee:2019apt,Banerjee:2020moh,Banerjee:2022oed}).
This whole construction is graded by the toric action which locally rescales $\mathscr{X}$ as $(x,y,z)\mapsto(e^{\texttt{h}_1}x,e^{\texttt{h}_2}y,e^{\texttt{h}_3}z)$ while preserving the volume form, which requires $\texttt{h}_1+\texttt{h}_2+\texttt{h}_3=0$.
Therefore, the complex plane that the quiver lattice is embedded into can be naturally identified with the equivariant weight plane parameterized by $\texttt{h}_{1,2}$.

\subsubsection{Canonical crystals and vacuum representations of BPS algebras}

For a given $(Q,W)$ pair,
a molten crystal in the original reference \cite{Ooguri:2008yb} consists of a subsets of atoms that are removed (or ``melted away") from a special 3D crystal following the ``melting rule". For the description and the reasoning of the melting rule, see \cite{Ooguri:2008yb,Yamazaki:2010fz}.

This special 3D crystal gives rise to the \emph{vacuum representation} of the BPS algebra and can actually be used to bootstrap the BPS algebra itself; for this bootstrap procedure, see \cite{Li:2020rij} for the rational case and \cite{Galakhov:2021vbo} for the trigonometric and elliptic cases. 
A non-vacuum representation can then be constructed by starting with a \emph{subcrystal} of this special crystal and using the molten crystals that are melted away from this subcrystal to span the corresponding representation. For the construction of these subcrystal representations see \cite{Galakhov:2021xum}, where we also named the special crystal that corresponds to the vacuum representation \emph{canonical crystal}.

Let us specify first a \emph{canonical} quiver framing as consisting of a single framing node $\ff$ with dimension 1 and an arrow $I_0$ connecting this node to any gauge quiver node $\fo$.
We denote the complex expectation value for the flavor field associated with $\ff$ as $u$ and call it the \emph{spectral parameter}.
To describe solutions to \eqref{fp} it is useful to incorporate a language of path operators:
\begin{align}
	\ldots\cdot\phi_{I_2}\cdot\phi_{I_1}\cdot\phi_{I_0}\,,
\end{align}
so that $I_1, I_2, \cdots$ form a path in $\mathscr{L}$ starting with any point projected to node $\fo$.
These path operators are subjected to equivalence relations following from the complex moment map and form a module of the quiver path algebra.
Monomials in this module are labeled by the endpoint of the path in the lattice and a discrete R-charge, forming in this way a 3d lattice---a lift of $\mathscr{L}$---actually a convex sublattice since all the R-charges could be chosen to be positive.
This sublattice was named \emph{canonical} crystal $\CC_0$ in \cite{Galakhov:2021xum} (in order to distinguish it from its subcrystals).  

Continuing the chemistry analogy we will call a point of the 3D lattice ``atom'' and denote it as $\Box$.
Any atom position can be projected to the weight lattice $\mathscr{L}$.
We will denote the complex parameter for the position of $\Box$ in $\mathscr{L}$ in the following way:
\begin{align}
    h_{\Box}:=\alpha_{\Box}\texttt{h}_1+\beta_{\Box}\texttt{h}_2\in \IC\,,
\end{align}
where $\alpha_{\Box}$, $\beta_{\Box}$ are atom coordinates in $\mathscr{L}$.
Having an atom position in $\mathscr{L}$, we can project it further to a node of quiver $Q$.
We call this projected node $a\in Q_0$ the \emph{color} of the atom.
When we want to emphasize that the atom $\Box$ has color $a$ we denote the atom as $\sqbox{$a$}$.

The fixed points \eqref{fp} correspond to subcrystals of $\CC_0$ satisfying a melting rule \cite{Ooguri:2008yb}---molten crystals.
An empty crystal, an empty QFT without fields, is also a valid crystal satisfying the melting rule.
For a molten crystal $\Kappa$ we could point out sets of atoms that could be added/removed to/from $\Kappa$ so that the resulting crystal satisfies the melting rule again:
\begin{align}
	\begin{split}
		{\rm Add}(\Kappa)&:=\left\{\Box\in\CC_0|\Kappa+\Box\mbox{ is molten}\right\}\,,\\
		{\rm Rem}(\Kappa)&:=\left\{\Box\in\CC_0|\Kappa-\Box\mbox{ is molten}\right\}\,.
	\end{split}
\end{align}

If a molten crystal $\Kappa$ is given, it is simple to restore an explicit solution to \eqref{fp}.
Correspondingly, molten crystals $\Kappa$ label classical vacua and BPS states of a D-brane system on $\mathscr{X}$ and form a module for the BPS algebra $\myY$.

The generators of the algebra $\myY$ acts on the crystal states by \cite{Li:2020rij}:\footnote{Relative to \cite{Li:2020rij}, the spectral parameter $z$ on the r.h.s.\ are shifted by the equivariant weight $u$ that is associated with the crystal representation.}
\begin{align}\label{crystal_rep}
	\begin{split}
		\psi^{(a)}(z)\, |\Kappa\rangle_{u}&= \Psi^{(a)}_{\Kappa}(z-u)\times|\Kappa \rangle_{u}\,,\\
		e^{(a)}(z)\,|\Kappa\rangle_{u}&= \sum\lm_{\sqbox{$a$}\in{\rm Add} \left(\Kappa\right)}   \frac{\left[\Kappa\to\Kappa+\sqbox{$a$}\right]}{z-\left(u+h_{\sqbox{$a$}}\right)} |\Kappa+\sqbox{$a$}\rangle_{u} \,,\\
		f^{(a)}(z)\, |\Kappa\rangle_{u}&= \sum\lm_{\sqbox{$a$}\in{\rm Rem} \left(\Kappa\right)}  \frac{\left[\Kappa\to\Kappa-\sqbox{$a$}\right]}{z-\left(u+h_{\sqbox{$a$}}\right)} |\Kappa-\sqbox{$a$}\rangle_{u} \,,\\
	\end{split}
\end{align}
where the eigenvalues of the Cartan generators $\Psi^{(a)}_{\Kappa}$ are:
\begin{align}\label{psi-eigenvalue}
    \Psi^{(a)}_{\Kappa}(z)=\varphi^{a\Leftarrow\ff}(z)\prod\lm_{b\in Q_0}\prod\lm_{\sqbox{$b$}\in\Kappa}\varphi^{a\Leftarrow b}\left(z-h_{\sqbox{$b$}}\right)\,,
\end{align}
with $\varphi^{a\Leftarrow\ff}(z)$ describing the contribution from the ground state of the representation;\footnote{For the vacuum representation, $\varphi^{a\Leftarrow\ff}(z)$ is simply $\frac{1}{z}$ (if we do not include any truncation factors), whose single pole $z=0$ corresponds to the leading atom of the canonical crystal.} 
and the matrix coefficients $\left[\Kappa\to\Kappa\pm\sqbox{$a$}\right]$ satisfy the following relations
\begin{align}\label{ef_matrix}
    \begin{split}
&[\Kappa+\sqbox{$a$}_1\to\Kappa+\sqbox{$a$}_1+\sqbox{$b$}_2\to\Kappa+\sqbox{$b$}_2]=\\
&\quad\quad\quad\quad\quad\quad\quad=(-1)^{|a||b|}[\Kappa+\sqbox{$a$}_1\to\Kappa\to\Kappa+\sqbox{$b$}_2] \;,\\
&\frac{[\Kappa\to\Kappa+\sqbox{$b$}_2\to\Kappa+\sqbox{$a$}_1+\sqbox{$b$}_2]}{[\Kappa\to\Kappa+\sqbox{$a$}_1\to\Kappa+\sqbox{$a$}_1+\sqbox{$b$}_2]}=(-1)^{|a||b|}\varphi^{a\Leftarrow b}\left(h_{\sqbox{$a$}_1}-h_{\sqbox{$b$}_2}\right),\\
&\frac{[\Kappa+\sqbox{$a$}_1+\sqbox{$b$}_2\to\Kappa+\sqbox{$b$}_2\to\Kappa]}{[\Kappa+\sqbox{$a$}_1+\sqbox{$b$}_2\to\Kappa+\sqbox{$a$}_1\to\Kappa]}=(-1)^{|a||b|}\varphi^{a\Leftarrow b}\left(h_{\sqbox{$a$}_1}-h_{\sqbox{$b$}_2}\right),\\
&(-1)^{|a|+1}
[\Kappa\to\Kappa+\sqbox{$a$}\to\Kappa]=\mathop{\rm Res}\lm_{t=h_{\sqbox{$a$}}}\,\Psi_{\Kappa}^{(a)}\left(t\right)\,,
\end{split}
\end{align}
using the short hand notation:
\begin{align}
    \left[\Kappa_1\to\Kappa_2\to\ldots\to\Kappa_k\right]:=\left[\Kappa_1\to\Kappa_2\right]\cdot\ldots\cdot\left[\Kappa_{k-1}\to\Kappa_k\right]\,.
\end{align}
One solution of these matrix coefficients  is \cite{Li:2020rij}:
\begin{equation}
[\Kappa\to\Kappa+\sqbox{$a$}]=\pm \sqrt{-(-1)^{|a|}\mathop{\rm Res}\lm_{t=h_{\sqbox{$a$}}}\,\Psi_{\Kappa}^{(a)}\left(t\right)}\,, \qquad [\Kappa\to\Kappa-\sqbox{$a$}]=\pm \sqrt{\mathop{\rm Res}\lm_{t=h_{\sqbox{$a$}}}\,\Psi_{\Kappa}^{(a)}\left(t\right)}
\end{equation} 
For the reasoning behind the signs and a way to fix the $\pm$'s see Section 6 of \cite{Li:2020rij} and for an explicit solution of these signs see Appendix D of  \cite{Galakhov:2021vbo}.

\subsubsection{Subcrystals and crystal representations with reduced dimensions}\label{sec:cry_dim}

For our current paper, we will need crystal reprsentations that are defined using certain 2D crystals, which can be viewed as \emph{subcrystals} of the canonical crystals. 
Let us now briefly review the construction of subcrystal representations, their translation to the \emph{framing} of the quiver, and their relation to the \emph{shift} of the quiver BPS algebras. For more detail, see \cite{Galakhov:2021xum}; for trigonometric version of the story see \cite{Galakhov:2021vbo, Noshita:2021ldl} and for the elliptic version see \cite{Galakhov:2021vbo}.

First of all, including other D-brane systems on $\mathscr{X}$ leads to a modification of the canonical framing and a change to the shape of the crystal ${}^{\sharp}\CC$ that defines the representation.
Since such crystals ${}^{\sharp}\CC$ can be viewed as a subcrystal of the original crystal of \cite{Li:2020rij}, we call their corresponding representations \emph{sub-crystal representations}, and they are non-vacuum representations, as opposed to the vacuum representation defined by the original crystals (named canonical crystal in \cite{Galakhov:2021xum}.

Once the shape of the subcrystal ${}^{\sharp}\CC$ is given,\footnote{A priori, there is no restriction to the shape of ${}^{\sharp}\CC$; however, we will only consider those that correspond to irreducible representations of the (shifted) BPS algebras, e.g.\ we will not consider disconnected ${}^{\sharp}\CC$.}
 all the molten crystal states $\Kappa$ of this representations are determined: they can only grow within the boundary of ${}^{\sharp}\CC$.
 The actions of the quiver BPS algebras on these molten crystal states $\Kappa$ are again given by \eqref{crystal_rep}, with the only change being the ground state factor $\varphi^{a\Leftarrow \ff}(z)$ in the Cartan eigenvalues \eqref{psi-eigenvalue}.

To determine $\varphi^{a\Leftarrow \ff}(z)$ for a given subcrystal ${}^{\sharp}\CC$, we can use the decomposition procedure of \cite{Galakhov:2021xum}.
The important observation is that any ${}^{\sharp}\CC$ can be constructed by decomposing ${}^{\sharp}\CC$ in terms of positive and negative $\CC_0$, placed at different positions.
The leading atoms of the positive $\CC_0$'s are called \emph{starters}, and those of the negative $\CC_0$'s are called either \emph{pausers} or \emph{stoppers}, depending whether they arise due to the intersections of positive $\CC_0$'s or they are introduced in order to stop the crystal from growing. For a detailed explanation of this decomposition procedure, see Section~3\ of \cite{Galakhov:2021xum}.

Once the sets of starter/pausers/stoppers of a subcrystal ${}^{\sharp}\CC$ are determined,\footnote{The set of starters and stopper are given by the shape of ${}^{\sharp}\CC$, whereas the set of pausers is determined once the starters are given.} we can immediately write down its ground state charge function \cite{Galakhov:2021xum}:
\begin{align}\label{psi_vac_0}
	\langle\varnothing|\psi^{(a)}(z)|\varnothing\rangle=\frac{\prod^{\mys^{(a)}_{-}}_{\beta=1}(z-\mathfrak{z}^{(a)}_{-\beta}) }{\prod^{\mys^{(a)}_{+}}_{\alpha=1}(z-\mathfrak{z}^{(a)}_{+\alpha})}=:\varphi^{a\Leftarrow \ff}(z)\,,\qquad a\in Q_0\,,
\end{align}
where $\{\mathfrak{z}^{(a)}_{+\alpha}\}$ corresponds to the set of the weights of all starters of color $a$, with $\mathfrak{s}^{(a)}_{+}$ the size of this set, whereas $\{\mathfrak{z}^{(a)}_{-\beta}\}$ corresponds to the set of the weights of all pausers (with multiplicity given by the order) and stoppers of color $a$, and with $\mathfrak{s}^{(a)}_{-}$ the size of this set. 
For the subcrystal ${}^{\sharp}\CC$, one can the define a ``shift" to capture the net degree of the ground state charge function defined in \eqref{psi_vac_0}:
\begin{align}\label{def-sa}
    \mathfrak{s}_a &\equiv \mathfrak{s}^{(a)}_{+} - \mathfrak{s}^{(a)}_{-} \,, 
    \\
   &=   ( \# \textrm{  starters} - \# \textrm{  stoppers}- \# \textrm{ pausers}) \textrm{ of color $a$}\,, 
\end{align}
for any $a\in Q_0$ \cite{Galakhov:2021xum}.\footnote{If we include truncation factors such as $(z+C)$ in the numerators of ground state charge function, as done e.g.\ in \cite{Prochazka:2015deb,Gaberdiel:2018nbs,Li:2020rij,Galakhov:2021xum} in order to study the truncations of these algebras, then this definition becomes $\mathfrak{s}_a\equiv   ( \# \textrm{  starters}- \# \textrm{  stoppers}- \# \textrm{ pausers}  - \# \textrm{  truncation factors}) \textrm{ of color $a$}$; for more on truncations see Section 7 of \cite{Li:2020rij} and Section 3.4 of \cite{Galakhov:2021xum}).}\label{footnotetruncation}

This then translates into the framing of the quiver.
For each subcrystal ${}^{\sharp}\CC$, the corresponding framing is still given by a single framing node $\ff$ of dimension 1, however, the arrows between the framing node $\ff$ and the gauge nodes acquire a richer structure. 
Each factor in the denominator of \eqref{psi_vac_0} (namely each starter) corresponds to an arrow $\alpha\in \{\ff\to a\}$ with weight $h_{\alpha}=\mathfrak{z}^{(a)}_{+\alpha}$, and each factor in the numerator of \eqref{psi_vac_0} (namely each pauser or stopper) corresponds to an arrow $\beta\in \{a\to \ff\}$ with weight $h_{\beta}=-\mathfrak{z}^{(a)}_{-\beta}$ \cite{Galakhov:2021xum}:
\begin{align}\label{psi_vac}
	\varphi^{a\Leftarrow \ff}(z)=\frac{\prod\lm_{I\in\{a\to\ff\}}(z+h_I)}{\prod\lm_{J\in\{\ff\to a\}}(z-h_J)},\quad a\in Q_0\,.
\end{align}
Finally, once the arrows and their weights for the sets
 $\{\ff\to a\}$ and $\{a \to \ff\}$ are determined, one can then add the appropriate correction terms to the super-potential.

In order to allow the subscrystal representations with non-zero shifts, one also needs to allow \emph{shifts} in the mode expansion of the Cartan generators:\footnote{Note that the convention of the mode expansion in \eqref{shift_psi_exp} is slightly different  from \cite{Galakhov:2021xum} (which have $k+1+\mathfrak{s}_a$ in the exponent of $z$). Namely, for \cite{Galakhov:2021xum}, the difference between non-chiral quivers and chiral quivers is captured by the summation range: $\mathbb{N}_0$ for non-chiral quivers and $\mathbb{Z}$ for chiral ones. In this paper, it is more conveninent to use the convention \eqref{shift_psi_exp} that capture both non-chiral and chiral quivers (with no difference even in the summation range), since we will need to contrast the non-chiral and chiral quivers. It is easy to convert to the convention of \cite{Galakhov:2021xum}, without changing any results of this paper. }
\begin{align}\label{shift_psi_exp}
	\psi^{(a)}(z)=\sum\lm_{k=\mathscr{S}_a}^{\infty}\frac{\psi_k^{(a)}}{z^{k}}\,,
\end{align}
with
\begin{align}
    \mathscr{S}_a = \begin{cases} \mathfrak{s}_a \qquad & \textrm{non-chiral quivers}\\
    -\infty \qquad &\textrm{chiral quivers}
    \end{cases}
\end{align}
For \emph{non-chiral} quivers that correspond to toric CY${}_3$ without compact four-cycles, $\mathscr{S}_a=\fs_a$ are constant  since the bond factors \eqref{bond_factor} are homogeneous.
For \emph{chiral} quivers corresponding to toric CY${}_3$ with compact four-cycles, the bond factors are non-homogeneous, therefore there is no bound on the modes in the expansion \eqref{shift_psi_exp}; in other words, the information of the shifts $ \mathfrak{s}_a$ defined for the subcrystal ${}^{\sharp}\CC$ representation via \eqref{def-sa} are lost at the level of the mode expansions of the Cartans \eqref{shift_psi_exp} for the chiral quiver. 

One can use the decomposition procedure of \cite{Galakhov:2021xum} to easily construct crystal modules of $\myY$ supported on crystals of reduced dimensions.
Starting from the canonical crystal $\CC_0$, by placing the stoppers immediately next to the starters, one can obtain a subcrystal ${}^{\sharp}\CC$ with the shape of a 2D (namely one-atom thin) layer of atoms or a 1D chain of atoms.
For a 2D crystal, because we need to stop the crystal from growing in the third direction, for each starter, we need (at least) one stopper (possibly with different colors). 
Therefore the shift of a 2D crystal cannot be positive.
Similarly, for a 1D crystal, we can only have one starter, and since we need to stop the crystal from growing in two directions, we also need (at least) two associate stoppers (possibly with different color); therefore, the shift of an 1D crystal is either $-1$ or $-2$ (depending on whether the 1D crystal is infinitely long or not).

\begin{figure}[htbp]
	\begin{center}
        \begin{tikzpicture}
				\node at (0,0) {$\begin{array}{c}
						\begin{tikzpicture}
							\draw[thick,postaction={decorate},decoration={markings, 
								mark= at position 0.75 with {\arrow{stealth}}}] (0,0) to[out=60,in=0] (0,1) to[out=180,in=120] (0,0);
							\node[above] at (0,1) {$\scriptstyle (Y,\texttt{h}_2)$};
							\begin{scope}[rotate=90]
								\draw[thick,postaction={decorate},decoration={markings, 
									mark= at position 0.75 with {\arrow{stealth}}}] (0,0) to[out=60,in=0] (0,1) to[out=180,in=120] (0,0);
								\node[left] at (0,1) {$\scriptstyle (X,\texttt{h}_1)$};
							\end{scope}
							\begin{scope}[rotate=270]
								\draw[thick,postaction={decorate},decoration={markings, 
									mark= at position 0.75 with {\arrow{stealth}}}] (0,0) to[out=60,in=0] (0,1) to[out=180,in=120] (0,0);
								\node[right] at (0,1) {$\scriptstyle (Z,\texttt{h}_3)$};
							\end{scope}
							\draw[thick,postaction={decorate},decoration={markings, 
								mark= at position 0.65 with {\arrow{stealth}}}] (0,-1.5) -- (0,0) node[pos=0.5,left] {$\scriptstyle (I,0)$};
							\begin{scope}[shift={(0,-1.5)}]
								\draw[fill=black!30!red] (-0.1,-0.1) -- (-0.1,0.1) -- (0.1,0.1) -- (0.1,-0.1) -- cycle;
							\end{scope}
							\draw[fill=\myblue] (0,0) circle (0.1);
							\node[below] at (0,-1.7) {$\scriptstyle W=\Tr\left(X[Y,Z]\right)$};
						\end{tikzpicture}
					\end{array}$};
				\node at (4.5,0) {$\begin{array}{c}
						\begin{tikzpicture}
							\draw[thick,postaction={decorate},decoration={markings, 
								mark= at position 0.75 with {\arrow{stealth}}}] (0,0) to[out=60,in=0] (0,1) to[out=180,in=120] (0,0);
							\node[above] at (0,1) {$\scriptstyle (Y,\texttt{h}_2)$};
							\begin{scope}[rotate=90]
								\draw[thick,postaction={decorate},decoration={markings, 
									mark= at position 0.75 with {\arrow{stealth}}}] (0,0) to[out=60,in=0] (0,1) to[out=180,in=120] (0,0);
								\node[left] at (0,1) {$\scriptstyle (X,\texttt{h}_1)$};
							\end{scope}
							\begin{scope}[rotate=270]
								\draw[thick,postaction={decorate},decoration={markings, 
									mark= at position 0.75 with {\arrow{stealth}}}] (0,0) to[out=60,in=0] (0,1) to[out=180,in=120] (0,0);
								\node[right] at (0,1) {$\scriptstyle (Z,\texttt{h}_3)$};
							\end{scope}
							\draw[thick,postaction={decorate},decoration={markings, 
								mark= at position 0.65 with {\arrow{stealth}}}] (0,-1.5) to[out=100,in=260] node[pos=0.5,left] {$\scriptstyle (I,0)$} (0,0);
							\draw[thick,postaction={decorate},decoration={markings, 
								mark= at position 0.65 with {\arrow{stealth}}}] (0,0) to[out=280,in=80] node[pos=0.5,right] {$\scriptstyle (J,-\texttt{h}_1)$} (0,-1.5);
							\begin{scope}[shift={(0,-1.5)}]
								\draw[fill=black!30!red] (-0.1,-0.1) -- (-0.1,0.1) -- (0.1,0.1) -- (0.1,-0.1) -- cycle;
							\end{scope}
							\draw[fill=\myblue] (0,0) circle (0.1);
							\node[below] at (0,-1.7) {$\scriptstyle W=\Tr\left(X[Y,Z]+{\color{\myblue} XJI}\right)$};
						\end{tikzpicture}
					\end{array}$};
				\node at (9.5,0) {$\begin{array}{c}
						\begin{tikzpicture}
							\tikzset{style0/.style={thin, fill=black!60!green}}
							\tikzset{style1/.style={thin, fill=\myblue}}
							\tikzset{style2/.style={thin, fill=black!30!red}}
							\draw[-stealth] (0,0) -- (0,3.2); 
							\node[left] at (0,3.2) {$\scriptstyle\texttt{h}_3$};
							\begin{scope}[rotate=120]
								\draw[-stealth] (0,0) -- (0,2); 
								\node[left] at (0,2) {$\scriptstyle\texttt{h}_1$};
							\end{scope}
							\begin{scope}[rotate=240]
								\draw[-stealth] (0,0) -- (0,2.5); 
								\node[right] at (0,2.5) {$\scriptstyle\texttt{h}_2$};
							\end{scope}
							\node[black!30!red] at (-0.5,-1.3) {\tiny MacMahon};
							\node[right,black!60!green] at (0.2,3) {\tiny vector};
							\node[right,\myblue] at (0.7,1.5) {\tiny Fock};
							\begin{scope}[scale=0.35]
							\draw[style0] (-0.707107,2.85774) -- (-0.707107,3.67423) -- (0.,3.26599) -- (0.,2.44949) -- cycle;
							\draw[style0] (-0.707107,3.67423) -- (-0.707107,4.49073) -- (0.,4.08248) -- (0.,3.26599) -- cycle;
							\draw[style0] (-0.707107,4.49073) -- (-0.707107,5.30723) -- (0.,4.89898) -- (0.,4.08248) -- cycle;
							\draw[style0] (-0.707107,5.30723) -- (-0.707107,6.12372) -- (0.,5.71548) -- (0.,4.89898) -- cycle;
							\draw[style0] (-0.707107,6.12372) -- (-0.707107,6.94022) -- (0.,6.53197) -- (0.,5.71548) -- cycle;
							\draw[style0] (-0.707107,6.94022) -- (-0.707107,7.75672) -- (0.,7.34847) -- (0.,6.53197) -- cycle;
							\draw[style1] (0.,1.63299) -- (0.,2.44949) -- (0.707107,2.04124) -- (0.707107,1.22474) -- cycle;
							\draw[style1] (0.,2.44949) -- (0.,3.26599) -- (0.707107,2.85774) -- (0.707107,2.04124) -- cycle;
							\draw[style1] (0.,3.26599) -- (0.,4.08248) -- (0.707107,3.67423) -- (0.707107,2.85774) -- cycle;
							\draw[style1] (0.,4.08248) -- (0.,4.89898) -- (0.707107,4.49073) -- (0.707107,3.67423) -- cycle;
							\draw[style1] (0.,4.89898) -- (0.,5.71548) -- (0.707107,5.30723) -- (0.707107,4.49073) -- cycle;
							\draw[style1] (0.707107,0.408248) -- (0.707107,1.22474) -- (1.41421,0.816497) -- (1.41421,0.) -- cycle;
							\draw[style1] (0.707107,1.22474) -- (0.707107,2.04124) -- (1.41421,1.63299) -- (1.41421,0.816497) -- cycle;
							\draw[style1] (0.707107,2.04124) -- (0.707107,2.85774) -- (1.41421,2.44949) -- (1.41421,1.63299) -- cycle;
							\draw[style1] (0.707107,2.85774) -- (0.707107,3.67423) -- (1.41421,3.26599) -- (1.41421,2.44949) -- cycle;
							\draw[style1] (0.707107,3.67423) -- (0.707107,4.49073) -- (1.41421,4.08248) -- (1.41421,3.26599) -- cycle;
							\draw[style1] (1.41421,-0.816497) -- (1.41421,0.) -- (2.12132,-0.408248) -- (2.12132,-1.22474) -- cycle;
							\draw[style1] (1.41421,0.) -- (1.41421,0.816497) -- (2.12132,0.408248) -- (2.12132,-0.408248) -- cycle;
							\draw[style1] (1.41421,0.816497) -- (1.41421,1.63299) -- (2.12132,1.22474) -- (2.12132,0.408248) -- cycle;
							\draw[style1] (1.41421,1.63299) -- (1.41421,2.44949) -- (2.12132,2.04124) -- (2.12132,1.22474) -- cycle;
							\draw[style1] (2.12132,-2.04124) -- (2.12132,-1.22474) -- (2.82843,-1.63299) -- (2.82843,-2.44949) -- cycle;
							\draw[style1] (2.12132,-1.22474) -- (2.12132,-0.408248) -- (2.82843,-0.816497) -- (2.82843,-1.63299) -- cycle;
							\draw[style1] (2.12132,-0.408248) -- (2.12132,0.408248) -- (2.82843,0.) -- (2.82843,-0.816497) -- cycle;
							\draw[style1] (2.82843,-2.44949) -- (2.82843,-1.63299) -- (3.53553,-2.04124) -- (3.53553,-2.85774) -- cycle;
							\draw[style1] (3.53553,-2.85774) -- (3.53553,-2.04124) -- (4.24264,-2.44949) -- (4.24264,-3.26599) -- cycle;
							\draw[style2] (-1.41421,0.816497) -- (-1.41421,1.63299) -- (-0.707107,1.22474) -- (-0.707107,0.408248) -- cycle;
							\draw[style2] (-1.41421,1.63299) -- (-1.41421,2.44949) -- (-0.707107,2.04124) -- (-0.707107,1.22474) -- cycle;
							\draw[style2] (-0.707107,0.408248) -- (-0.707107,1.22474) -- (0.,0.816497) -- (0.,0.) -- cycle;
							\draw[style2] (0.,-0.816497) -- (0.,0.) -- (0.707107,-0.408248) -- (0.707107,-1.22474) -- cycle;
							\draw[style2] (-2.12132,-0.408248) -- (-2.12132,0.408248) -- (-1.41421,0.) -- (-1.41421,-0.816497) -- cycle;
							\draw[style2] (-1.41421,-0.816497) -- (-1.41421,0.) -- (-0.707107,-0.408248) -- (-0.707107,-1.22474) -- cycle;
							\draw[style2] (-0.707107,-2.04124) -- (-0.707107,-1.22474) -- (0.,-1.63299) -- (0.,-2.44949) -- cycle;
							\draw[style2] (0.,-2.44949) -- (0.,-1.63299) -- (0.707107,-2.04124) -- (0.707107,-2.85774) -- cycle;
							\draw[style2] (-2.12132,-2.04124) -- (-2.12132,-1.22474) -- (-1.41421,-1.63299) -- (-1.41421,-2.44949) -- cycle;
							\draw[style2] (-3.53553,-2.04124) -- (-3.53553,-1.22474) -- (-2.82843,-1.63299) -- (-2.82843,-2.44949) -- cycle;
							\draw[style0] (0.707107,6.12372) -- (0.707107,6.94022) -- (0.,6.53197) -- (0.,5.71548) -- cycle;
							\draw[style0] (0.707107,6.94022) -- (0.707107,7.75672) -- (0.,7.34847) -- (0.,6.53197) -- cycle;
							\draw[style1] (1.41421,4.89898) -- (1.41421,5.71548) -- (0.707107,5.30723) -- (0.707107,4.49073) -- cycle;
							\draw[style1] (2.12132,2.85774) -- (2.12132,3.67423) -- (1.41421,3.26599) -- (1.41421,2.44949) -- cycle;
							\draw[style1] (2.12132,3.67423) -- (2.12132,4.49073) -- (1.41421,4.08248) -- (1.41421,3.26599) -- cycle;
							\draw[style1] (2.82843,0.816497) -- (2.82843,1.63299) -- (2.12132,1.22474) -- (2.12132,0.408248) -- cycle;
							\draw[style1] (2.82843,1.63299) -- (2.82843,2.44949) -- (2.12132,2.04124) -- (2.12132,1.22474) -- cycle;
							\draw[style1] (3.53553,-1.22474) -- (3.53553,-0.408248) -- (2.82843,-0.816497) -- (2.82843,-1.63299) -- cycle;
							\draw[style1] (3.53553,-0.408248) -- (3.53553,0.408248) -- (2.82843,0.) -- (2.82843,-0.816497) -- cycle;
							\draw[style1] (4.94975,-2.85774) -- (4.94975,-2.04124) -- (4.24264,-2.44949) -- (4.24264,-3.26599) -- cycle;
							\draw[style2] (0.,1.63299) -- (0.,2.44949) -- (-0.707107,2.04124) -- (-0.707107,1.22474) -- cycle;
							\draw[style2] (0.707107,0.408248) -- (0.707107,1.22474) -- (0.,0.816497) -- (0.,0.) -- cycle;
							\draw[style2] (1.41421,-0.816497) -- (1.41421,0.) -- (0.707107,-0.408248) -- (0.707107,-1.22474) -- cycle;
							\draw[style2] (2.12132,-2.04124) -- (2.12132,-1.22474) -- (1.41421,-1.63299) -- (1.41421,-2.44949) -- cycle;
							\draw[style2] (0.,-0.816497) -- (0.,0.) -- (-0.707107,-0.408248) -- (-0.707107,-1.22474) -- cycle;
							\draw[style2] (1.41421,-2.44949) -- (1.41421,-1.63299) -- (0.707107,-2.04124) -- (0.707107,-2.85774) -- cycle;
							\draw[style2] (-0.707107,-2.04124) -- (-0.707107,-1.22474) -- (-1.41421,-1.63299) -- (-1.41421,-2.44949) -- cycle;
							\draw[style2] (-2.12132,-2.04124) -- (-2.12132,-1.22474) -- (-2.82843,-1.63299) -- (-2.82843,-2.44949) -- cycle;
							\draw[style0] (0.,8.16497) -- (0.707107,7.75672) -- (0.,7.34847) -- (-0.707107,7.75672) -- cycle;
							\draw[style1] (0.707107,6.12372) -- (1.41421,5.71548) -- (0.707107,5.30723) -- (0.,5.71548) -- cycle;
							\draw[style1] (1.41421,4.89898) -- (2.12132,4.49073) -- (1.41421,4.08248) -- (0.707107,4.49073) -- cycle;
							\draw[style1] (2.12132,2.85774) -- (2.82843,2.44949) -- (2.12132,2.04124) -- (1.41421,2.44949) -- cycle;
							\draw[style1] (2.82843,0.816497) -- (3.53553,0.408248) -- (2.82843,0.) -- (2.12132,0.408248) -- cycle;
							\draw[style1] (3.53553,-1.22474) -- (4.24264,-1.63299) -- (3.53553,-2.04124) -- (2.82843,-1.63299) -- cycle;
							\draw[style1] (4.24264,-1.63299) -- (4.94975,-2.04124) -- (4.24264,-2.44949) -- (3.53553,-2.04124) -- cycle;
							\draw[style2] (-0.707107,2.85774) -- (0.,2.44949) -- (-0.707107,2.04124) -- (-1.41421,2.44949) -- cycle;
							\draw[style2] (0.,1.63299) -- (0.707107,1.22474) -- (0.,0.816497) -- (-0.707107,1.22474) -- cycle;
							\draw[style2] (0.707107,0.408248) -- (1.41421,0.) -- (0.707107,-0.408248) -- (0.,0.) -- cycle;
							\draw[style2] (1.41421,-0.816497) -- (2.12132,-1.22474) -- (1.41421,-1.63299) -- (0.707107,-1.22474) -- cycle;
							\draw[style2] (-1.41421,0.816497) -- (-0.707107,0.408248) -- (-1.41421,0.) -- (-2.12132,0.408248) -- cycle;
							\draw[style2] (-0.707107,0.408248) -- (0.,0.) -- (-0.707107,-0.408248) -- (-1.41421,0.) -- cycle;
							\draw[style2] (0.,-0.816497) -- (0.707107,-1.22474) -- (0.,-1.63299) -- (-0.707107,-1.22474) -- cycle;
							\draw[style2] (0.707107,-1.22474) -- (1.41421,-1.63299) -- (0.707107,-2.04124) -- (0.,-1.63299) -- cycle;
							\draw[style2] (-2.12132,-0.408248) -- (-1.41421,-0.816497) -- (-2.12132,-1.22474) -- (-2.82843,-0.816497) -- cycle;
							\draw[style2] (-1.41421,-0.816497) -- (-0.707107,-1.22474) -- (-1.41421,-1.63299) -- (-2.12132,-1.22474) -- cycle;
							\draw[style2] (-2.82843,-0.816497) -- (-2.12132,-1.22474) -- (-2.82843,-1.63299) -- (-3.53553,-1.22474) -- cycle;
							\end{scope}
						\end{tikzpicture}
					\end{array}$};
				\node at (0,-2.65) {(a)};
				\node at (4.5,-2.65) {(b)};
				\node at (9.5,-2.65) {(c)};
				\draw (2.25, -2) -- (2.25,2) (6.75, -2) -- (6.75,2);
			\end{tikzpicture}
	\end{center}
	\caption{(a) Quiver for $\IC^3$, (b) quiver for $\IC^3$ with D4 wrapping $\IC^2$, (c) vector, Fock and MacMahon modules}\label{fig:MacMahon}
\end{figure}
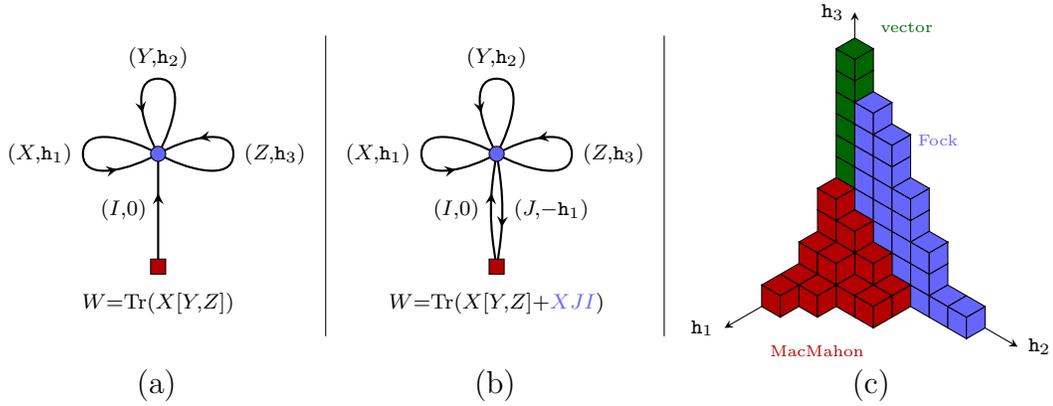

A simple yet dramatic example can be given in the same model of $\IC^3$ parameterized by $(x,y,z)$ (see Figure~\ref{fig:MacMahon}(a)).
A D4 brane wrapping a divisor $\IC^2$ plane spanned by, say, $y$ and $z$, modifies the quiver framing and superpotential (see Figure~\ref{fig:MacMahon}(b)).
As a result the original molten crystals represented by plane partitions and forming \emph{MacMahon} modules are restricted to 2D layer integer partitions (see Figure~\ref{fig:MacMahon}(c)) forming \emph{Fock} modules.
One could go further and restrict to 1D chains---\emph{vector} representations.\footnote{Note that the vector representation has negative shift $\mathfrak{s}=-1$.}

In applications to the integrable models, $R$-matrices and the BAE, the majority of the literature \cite{Feigin:2013fga,feigin2017finite,Garbali:2021qko} works with Fock representations, however MacMahon $R$-matrix constructions are known \cite{Awata:2018svb}.
From our analysis we see no apparent contradiction in implementing 3D crystals in the $R$-matrix construction, however on the level of the BAE we expect certain difficulites.

The most drastic difference between 3D and 2D crystals comes from the fact that in a 3D crystal multiple atoms having distinct R-charge coordinate values may be arranged over the same projection to the weight lattice \eqref{lattice}, whereas in a 2D slice distinct atoms $\Box_1$ and $\Box_2$ always have distinct weights:
\begin{align}
	\mbox{in a 2D crystal }\;h_{\Box_1}\neq h_{\Box_2}\;\mbox{ if }\;\Box_1\neq \Box_2\,.
\end{align}

So in constructing a self-consisting set of BAE we expect the following difficulties:
\begin{itemize}
	\myitem First, since the Bethe vectors are anti-symmetric upon exchanging of Bethe roots (see e.g.\cite[Section~2]{Staudacher:2010jz}), when we have coincident Bethe roots, the resulting Bethe vectors would vanish.
	\myitem 3D molten crystals count cohomologies of (in general) singular quiver varieties (e.g. for $\IC^3$ it is the moduli space of  Hilbert schemes ${\rm Hilb}^n(\IC^3)$ \cite{Rapcak:2018nsl}).
	For 2D crystals there is a chance that the resulting quiver variety is smooth.
	In particular, to cut out 2D subcrystals in the case of $\myY(\widehat{\fg\fl}_m)$ one has to modify the framing and superpotential in such a way that the resulting effective QFT acquires a supersymmetry enhancement, and the resulting quiver variety is a smooth Nakajima variety (see e.g. \cite{2009arXiv0905.0686G}).
	The singularities of the quiver moduli spaces may break the Higgs-Coulomb duality, which is needed in order to associate Bethe roots with vacua of a 2D theory in the Bethe/gauge correspondence (see Section~\ref{sec:Bethe-gauge}). 
	\myitem In Section~\ref{sec:BCtoCrystal}, we associate the locations of the atoms in the weight plane \eqref{lattice} to the solutions to the BAE (namely the Bethe roots) in the large volume limit.
	Bethe roots are further identified with the expectation values of the scalars $\sigma_i^{(a)}$ in classical vacua that contribute to the partition function \eqref{D_2_pf} as saddle points.
	However the integral contains a Vandermonde determinant which becomes singular in the case of coincident $\sigma_i^{(a)}$'s and serves as a repelling term.
	The 3D crystal saddle points are unstable as a result.
\end{itemize}
We hope to return to these issues elsewhere.

\subsubsection{Crystal chains and naive tensor representations}\label{sec:cry_chain}

\begin{figure}[htbp]
	\begin{center}
		\begin{tikzpicture}
			\node at (0,0) {$\begin{array}{c}
					\begin{tikzpicture}
						\draw[postaction={decorate},
						decoration={markings, mark= at position 0.6 with {\arrow{stealth}}}] (-0.7,0) to[out=20,in=160] (0.7,0);
						\draw[postaction={decorate},
						decoration={markings, mark= at position 0.6 with {\arrow{stealth}}}] (0.7,0) to[out=200,in=340] (-0.7,0);
						\draw[postaction={decorate},
						decoration={markings, mark= at position 0.85 with {\arrow{stealth}}}] (0.7,0) to[out=0,in=315] (1,0.3) to[out=135,in=90] (0.7,0);
						\draw[postaction={decorate},
						decoration={markings, mark= at position 0.6 with {\arrow{stealth}}}] (0,-2) -- (-0.7,0);
						\draw[postaction={decorate},
						decoration={markings, mark= at position 0.6 with {\arrow{stealth}}}] (0.7,0) -- (0,-2);
						\draw[postaction={decorate},
						decoration={markings, mark= at position 0.6 with {\arrow{stealth}}}] (1,-2) -- (0.7,0);
						\draw[fill=\myblue] (-0.7,0) circle (0.08) (0.7,0) circle (0.08);
						\begin{scope}[shift={(0,-2)}]
							\draw[fill=white] (-0.15,-0.15) -- (-0.15,0.15) -- (0.15,0.15) -- (0.15,-0.15) -- cycle; 
							\node {$\scriptstyle 3$};
							\node[below] at (0,-0.15) {$\scriptstyle (u_1,u_2,u_3)$};
						\end{scope}
						\begin{scope}[shift={(1,-2)}]
							\draw[fill=white] (-0.15,-0.15) -- (-0.15,0.15) -- (0.15,0.15) -- (0.15,-0.15) -- cycle; 
							\node {$\scriptstyle 1$};
							\node[right] at (0.15,0) {$\scriptstyle (u_4)$};
						\end{scope}
					\end{tikzpicture}
				\end{array}$};
			\node at (6,0) {$\begin{array}{c}
					\begin{tikzpicture}
						\draw[postaction={decorate},
						decoration={markings, mark= at position 0.6 with {\arrow{stealth}}}] (-0.7,0) to[out=20,in=160] (0.7,0);
						\draw[postaction={decorate},
						decoration={markings, mark= at position 0.6 with {\arrow{stealth}}}] (0.7,0) to[out=200,in=340] (-0.7,0);
						\draw[postaction={decorate},
						decoration={markings, mark= at position 0.85 with {\arrow{stealth}}}] (0.7,0) to[out=0,in=315] (1,0.3) to[out=135,in=90] (0.7,0);
						\draw[postaction={decorate},
						decoration={markings, mark= at position 0.6 with {\arrow{stealth}}}] (-1,-2) -- (-0.7,0);
						\draw[postaction={decorate},
						decoration={markings, mark= at position 0.8 with {\arrow{stealth}}}] (0.7,0) -- (-1,-2);
						\draw[postaction={decorate},
						decoration={markings, mark= at position 0.8 with {\arrow{stealth}}}] (0,-2) -- (-0.7,0);
						\draw[postaction={decorate},
						decoration={markings, mark= at position 0.8 with {\arrow{stealth}}}] (0.7,0) -- (0,-2);
						\draw[postaction={decorate},
						decoration={markings, mark= at position 0.8 with {\arrow{stealth}}}] (1,-2) -- (-0.7,0);
						\draw[postaction={decorate},
						decoration={markings, mark= at position 0.6 with {\arrow{stealth}}}] (0.7,0) -- (1,-2);
						\draw[postaction={decorate},
						decoration={markings, mark= at position 0.6 with {\arrow{stealth}}}] (2,-2) -- (0.7,0);
						\draw[fill=\myblue] (-0.7,0) circle (0.08) (0.7,0) circle (0.08);
						\begin{scope}[shift={(-1,-2)}]
							\draw[fill=white] (-0.15,-0.15) -- (-0.15,0.15) -- (0.15,0.15) -- (0.15,-0.15) -- cycle; 
							\node {$\scriptstyle 1$};
							\node[below] at (0,-0.15) {$\scriptstyle (u_1)$};
						\end{scope}
						\begin{scope}[shift={(0,-2)}]
							\draw[fill=white] (-0.15,-0.15) -- (-0.15,0.15) -- (0.15,0.15) -- (0.15,-0.15) -- cycle; 
							\node {$\scriptstyle 1$};
							\node[below] at (0,-0.15) {$\scriptstyle (u_2)$};
						\end{scope}
						\begin{scope}[shift={(1,-2)}]
							\draw[fill=white] (-0.15,-0.15) -- (-0.15,0.15) -- (0.15,0.15) -- (0.15,-0.15) -- cycle; 
							\node {$\scriptstyle 1$};
							\node[below] at (0,-0.15) {$\scriptstyle (u_3)$};
						\end{scope}
						\begin{scope}[shift={(2,-2)}]
							\draw[fill=white] (-0.15,-0.15) -- (-0.15,0.15) -- (0.15,0.15) -- (0.15,-0.15) -- cycle; 
							\node {$\scriptstyle 1$};
							\node[below] at (0,-0.15) {$\scriptstyle (u_4)$};
						\end{scope}
					\end{tikzpicture}
				\end{array}$};
			\node at (3,0) {$=$};
		\end{tikzpicture}
	\end{center}
	\caption{Multidimensional framing}\label{fig:framings}
\end{figure}
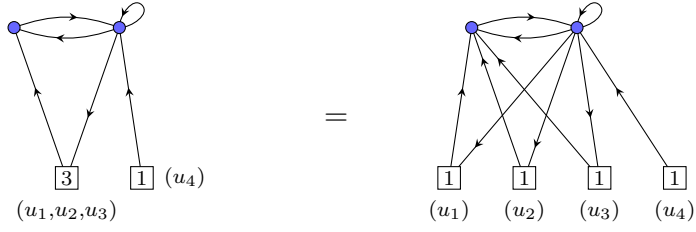

One of the first ingredients needed for the construction of the spin chain is a representation of the BPS algebra.
For our quiver BPS algebras $\myY$, we can use the crystal-melting representations of the algebra discussed in \cite{Li:2020rij,Galakhov:2021xum}.
Here a representation $\mathcal{F}({}^{\sharp}\CC)$ is labeled by the choice of a subcrystal ${}^{\sharp}\CC$ of the canonical crystal,
which in turn can be translated into the choice of the framed quiver.
A more generic quiver framing reduces to a collection of framings $\ff_i$ by a single framing node with dimension 1 (see Figure~\ref{fig:framings}).

The representation space is spanned by states 
\begin{align}
\mathcal{F}({}^{\sharp}\CC): \quad	 \left\{ |\Kappa,{}^{\sharp}\CC\rangle_{u}  \right\}\;,
\end{align}
where $u$ denotes the spectral parameter. 

To construct spin chains we consider a representation  $\mathcal{F}({}^{\sharp}\CC_i)$
for each site $i$, and consider the tensor product $\otimes_i \mathcal{F}({}^{\sharp}\CC_i)$ of the representations.
The resulting crystal can be represented by a composite crystal consisting of mutually independent molten crystals $\Kappa_i$ with centers-of-mass in respective $u_i$ in a common complex plane of weights:
\begin{align}\label{cry_chain}
\begin{array}{c}
	\begin{tikzpicture}
	\begin{scope}[shift={(-2.5,-0.7)}]
	\draw[-stealth] (0,0) -- (0.8,0);
	\draw[-stealth] (0,0) -- (0,0.8);
	\node[right] at (0.8,0){$\scriptstyle \texttt{h}_1$};
	\node[above] at (0,0.8){$\scriptstyle \texttt{h}_2$};
	\end{scope}
		\begin{scope}[scale=0.6]
		\draw (1,0) -- (-1,0) (-0.5, -0.866025) -- (0.5, 0.866025) (0.5, -0.866025) -- (-0.5, 0.866025) (1,0) -- (0.5, 0.866025) -- (-0.5, 0.866025) -- (-1,0) -- (-0.5, -0.866025) -- (0.5, -0.866025) -- cycle;
		\draw[fill=white] (0,0) circle (0.1) (1,0) circle (0.1) (0.5, 0.866025) circle (0.1) (-0.5, 0.866025) circle (0.1) (-1,0) circle (0.1) (-0.5, -0.866025) circle (0.1) (0.5, -0.866025) circle (0.1);
		\end{scope}
		\draw[-stealth] (0.5,-0.9) to[out=150,in=270] (0,-0.2);
		\node[right] at (0.5,-1) {$\scriptstyle u_1$};
		\node[left] at (-0.5,0.4) {$\scriptstyle \Kappa_1$};
		\begin{scope}[shift={(2,0)}]
			\begin{scope}[scale=0.6]
				\draw (1,0) -- (-1,0) (-0.5, -0.866025) -- (0.5, 0.866025) (0.5, -0.866025) -- (-0.5, 0.866025) (1,0) -- (0.5, 0.866025) -- (-0.5, 0.866025) -- (-1,0) -- (-0.5, -0.866025) -- (0.5, -0.866025) -- cycle;
				\draw[fill=white] (0,0) circle (0.1) (1,0) circle (0.1) (0.5, 0.866025) circle (0.1) (-0.5, 0.866025) circle (0.1) (-1,0) circle (0.1) (-0.5, -0.866025) circle (0.1) (0.5, -0.866025) circle (0.1);
			\end{scope}
			\draw[-stealth] (0.5,-0.9) to[out=150,in=270] (0,-0.2);
			\node[right] at (0.5,-1) {$\scriptstyle u_2$};
			\node[left] at (-0.5,0.4) {$\scriptstyle \Kappa_2$};
		\end{scope}
		\begin{scope}[shift={(6,0)}]
			\begin{scope}[scale=0.6]
				\draw (1,0) -- (-1,0) (-0.5, -0.866025) -- (0.5, 0.866025) (0.5, -0.866025) -- (-0.5, 0.866025) (1,0) -- (0.5, 0.866025) -- (-0.5, 0.866025) -- (-1,0) -- (-0.5, -0.866025) -- (0.5, -0.866025) -- cycle;
				\draw[fill=white] (0,0) circle (0.1) (1,0) circle (0.1) (0.5, 0.866025) circle (0.1) (-0.5, 0.866025) circle (0.1) (-1,0) circle (0.1) (-0.5, -0.866025) circle (0.1) (0.5, -0.866025) circle (0.1);
			\end{scope}
			\draw[-stealth] (0.5,-0.9) to[out=150,in=270] (0,-0.2);
			\node[right] at (0.5,-1) {$\scriptstyle u_n$};
			\node[left] at (-0.5,0.4) {$\scriptstyle \Kappa_n$};
		\end{scope}
		\node at (4,0) {$\ldots$};
	\end{tikzpicture}
\end{array}
\end{align}
Each $\Kappa_i$ grows within boundaries of ${}^{\sharp}\CC_i$.
We have chosen a parameterization of complex scalar expectation values associated with $\ff_i$ and denote them as $u_i$.
Note that we assume that those crystals do not interfere with each other, even if 
the projections of the crystals onto the plane become large enough and overlap with each other.

As it was proposed in \cite{Galakhov:2020vyb} (see also \cite{Galakhov:2021vbo}), such a geometric picture finds a natural identification with a tensor product of representations associated with subcrystals ${}^{\sharp}\CC_i$.
We denote a vector in such a representation associated with a disjoint union of molten crystals $\bigsqcup\lm_i\Kappa_i$ as
\begin{align}
	|\Kappa_1,{}^{\sharp}\CC_1\rangle_{u_1}\otimes |\Kappa_2,{}^{\sharp}\CC_2\rangle_{u_2}\otimes\ldots\otimes |\Kappa_n,{}^{\sharp}\CC_n\rangle_{u_n}\,.
\end{align}
When concrete subcrystals ${}^{\sharp}\CC_i$ are irrelevant or apparent from the context, we will omit these letters in the notation.
We call this representation a \emph{crystal chain} by analogy with spin chains.

Following \cite{Galakhov:2021vbo} we define a \emph{naive}\footnote{We call these representations ``naive" since those invoke a simple generalization of matrix coefficient structures \eqref{crystal_rep} to crystal chains.
The mutual atom exchange processes among distinct crystals are not taken into account.} tensor product representation of the quiver BPS algebra on a chain of crystals as 
\begin{align}\label{chain_naiv_co}
	\begin{split}
	\bDelta_0^{(n)}(\psi(z))\,\bigotimes\lm_{i=1}^n|\Kappa_i\rangle_{u_i}=&\prod\lm_i\Psi_{\Kappa_i}(z-u_i)\times\bigotimes\lm_{i}|\Kappa_i\rangle_{u_i}\,,\\
		\bDelta_0^{(n)}(e(z))\,\bigotimes\lm_{i=1}^n|\Kappa_i\rangle_{u_i}=&\sum\lm_i\sum\lm_{\Box\in{\rm Add}\left(\Kappa_i\right)}\prod\lm_{j<i}\Psi_{\Kappa_j}\left(u_i+h_{\Box}-u_j\right)\times \frac{\left[\Kappa_i\to\Kappa_i+\Box\right]}{z-\left(u_i+h_{\Box}\right)}\times\\
		&\bigotimes\lm_{j<i}|\Kappa_j\rangle_{u_j}\otimes |\Kappa_i+\Box\rangle_{u_i}\otimes \bigotimes\lm_{k>i}|\Kappa_k\rangle_{u_k}\,,\\
		\bDelta_0^{(n)}(f(z))\,\bigotimes\lm_{i=1}^n|\Kappa_i\rangle_{u_i}=&\sum\lm_i\sum\lm_{\Box\in{\rm Rem}\left(\Kappa_i\right)}\prod\lm_{k>i}\Psi_{\Kappa_k}\left(u_i+h_{\Box}-u_k\right)\times \frac{\left[\Kappa_i\to\Kappa_i-\Box\right]}{z-\left(u_i+h_{\Box}\right)}\times\\
		&\bigotimes\lm_{j<i}|\Kappa_j\rangle_{u_j}\otimes |\Kappa_i-\Box\rangle_{u_i}\otimes \bigotimes\lm_{k>i}|\Kappa_k\rangle_{u_k}\,,\\
	\end{split}
\end{align}
where we have denoted the generators of the algebra $a\in \myY$ as $\bDelta_0^{(n)}(a)$ to indicate that this is a representation on $\myY^{\otimes n}$.
In what follows we will often consider a chain that consists of only two sites,  and will denote the corresponding $\bDelta_0^{(2)}$ simply as $\bDelta_0$.
For later purposes it is useful to introduce short-hand notations for \eqref{chain_naiv_co}:
	\begin{align}\label{chain_naiv_co_2}
		\begin{split}
			\bDelta_0e&=e\otimes 1+\psi\overset{\scriptstyle\rightarrow}{\otimes}e \,,\\
			\bDelta_0f&=1\otimes f+f\overset{\scriptstyle\leftarrow}{\otimes}\psi \,,\\
			\bDelta_0\psi&=\psi\otimes\psi\,.
		\end{split}
	\end{align}

\subsection{Integrable models}\label{sec:integrable_review}

\subsubsection{\texorpdfstring{$R$-matrices and Yang-Baxter equations}{R-matrices and Yang-Baxter equations}}

As can be seen in the example of \eqref{chain_naiv_co}, 
inside the tensor product the factors may enter in a nonequivalent way:
physically, when we bring the crystals together and obtain a tensor product,
it matters which crystal goes first and which goes next into the tensor product.
Tensor powers with an arbitrary ordering of factor representations, however, are expected to be isomorphic representations of the algebra.
An isomorphism (intertwiner) between two-site chains with opposite orderings is identified with an $R$-matrix:
\begin{align}\label{R-matrix}
	R:\quad \CF_{u_1}\otimes \CF_{u_2}\longrightarrow \CF_{u_2}\otimes \CF_{u_1}\,.
\end{align}

Note that throughout this paper, the $R$-matrix  \eqref{R-matrix} is defined to permute the tensor factors as in \eqref{R-matrix}, as opposed to the conventional definition (see e.g.\ \cite{2018arXiv180407350S}) in which the $R$-matrix map $\CF_{u_1}\otimes \CF_{u_2}$ to $\CF_{u_1}\otimes \CF_{u_2}$:
\begin{align}\label{eq:convRmatrix}
     R^{(\mbox{\tiny conv.})}:\quad \CF_{u_1}\otimes \CF_{u_2}\longrightarrow \CF_{u_1}\otimes \CF_{u_2}\,,
\end{align}
which is related to our definition via 
\begin{align}
     R_{12}^{(\mbox{\tiny ours})}=P_{12}\cdot R_{12}^{(\mbox{\tiny conv.})}\,,
\end{align}
where $P_{12}$ is the permutation operator acting on the two factors in the tensor product.
The $R$-matrix defined as in \eqref{R-matrix} (i.e.\ with the permutation) is sometimes called the \emph{twisted} $R$-matrix.
In this paper, we use the twisted $R$-matrix \eqref{R-matrix} instead of the conventional one \eqref{eq:convRmatrix} because in the context of the Gauge/Bethe correspondence, it is the former one that corresponds directly to the Janus interface, see Section~\ref{sec:interface}.
Throughout this paper, we will omit the adjective ``twisted" in front of the $R$-matrix.

A priori, $R$ is a function of the two spectral parameters, namely $R=R(u_1,u_2)$.
However, throughout this paper, since the ``center-of-mass coordinate" $(u_1+u_2)/2$ corresponds to the fugacity of the global $U(1)$ flavor symmetry of underlying physical system, $R$ becomes a 
function of $u_1-u_2$. 
For the rational/trigonometric/elliptic case, $R(u_1,u_2)$ is a rational function of $(u_1-u_2)$/$\sinh\beta(u_1-u_2)$/$\theta(u_1-u_2)$. For the physical origin of these three factors, see \cite{Galakhov:2021vbo}.

For a tensor product it is simple to generalize this two-site $R$-matrix to an $R$-matrix that permutes the two neighboring factors at position $i$ and $(i+1)$ in the tensor product:
\begin{align}
	\begin{split}
	R_{i,i+1}:\quad &\ldots\otimes \CF_{u_{i-1}}\otimes \CF_{u_i}\otimes \CF_{u_{i+1}}\otimes \CF_{u_{i+2}}\otimes \ldots\longrightarrow\\
	&\ldots\otimes \CF_{u_{i-1}}\otimes \CF_{u_{i+1}}\otimes \CF_{u_i}\otimes \CF_{u_{i+2}}\otimes \ldots\,,\\
	R_{i,i+1}=&\ldots \otimes 1\otimes R\otimes 1\otimes \ldots\,.
	\end{split}
\end{align}

In what follows we will often use a graphical language.
In this language, the tensor factors are depicted as straight strands flowing from the right to the left and ordered according to the ordering in the tensor product, such that the first element corresponds to the top strand.
The matrix element of the operator $R$ is depicted as a simple braid by the following diagram: 
\begin{align}\label{Rm_graph}
	\begin{aligned}
		&{}_{u_2}\langle \Kappa'_2|\otimes{}_{u_1}\langle K'_1|\;R(u_{1}- u_{2})\;|\Kappa_1\rangle_{u_1}\otimes|\Kappa_2\rangle_{u_2}=\begin{array}{c}
			\begin{tikzpicture}[yscale=-1]
				\draw[thick] (-0.8,-0.5) to[out=0,in=225] (0,0) to[out=45,in=180] (0.8,0.5);
				\begin{scope}[yscale=-1]
					\draw[thick] (-0.8,-0.5) to[out=0,in=225] (0,0) to[out=45,in=180] (0.8,0.5);
				\end{scope}
				\draw[-stealth] (0.5,0.7) -- (-0.5,0.7);
				\node[below] at (-0.5,0.7) {$\scriptstyle s$};
				\node[right] at (0.8,-0.5) {$\scriptstyle (\Kappa_1,u_1)$};
				\node[right] at (0.8,0.5) {$\scriptstyle (\Kappa_2,u_2)$};
				\node[left] at (-0.8,-0.5) {$\scriptstyle (\Kappa_2',u_2)$};
				\node[left] at (-0.8,0.5) {$\scriptstyle (\Kappa'_1,u_1)$};
			\end{tikzpicture}
		\end{array}
	\end{aligned}
\end{align}
where the adiabatic flow is directed from the right to the left.

Since an $R$-matrix permutes the factors in a tensor product, it should give a representation of the permutation group; therefore we impose the following constraints on its matrix elements.
\begin{enumerate}
    \item The $R$-matrix $R$ should satisfy the Yang-Baxter equations (YBE),
\begin{align}
	R_{12}(u_{12}) \, R_{13}(u_{13}) \, R_{23}(u_{23})= R_{23}(u_{23}) \, R_{13}(u_{13}) \, R_{12}(u_{12})
\end{align}
with $u_{ij}\equiv u_i -u_j$. These equations can be represented by the following graph:
\begin{align}\label{YBE}
	\begin{array}{c}
		\begin{tikzpicture}[xscale=0.6,yscale=-0.8]
			\draw[thick] (1.2,-1.5) to[out=180,in=0] (-1.2,0.5) (1.2,-0.5) to[out=180,in=0] (-1.2,-0.5) (1.2,0.5) to[out=180,in=0] (-1.2,-1.5);
			\node[right] at (1.2,-1.5) {$\scriptstyle u_1$};
			\node[right] at (1.2,-0.5) {$\scriptstyle u_2$};
			\node[right] at (1.2,0.5) {$\scriptstyle u_3$};
			\node[left] at (-1.2,-1.5) {$\scriptstyle u_3$};
			\node[left] at (-1.2,-0.5) {$\scriptstyle u_2$};
			\node[left] at (-1.2,0.5) {$\scriptstyle u_1$};
		\end{tikzpicture}
	\end{array}=
	\begin{array}{c}
		\begin{tikzpicture}[xscale=0.5,yscale=-0.8]
			\draw[thick] (-0.8,-0.5) to[out=0,in=225] (0,0) to[out=45,in=180] (0.8,0.5);
			\begin{scope}[yscale=-1]
				\draw[thick] (-0.8,-0.5) to[out=0,in=225] (0,0) to[out=45,in=180] (0.8,0.5);
			\end{scope}
			\begin{scope}[shift={(-1.6,-1)}]
				\draw[thick] (-0.8,-0.5) to[out=0,in=225] (0,0) to[out=45,in=180] (0.8,0.5);
				\begin{scope}[yscale=-1]
					\draw[thick] (-0.8,-0.5) to[out=0,in=225] (0,0) to[out=45,in=180] (0.8,0.5);
				\end{scope}
			\end{scope} 
			\begin{scope}[shift={(1.6,-1)}]
				\draw[thick] (-0.8,-0.5) to[out=0,in=225] (0,0) to[out=45,in=180] (0.8,0.5);
				\begin{scope}[yscale=-1]
					\draw[thick] (-0.8,-0.5) to[out=0,in=225] (0,0) to[out=45,in=180] (0.8,0.5);
				\end{scope}
			\end{scope} 
			\draw[thick] (-0.8,-1.5) -- (0.8,-1.5) (-0.8,0.5) -- (-2.4,0.5) (0.8,0.5) -- (2.4,0.5);
			\node[right] at (2.4,-1.5) {$\scriptstyle u_1$};
			\node[right] at (2.4,-0.5) {$\scriptstyle u_2$};
			\node[right] at (2.4,0.5) {$\scriptstyle u_3$};
			\node[left] at (-2.4,-1.5) {$\scriptstyle u_3$};
			\node[left] at (-2.4,-0.5) {$\scriptstyle u_2$};
			\node[left] at (-2.4,0.5) {$\scriptstyle u_1$};
		\end{tikzpicture}
	\end{array}=\begin{array}{c}
		\begin{tikzpicture}[xscale=0.5,yscale=0.8]
			\draw[thick] (-0.8,-0.5) to[out=0,in=225] (0,0) to[out=45,in=180] (0.8,0.5);
			\begin{scope}[yscale=-1]
				\draw[thick] (-0.8,-0.5) to[out=0,in=225] (0,0) to[out=45,in=180] (0.8,0.5);
			\end{scope}
			\begin{scope}[shift={(-1.6,-1)}]
				\draw[thick] (-0.8,-0.5) to[out=0,in=225] (0,0) to[out=45,in=180] (0.8,0.5);
				\begin{scope}[yscale=-1]
					\draw[thick] (-0.8,-0.5) to[out=0,in=225] (0,0) to[out=45,in=180] (0.8,0.5);
				\end{scope}
			\end{scope} 
			\begin{scope}[shift={(1.6,-1)}]
				\draw[thick] (-0.8,-0.5) to[out=0,in=225] (0,0) to[out=45,in=180] (0.8,0.5);
				\begin{scope}[yscale=-1]
					\draw[thick] (-0.8,-0.5) to[out=0,in=225] (0,0) to[out=45,in=180] (0.8,0.5);
				\end{scope}
			\end{scope} 
			\draw[thick] (-0.8,-1.5) -- (0.8,-1.5) (-0.8,0.5) -- (-2.4,0.5) (0.8,0.5) -- (2.4,0.5);
			\node[right] at (2.4,-1.5) {$\scriptstyle u_3$};
			\node[right] at (2.4,-0.5) {$\scriptstyle u_2$};
			\node[right] at (2.4,0.5) {$\scriptstyle u_1$};
			\node[left] at (-2.4,-1.5) {$\scriptstyle u_1$};
			\node[left] at (-2.4,-0.5) {$\scriptstyle u_2$};
			\node[left] at (-2.4,0.5) {$\scriptstyle u_3$};
		\end{tikzpicture}
	\end{array}\,,
\end{align}
\item
Furthermore, the $R$-matrix satisfies the unitarity constraint:
\begin{align}
	R_{12}(u_{12})\, R_{21}(u_{21})=1\,,
\end{align}
which can be represented as
\begin{align}\label{unitarity}
	\begin{array}{c}
		\begin{tikzpicture}[xscale=0.6,yscale=-0.8]
			\draw[thick] (-0.8,-0.5) to[out=0,in=225] (0,0) to[out=45,in=180] (0.8,0.5);
			\begin{scope}[yscale=-1]
				\draw[thick] (-0.8,-0.5) to[out=0,in=225] (0,0) to[out=45,in=180] (0.8,0.5);
			\end{scope}
			\begin{scope}[shift={(-1.6,0)}]
				\draw[thick] (-0.8,-0.5) to[out=0,in=225] (0,0) to[out=45,in=180] (0.8,0.5);
				\begin{scope}[yscale=-1]
					\draw[thick] (-0.8,-0.5) to[out=0,in=225] (0,0) to[out=45,in=180] (0.8,0.5);
				\end{scope}
			\end{scope} 
			\node[right] at (0.8,-0.5) {$\scriptstyle u_1$};
			\node[right] at (0.8,0.5) {$\scriptstyle u_2$};
			\node[left] at (-2.4,-0.5) {$\scriptstyle u_1$};
			\node[left] at (-2.4,0.5) {$\scriptstyle u_2$};
		\end{tikzpicture}
	\end{array}=\begin{array}{c}
		\begin{tikzpicture}[xscale=0.6,yscale=-0.8]
			\draw[thick] (-1,-0.5) -- (1,-0.5) (-1,0.5) -- (1,0.5);
			\node[right] at (1,-0.5) {$\scriptstyle u_1$};
			\node[right] at (1,0.5) {$\scriptstyle u_2$};
			\node[left] at (-1,-0.5) {$\scriptstyle u_1$};
			\node[left] at (-1,0.5) {$\scriptstyle u_2$};
		\end{tikzpicture}
	\end{array}\,.
\end{align}
\end{enumerate}
\subsubsection{Transfer matrices, Bethe vectors, and Bethe ansatz equations} 

Let us consider a tensor product of $n$ crystal modules $\CF_{u_i}$ associated with some fixed framing $\ff$:
\begin{align}
	\mathscr{F}:=\CF_{u_1}\otimes \CF_{u_2}\otimes \ldots\otimes \CF_{u_n}\,.
\end{align}

To compute traces we introduce a grading operator acting on crystal representations as:
\begin{align}
	{\bf s}|\Kappa\rangle=\left(\prod\lm_{a\in Q_0}\fq_a^{|\Kappa^{(a)}|}\right)|\Kappa\rangle\,,
\end{align}
for some fugacities $|\fq_a|<1$.
This operator allows one to introduce the notion of a character on a crystal representation, so that
\begin{equation}
\Tr'\,1=\Tr\,{\bf s}=\sum\lm_{\Kappa}\fq_a^{|\Kappa^{(a)}|}
\end{equation}
represents a partition function associated with the crystal module---a generating function for the corresponding Donaldson-Thomas invariants.
One expects this partition function to be convergent inside the unit ball $|\fq_a|<1$. 
In particular, for the case of $\myY(\widehat{\fg\fl}_1)$ there is a single color $\fq_a=\fq$ and we have the canonical results: 
\begin{equation}
\Tr_{\rm Fock}'\,1=\prod\lm_{k\geq 1}\frac{1}{\left(1-\fq^k\right)},\quad \Tr_{\rm MacMahon}'\,1=\prod\lm_{k\geq 1}\frac{1}{\left(1-\fq^k\right)^k}\,.
\end{equation}
Graphically we can denote the action of the $\bf s$ operator in the following way:
\begin{align}
	{\bf s}=\begin{array}{c}
		\begin{tikzpicture}
			\draw[thick] (-0.5,0) -- (0.5,0);
			\draw[fill=white] (0,0) circle (0.08);
		\end{tikzpicture}
	\end{array}\,.
\end{align}

The $R$-matrices only reshuffle atoms between crystals and intertwine the grading operator. 
As a result, we derive the following relation between the $\bf s$-operators and the $R$-matrices:
\begin{align}\label{s-R}
	\begin{array}{c}
		\begin{tikzpicture}[xscale=0.6,yscale=0.6]
			\draw[thick] (-0.8,-0.5) to[out=0,in=225] (0,0) to[out=45,in=180] (0.8,0.5);
			\begin{scope}[yscale=-1]
				\draw[thick] (-0.8,-0.5) to[out=0,in=225] (0,0) to[out=45,in=180] (0.8,0.5);
			\end{scope}
			\draw[thick] (0.8,0.5) -- (1.2,0.5) (0.8,-0.5) -- (1.2,-0.5);
			\draw[fill=white] (0.8,0.5) circle (0.1) (0.8,-0.5) circle (0.1);
		\end{tikzpicture}
	\end{array}=\begin{array}{c}
		\begin{tikzpicture}[xscale=0.6,yscale=0.6]
			\draw[thick] (-0.8,-0.5) to[out=0,in=225] (0,0) to[out=45,in=180] (0.8,0.5);
			\begin{scope}[yscale=-1]
				\draw[thick] (-0.8,-0.5) to[out=0,in=225] (0,0) to[out=45,in=180] (0.8,0.5);
			\end{scope}
			\begin{scope}[xscale=-1]
				\draw[thick] (0.8,0.5) -- (1.2,0.5) (0.8,-0.5) -- (1.2,-0.5);
				\draw[fill=white] (0.8,0.5) circle (0.1) (0.8,-0.5) circle (0.1);
			\end{scope}
		\end{tikzpicture}
	\end{array}\,.
\end{align}

The transfer matrix
\begin{align}
	T(z):\quad \mathscr{F}\longrightarrow \mathscr{F}
\end{align}
can then be constructed in the canonical way as an iterated braiding with an auxiliary module $\CF_z$, whose strand we label as 0$^{\textrm{th}}$:
\begin{align}\label{transfer_m}
	\begin{split}
	T_{\bf s}(z):&=\Tr_{\CF_z}\;\left({\bf s}_{\CF_z}\otimes 1_{\mathscr{F}}\right)\,R_{0n}(z-u_n)\ldots R_{01}(z-u_1)=\\
	&={\color{black!20!blue}\sum\lm_{\Kappa}}\begin{array}{c}
		\begin{tikzpicture}
			\draw[thick] (0,0) -- (3,0);
			\draw[dashed] (0,-0.3) -- (3,-0.3);
			\draw[thick] (0,-0.6) -- (3,-0.6);
			\draw[thick] (0,-0.9) -- (3,-0.9);
			\node[right] at (3,0) {$\scriptstyle u_n$};
			\node[right] at (3,-0.6) {$\scriptstyle u_2$};
			\node[right] at (3,-0.9) {$\scriptstyle u_1$};
			\node[left] at (0,0) {$\scriptstyle u_n$};
			\node[left] at (0,-0.6) {$\scriptstyle u_2$};
			\node[left] at (0,-0.9) {$\scriptstyle u_1$};
			\draw[thick, black!20!blue] (3,-1.4) -- (2.5,-1.4) to[out=180,in=0] (0.5,0.5) -- (0,0.5);
			\node[right, black!20!blue] at (3,-1.4) {$\scriptstyle (\Kappa,z)$};
			\node[left, black!20!blue] at (0,0.5) {$\scriptstyle (\Kappa,z)$};
			\draw[black!20!blue,fill=white] (0.25,0.5) circle (0.08);
		\end{tikzpicture}
	\end{array}\,.
	\end{split}
\end{align}

Applying the YBE \eqref{YBE}, the unitarity constraint \eqref{unitarity}, and the relation \eqref{s-R}, one can show that the transfer matrices with different values of the spectral parameter commute:
\begin{align}
	\left[T(z_1),T(z_2)\right]=0\,.
\end{align}
This is the hallmark for integrability since the expansion coefficients $I_k$ of the transfer matrix at $z\to \infty$:
\begin{align}
    T(z)=\frac{I_0}{z}+\frac{I_1}{z^2}+\frac{I_2}{z^3}+\ldots
\end{align}
form an infinite set of mutually commuting integrals of motion for the integrable model in question.

A canonical way to solve this system is to use the Bethe vectors $|B\rangle$ to solve the eigen-value problem for the transfer matrix:
\begin{align}\label{eigen_value}
	T(z)|B\rangle=\Lambda(z)|B\rangle\,,
\end{align}
where the Bethe vector $|B\rangle$  is independent of $z$.
In this procedure, the canonical way of constructing the Bethe vectors (see e.g. \cite{Faddeev:1996iy,2018arXiv180407350S}) is via the algebraic Bethe ansatz, where $|B\rangle$ is generated from the lowest weight vector in the module $\mathscr{F}$ by applying the off-diagonal $R$-matrix elements.
The consistency condition of such an ansatz with \eqref{eigen_value} leads to a set of equations on the spectral parameters of the $R$-matrices, called the \emph{Bethe Ansatz Equations} (BAE).

\subsection{Gauge/Bethe correspondence} \label{sec:Bethe-gauge}

In this paper we will be interested in a 2D $\CN=(2,2)$ quiver gauge theory associated with a toric CY${}_3$.
(We will later in Section~\ref{sec:trig} discuss trigonometric models associated with 3D $\CN=2$ quiver gauge theories on $S^1$.)
As we have described already, the gauge/matter content of this theory is defined by the quiver $Q$.
To each node $a\in Q_0$, we associate a gauge group $U(N_a)$, where $N_a$ is the corresponding dimension, and to each arrow $I: a\rightarrow b$, we associate a bi-fundamental field with respect to $\overline{U(N_a)}\times U(N_b)$.

The twisted superpotential of the theory is given by
\begin{align}\label{W_LG}
	\begin{split}
    &\myW(\vec\sigma)=\I \sum\lm_{a\in Q_0}\sum\lm_{i=1}^{N_a}t_a\sigma_i^{(a)}+\sum\lm_{(I:a\to b)\in Q_1}\sum\lm_{i=1}^{N_a}\sum\lm_{j=1}^{N_b}{\bf w}\left(\sigma_j^{(b)}-\sigma_i^{(a)}-h_I\right)\,,\\
	&{\bf w}(\sigma)=\sigma\left(\log\,\sigma-1\right)\,.
	\end{split}
\end{align}
on which we will comment further in Section~\ref{sec:Higgs-Coulomb}.
The vacuum equation \eqref{eq.BAE} for the twisted superpotential $\myW$ is a set of algebraic equations
\begin{align}
   \textrm{exp}[\p_{\sigma^{(a)}_i}\myW]=1
\end{align}
for the expectation values of the twisted chiral fields $\sigma^{(a)}_i$. After some algebra the equations can be rewritten 
into the following form:
\begin{tcolorbox}[ams equation]
\begin{split}\label{BAE_factor}
&1=\BAE^{(a)}_i\left(\vec \sigma,\vec u,\vec \fq\right):=\fq_a^{-1}\prod\lm_{\substack{1\leq j\leq N_a\\ j \neq i}}\varphi^{a\Leftarrow a}\left(\sigma_i^{(a)}-\sigma_j^{(a)}\right)\times\\
&\qquad\qquad\qquad\times\prod\lm_{\substack{b\in Q_0\\ b\neq a}}\prod\lm_{k=1}^{N_b}\varphi^{a\Leftarrow b}\left(\sigma_i^{(a)}-\sigma_k^{(b)}\right)\prod\lm_{\ff}\varphi^{a\Leftarrow\ff}\left(\sigma_i^{(a)}-u_{\ff}\right)\,,
\end{split}
\end{tcolorbox}
\noindent where the function $\varphi$ is the bond factor defined in \eqref{bond_factor}, the product in the last term runs over all the 1-dimensional quiver framings (see Figure~\ref{fig:framings}, for the bonding of gauge and framing quiver nodes see \eqref{psi_vac}), and $\fq_a:=e^{\I t_a}$.
For the vacuum configuration of $\sigma$'s equation \eqref{BAE_factor} is satisfied for all $a\in Q_0$, $i=1,\ldots,N_a$.

Equations \eqref{BAE_factor} have a form reminiscent of the Bethe ansatz equations.
Indeed the Gauge/Bethe correspondence \cite{Nekrasov:2009uh,Nekrasov:2009ui}
states that the equations \eqref{BAE_factor} can be identified with the 
BAE of some integrable model. As emphasized in the introduction,
this is a highly non-trivial statement. The goal of this paper is to 
check this statement for a large class of supersymmetric gauge theories 
associated with toric Calabi-Yau three-folds.

\section{Integrable models from quiver BPS algebras} \label{sec:coproduct}

In Section~\ref{sec:cry_rep} we discussed crystal-melting representations of the quiver Yangians.
In crystal chain language, this represents a single site of the crystal chain.
To construct a crystal chain (and to further discuss the $R$-matrix and BAE) we need to combine these representations into 
suitable tensor-product representations.  As we will explain further below, one of the systematic methods to achieve this is to 
consider a coproduct in the algebra. It turns out, however, that it is non-trivial to identify the coproduct
relevant for our BAE.

In this section, we will explain how to define a non-trivial coproduct structure for the quiver BPS algebras. 
We will focus on the rational case in this section, and postpone the discussion of the trigonometric case to Section~\ref{sec:trig}.

Our discussion in the following partly depends on QFT considerations.
Since we have in mind readers with mixed backgrounds,
in this section we will try to make the presentation understandable
without detailed knowledge of QFT, and postpone the discussions from the gauge-theory viewpoints to Section~\ref{sec:gauge_deriv}. Consequently,
we will state a few assumptions in this section,
which will be better motivated further in Section~\ref{sec:gauge_deriv}.
Note that even without going into the details of the gauge-theory discussions in Section~\ref{sec:gauge_deriv},
the fact that we have successfully reproduced 
BAE for unshifted quiver Yangians for non-chiral quivers in Section~\ref{sec:BAE}
provides strong evidence for the validity of our discussions. 
We will also provide further motivations when we discuss the trigonometric case in Section~\ref{sec:trig}.

\subsection{Non-diagonal coproducts of quiver BPS algebras} \label{sec:co-mul}

Let us denote the quiver BPS algebra as $\myY$. We are going to consider a crystal chain associated with crystal-melting representations of this algebra,
and further discuss the Bethe ansatz equations.

As stated already, we are interested in coproducts of the algebra,
since the coproduct allows one to construct chains of crystals as tensor powers of old ones.
For some quiver BPS algebras the coproducts are already in the literature:
for example, toroidal quiver BPS algebras have a known coproduct,
and in the rational/Yangian case the simplest algebra---the affine Yangian $\myY(\widehat{\mathfrak{gl}}_1)$---is known to have a coproduct.\footnote{Proposals for coproducts for more exotic instances of algebras like $\myY(\widehat{\fg\fl}_n)$, $\myY(\widehat{\fs\fl}_{m|n})$, $\myY(\widehat{\fg\fl}_{m|n})$ could be found in  recent papers \cite{guay2018coproduct,2019arXiv191106666U,Bao:2022fpk} respectively.} 
We also expect the quiver BPS algebra to have a coproduct---an algebra homomorphism:
	\begin{align}
		\Delta:\quad \myY\longrightarrow \myY\otimes \myY\,.
	\end{align}
When the coproduct $\Delta$ satisfies the co-associativity condition
	\begin{align}
		(\Delta\otimes 1)\circ \Delta=(1\otimes \Delta )\circ \Delta\,,
	\end{align}
the iterated coproduct:
	\begin{align}
		\begin{split}
		\Delta^{(n)}:\quad &\myY\longrightarrow \myY^{\otimes n}\,,\\
		&\ldots\circ(1\otimes 1\otimes\Delta)\circ(1\otimes \Delta)\circ\Delta\,,
		\end{split}
	\end{align}
is actually independent of the order in which the products are taken.
	
Let us also consider a representation:
	\begin{align}
		{\rm Rep}:\quad \myY\otimes \myY\longrightarrow {\rm End}\left(\CF\otimes \CF\right)\,,
	\end{align}
	where $\CF$ are $\myY$-modules associated with some subcrystals ${}^{\sharp}\CC_{1,2}$.
We can now evaluate the coproduct in the representation, to obtain a new structure $\bDelta$:
	\begin{align}\label{copr_rep_diag}
		\begin{array}{c}
			\begin{tikzpicture}
				\node(A) at (0,0) {$\myY$};
				\node(B) at (3,0) {$\myY\otimes \myY$};
				\node(C) at (6,0) {${\rm End}\left(\CF\otimes \CF\right)$};
				\draw[-stealth] (A.east) -- (B.west) node[pos=0.5,above] {$\scriptstyle \Delta$};
				\draw[-stealth] (B.east) -- (C.west) node[pos=0.5,above] {$\scriptstyle {\rm Rep}$};
				\draw[-stealth] (A.south) to[out=330,in=180] (3,-1) to[out=0,in=210] (C.south);
				\node[below] at (3,-1) {$\scriptstyle \bDelta={\rm Rep}\,\circ\,\Delta$};
			\end{tikzpicture}
		\end{array}
	\end{align}

In what follows we will say that a structure $\bSigma\in {\rm End}(\CF\otimes\CF)$ \emph{factorizes} if
\begin{equation}\label{split}
    \exists\; \Sigma\in \myY\otimes\myY\quad\mbox{s.t.}\quad \bSigma={\rm Rep}\circ \Sigma\,,
\end{equation}
otherwise we say $\bSigma$ \emph{does not factorize}.

The naive crystal chain representation \eqref{chain_naiv_co} is a natural candidate for the coproduct representation of the quiver BPS algebra, since $\bDelta_0$ is an algebra homomorphism by construction.
However if we look closer at the structure of \eqref{chain_naiv_co} we will find that $\bDelta_0$ does not factorize as in \eqref{copr_rep_diag}.
Indeed, for example, $\bDelta_0e(z)$ has an element that has a phase $\Psi_{\Kappa_1}(h_{\Box}+u_2-u_1)$, where $\Kappa_1$ is the first crystal in the crystal chain and $h_{\Box}+u_2$ is the coordinate of the added atom to crystal $\Kappa_2$ in the weight plane.
So there is no way to untangle operators from the first and the second factor in the coproduct (see details in Appendix~\ref{sec:splitting}). 
Thus we conclude that $\bDelta_0$ \emph{does not factorize}. 

In addition to this problem, we will later see that the $R$-matrix $R^{(0)}$ originating from $\bDelta_0$
does not give rise to the vacuum equation for gauge theories (see \eqref{R-diag} and \eqref{recurrence}), and hence is not relevant for the Gauge/Bethe correspondence.

We therefore conclude that the coproduct $\Delta$ relevant for the Gauge/Bethe correspondence,
when evaluated as  $\bDelta: \myY \to  {\rm End}\left(\CF\otimes \CF \right)$  on crystal chains,
is different from $\bDelta_0$.
	Note that \emph{both} $\bDelta_0$ and $\bDelta$ are valid representations of $\myY$ on a crystal chain consisting of two sites. 
	It is not a simple task to invent a new representation, therefore it seems to be natural to assume that $\bDelta_0$ and $\bDelta$ are \emph{isomorphic} representations:
\begin{assumption}
  $\bDelta_0$ and $\bDelta$ are isomorphic representations.
  In other words, we have 
  	\begin{align}\label{iso}
		\bDelta \cdot U=U\cdot \bDelta_0\,,
	\end{align}
     where $U$ is a function of the spectral parameters $u_1$ and $u_2$ that correspond to the two factors $\CF$ in the tensor product.
\end{assumption}	
From a representation-theory perspective, one expects such a decomposition of the coproduct 
from the Gauss decomposition of the universal $R$-matrix along the lines of \cite{MR1134942}.
We will provide further motivations for this assumption in Section~\ref{sec:stab_bas}.

Let us assign the following degrees for the generators in $\myY\otimes \myY$:
	\begin{align}\label{degrees}
		\begin{array}{lllll}
			{\rm deg}\,\left(e\otimes 1\right) =-1,	& & {\rm deg}\,\left(f\otimes 1\right) =+1,	& & {\rm deg}\,\left(\psi\otimes 1 \right)=0\,;\\
			{\rm deg}\,\left(1\otimes e\right) =+1,	& & {\rm deg}\,\left(1\otimes f\right) =-1,	& & {\rm deg}\,\left(1\otimes \psi \right)=0\,.\\
		\end{array}
	\end{align}
	If we order the vectors in $\CF\otimes \CF$ by the number of atoms in the second factor, this grading corresponds to, in the matrix representation of the operator, how far a nonzero matrix element is from the main diagonal.
	Having established the grading we could employ a filtration on expressions in $\myY^{\otimes 2}=\myY\otimes \myY$.
	In what follows we will say that some expression is defined modulo $\myY^{\otimes 2}_k$ implying that we consider an expansion up to degree $k$.
	
For physical reasons we discuss in Section~\ref{sec:stab_bas}, the matrix $U$ is expected to be \emph{lower}-triangular. Since this is important for our subsequent discussions, let us state this as an assumption:	
\begin{assumption}
  The matrix $U$ is lower-triangular.
  \end{assumption}	
	
Since $U$ gives a homomorphism of representations, all the eigenvalues of $U$ are equal to 1.
Summarizing, one could reflect these facts about the map $U$ in the following expansion:
	\begin{align}\label{U_op}
		U=1\otimes 1+\sum\lm_{k=1}^{\infty}S_k,\quad {\rm deg}\,S_k=2k\,.
	\end{align}
	
Defining a truncated $U$ matrix by:
	\begin{align}
		U_n \equiv 1\otimes 1+\sum\lm_{k=1}^n S_k\,,
	\end{align} 
we have:
	\begin{align}
		U^{-1}=U_n^{-1}-S_{n+1}\; {\rm mod}\; \myY_{2n+4}^{\otimes 2}\,.
	\end{align}

Using \eqref{iso} and \eqref{chain_naiv_co_2} we derive the following equations:
	\begin{align}
		\begin{split}
			\bDelta e =U_n\cdot\bDelta_0e\cdot U_n^{-1}+\left[S_{n+1},e\otimes 1\right]\;{\rm mod}\;\myY_{2n+3}^{\otimes 2}\,,\\
			\bDelta f =U_n\cdot\bDelta_0f\cdot U_n^{-1}+\left[S_{n+1},1\otimes f\right]\;{\rm mod}\;\myY_{2n+3}^{\otimes 2}\,,\\
			\bDelta\psi=U_n\cdot\bDelta_0\psi\cdot U_n^{-1}+\left[S_{n+1},\psi\otimes \psi\right]\;{\rm mod}\;\myY_{2n+4}^{\otimes 2}\,.
		\end{split}
	\end{align}
	
Requiring that $\bDelta$ originates from a true coproduct $\Delta$  (so that we can explicitly factorize $\bDelta={\rm Rep}\circ \Delta$) leads to a set of recurrence relations for $S_n$:
	\begin{align}
		\begin{split}
			U_n\cdot\bDelta_0 e\cdot U_n^{-1}\Big|_{{\rm deg}=2n+1}+\left[S_{n+1},e\otimes 1\right]\quad\mbox{factorizes}\,,\\
			U_n\cdot\bDelta_0 f\cdot U_n^{-1}\Big|_{{\rm deg}=2n+1}+\left[S_{n+1},1\otimes f\right]\quad\mbox{factorizes}\,,\\
			U_n\cdot\bDelta_0 \psi\cdot U_n^{-1}\Big|_{{\rm deg}=2n+2}+\left[S_{n+1},\psi\otimes \psi\right]\quad\mbox{factorizes}\,.
		\end{split}
	\end{align}
	
If we can find a solution to these equations such that the resulting $\Delta$ is associative, we have constructed a true coproduct structure on the quiver Yangians.
Let us write down the first few levels explicitly.
	
	For level 1 we have:
	\begin{align}\label{cop_lvl1}
		\begin{split}
			\psi\overset{\scriptstyle\rightarrow}{\otimes}e+\left[S_{1},e\otimes 1\right]\quad\mbox{factorizes}\,,\\
			f\overset{\scriptstyle\leftarrow}{\otimes}\psi+\left[S_{1},1\otimes f\right]\quad\mbox{factorizes}\,,\\
			\left[S_{1},\psi\otimes \psi\right]\quad\mbox{factorizes}\,.
		\end{split}
	\end{align}
	
For level 2 we have:
	\begin{align}
		\begin{split}
			\left[S_1,\psi\overset{\scriptstyle\rightarrow}{\otimes}e\right]+\left[e\otimes 1,S_1\right]S_1+\left[S_2,e\otimes 1\right]\quad\mbox{factorizes}\,,\\
			\left[S_1,f\overset{\scriptstyle\leftarrow}{\otimes}\psi\right]+\left[1\otimes f,S_1\right]S_1+\left[S_2,1\otimes f\right]\quad\mbox{factorizes}\,,\\
			\left[\psi\otimes \psi,S_1\right]S_1+\left[S_2,\psi\otimes \psi\right]\quad\mbox{factorizes}\,.\\
		\end{split}
	\end{align}
	
It is simple to derive $S_1$ satisfying these conditions for example from a QFT consideration as in Section~\ref{sec:corrections}:
	\begin{align}\label{1-solit}
		\begin{split}
		S_1\,|\Kappa_1\rangle_{u_1}\otimes|\Kappa_2\rangle_{u_2}&=\sum\lm_{a\in Q_0} (-1)^{|a|+1}
		\sum\lm_{\sqbox{$a$}_1\in{\rm Rem}(\Kappa_1)}\sum\lm_{\sqbox{$a$}_2\in{\rm Add}(\Kappa_2)}\\
		&\frac{\left[\Kappa\to\Kappa-\sqbox{$a$}_1\right]\left[\Kappa'\to\Kappa'+\Box'\right]}{z+h_{\Box}-h_{\Box'}}|\Kappa-\Box\rangle_{x_1}\otimes|\Kappa'+\Box'\rangle_{x_2}\,.
		\end{split}
	\end{align}
	While we have QFT motivations for this expression, we can state this as an assumption for more mathematically-oriented readers:
	\begin{assumption}
$S_1$ (in the expansion \eqref{U_op} of $U$) is given by \eqref{1-solit}.
  \end{assumption}	
	
The computation of higher corrections $S_{k\geq 2}$ is rather involved.
	We compute all the corrections $S_k$ for the known coproduct structure for $Y(\widehat{\fg\fl}_1)$ in Appendix~\ref{app:solit_corr_gl_1}, and we compute $S_2$ and the coproduct for $Y(\widehat{\fg\fl}_{m|n})$ up to $\myY_5^{\otimes 2}$ terms in Appendix~\ref{app:solit_corr_gl_m|n}.
	
	If we apply the explicit form \eqref{1-solit} of $S_1$ we derive:\footnote{Here we apply the canonical decomposition of a rational function:
	$$
	f(z):=\frac{\prod\lm_{i=1}^m(z-p_i)}{\prod\lm_{j=1}^n(z-q_j)}=\sum\lm_{k=0}^{m-n}f_{-k}z^k+\sum\lm_{j=1}^n\frac{\mathop{\rm res}\lm_{w=q_j}f(w)}{z-q_j},\mbox{ where } f_{-k}=-\mathop{\rm res}\lm_{w=\infty} f(w)/w^{k+1}\,.
	$$
	}
	\begin{align}\label{coprod}
	\begin{split}
		\Delta e(z)&=\Delta_1e(z)-\sum\lm_{k\geq 1}\sum\lm_{j=0}^{k-1}z^{k-1-j}\psi_{-k}\otimes e_j\quad{\rm mod}\,\myY_3^{\otimes 2}\,,\\
		\Delta f(z)&=\Delta_1f(z)-\sum\lm_{k\geq 1}\sum\lm_{j=0}^{k-1}z^{k-1-j}f_j\otimes \psi_{-k}\quad{\rm mod}\,\myY_3^{\otimes 2}\,,\\
		\Delta \psi(z)&=\Delta_1\psi(z)\quad{\rm mod}\,\myY_2^{\otimes 2}\,,
	\end{split}
	\end{align}
	where $\Delta_1$ is defined by
	\begin{align}\label{Delta_1}
		\begin{split}
			\Delta_1 e(z)&=e(z)\otimes 1+\psi(z)\otimes e(z)\quad \in \myY\otimes \myY\,,\\
			\Delta_1 f(z)&=1\otimes f(z)+f(z)\otimes \psi(z)\quad \in \myY\otimes \myY\,,\\
			\Delta_1\psi(z)&=\psi(z)\otimes \psi(z)\quad \in \myY\otimes \myY\,.
		\end{split}
	\end{align}

The expression for $\Delta_1$ is reminiscent of the coproducts for quantum groups and related algebras in the literature. 
We note, however, that $\Delta_1$ is not an algebra homomorphism for quiver Yangians, see Appendix~\ref{sec:app:Delta1}.
This is an interesting subtlety in our discussion of the rational case.
The situation is different for the trigonometric case, namely quantum toroidal quiver BPS algebras (which we will discuss in more detail in Section~\ref{sec:trig} and Appendix~\ref{sec:splitting}):
there the trigonometric analogue for $\Delta_1$, which we denote as  $\ddot \Delta_1$ in the notation of Section~\ref{sec:trig}, is a legitimate coproduct, and 
the trigonometric analogue of $\bDelta_0$, $\ddot \bDelta_0$,
factorizes as $\ddot\bDelta_0={\rm Rep}\circ \ddot\Delta_1$ \cite{feigin2017finite}.
In this paper we will nevertheless continue to call  $\ddot\bDelta_0$ ``naive'', since there is another coproduct structure related to $\bDelta_0$ by a Miki automorphism (as will be discussed further in Appendix~\ref{app:Miki}). We expect that in the degeneration to the rational case the Miki automorphism in the trigonometric case reduces to our map $U$.

Before we end this subsection, let us summarize the properties of various coproducts that we have encountered; we have also included their trigonometric counterparts for comparison:\footnote{The ``n/a'' (not applicable) are there since the question of factorization only concerns $\bDelta_0$ and $\bDelta$,  see \eqref{copr_rep_diag} and \eqref{split}, whereas $\Delta_1$ and $\Delta$ are  elements of $\myY\otimes\myY$ by definition.
}
\begin{center}\label{table:comparison}\renewcommand{\arraystretch}{1.5}
    \begin{tabular}{ | c | c | c | c | c|c|c |c|c|}
    \hline
     & $\bDelta_{0}$ & $\ddot\bDelta_{0}$ & $\Delta_{1}$ & $\ddot\Delta_{1}$& $\bDelta$ & $\ddot\bDelta$ & $\Delta$ & $\ddot\Delta$\\ \hline
 \textrm{factorize?}    & $\mytimes$ & \checkmark & n/a & n/a & \checkmark &\checkmark & n/a & n/a
     \\ \hline
     \textrm{algebra homomorphism?}    & \checkmark & \checkmark & $\mytimes$ & \checkmark& \checkmark & \checkmark & \checkmark & \checkmark
     \\ \hline
      \textrm{non-diagonal?}    & $\mytimes$ & $\mytimes$ & $\mytimes$ & $\mytimes$ & \checkmark & \checkmark & \checkmark & \checkmark
     \\ \hline
    \end{tabular}
\end{center}

\subsection{\texorpdfstring{(Twisted) $R$-matrices}{(Twisted) R-matrices}} \label{sec:Rmatrices}

Now we can discuss the (twisted) $R$-matrices.
It is expected that the $R$-matrix intertwines the coproduct structure:\footnote{
Note that since the $R$-matrix used in this paper is actually the so-called twisted $R$-matrix (see definition \eqref{R-matrix}), the relation between our twisted $R$-matrix and the coproduct is given by \eqref{intertw}. 
As a comparison, the $R$-matrix defined conventionally (see definition \eqref{eq:convRmatrix}) is related to the coproduct by 
$$
	R^{(\mbox{\tiny conv.})}\cdot \Delta=\Delta^{\rm op}\cdot R^{(\mbox{\tiny conv.})}\,,
$$
in which
$
	\Delta^{\rm op}\equiv P\cdot \Delta\cdot P\,,
$
where $P$ is the permutation operator acting on the two modules.
}
\begin{align}\label{intertw}
	R\cdot \Delta=\Delta\cdot R\,.
\end{align}

Let us note that the naive crystal chain representation \eqref{chain_naiv_co} also acts on tensor factors in a nonequivalent way.
Therefore we can construct a ``naive" $R$-matrix $R^{(0)}$ intertwining $\bDelta_0$:
\begin{align}\label{intertw_naive}
	R^{(0)}\cdot \bDelta_0=\bDelta_0\cdot R^{(0)}\,.
\end{align}
It is quite simple to derive an explicit expression for $R^{(0)}$ from \eqref{intertw_naive} in a crystal representation:
\begin{align}\label{R-diag}
	R^{(0)}(u_{12})\,|\Kappa_1\rangle_{u_1}\otimes|\Kappa_2\rangle_{u_2}=\rho_{\Kappa_1,\Kappa_2}(u_{12})\,|\Kappa_2\rangle_{u_2}\otimes|\Kappa_1\rangle_{u_1}\,,
\end{align}
where the scalar phase $\rho$ satisfies the recurrent relations:
\begin{align}\label{recurrence}
	\begin{split}
	\rho_{\Kappa_1+\sqbox{$a$},\Kappa_2}(z)&=\Psi^{(a)}_{\Kappa_2}\left(z+h_{\sqbox{$a$}}\right)\,\rho_{\Kappa_1,\Kappa_2}(z)\,,\\ \rho_{\Kappa_1,\Kappa_2+\sqbox{$a$}}(z)&=\Psi^{(a)}_{\Kappa_1}(h_{\sqbox{$a$}}-z)^{-1}\rho_{\Kappa_1,\Kappa_2}(z)\,.	\end{split}
\end{align}
This $R^{(0)}$ satisfies YBE \eqref{YBE} and the unitarity constraint \eqref{unitarity}.

From the relation between $\Delta$ and $\Delta_0$, namely
\begin{align}
    \bDelta=U(z)\cdot \bDelta_0\cdot U(z)^{-1}\,,
\end{align}
we see that the $R$-matrix $R$ and the naive $R$-matrix $R^{(0)}$ are related in the following way:
\begin{tcolorbox}[ams equation]
\label{main_R}
	R(z)=U(-z)\cdot R^{(0)}(z)\cdot U(z)^{-1}\,.
\end{tcolorbox}
	
\subsection{Lax operators in terms of quiver BPS algebras} \label{sec:LaxOperators}

Having the $R$-matrix expression we can construct the so-called $L$-operators acting on a representation by fixing two of four legs in the $R$-matrix.
In other words, these operators are constructed as a braiding of the unknown crystal with the known one:
\begin{align}\label{Lax_graph}
	\CL_{\Kappa',\Kappa}^{(\mya)}(u):=\begin{array}{c}
		\begin{tikzpicture}
			\draw[ultra thick] (0.8,0) -- (-0.8,0);
			\draw[thick] (0.8,0.4) to[out=180, in=0] (-0.8,-0.4);
			\node[right] at (0.8,0.4) {$\scriptstyle (\Kappa,u)$};
			\node[left] at (-0.8,-0.4) {$\scriptstyle (\Kappa',u)$};
		\end{tikzpicture}
	\end{array}
\end{align}
One can translate this pictorial notation into a relation between matrix elements of the Lax operator $\CL_{\Kappa',\Kappa}$ and those of the $R$-matrices as:
\begin{align}
	{}_{v}\langle P_2|\CL_{\Kappa',\Kappa}(u)|P_1\rangle_v:={}_v\langle P_2|\otimes{}_u\langle\Kappa'|\,R(u-v)\,|\Kappa\rangle_u\otimes|P_1\rangle_{v}
	\,.
\end{align}

The graphical depiction of the Lax operators \eqref{Lax_graph} suggests a natural definition of algebraic operators acting on crystal chains.
In the graphical notation we identify the tensor powers with sheaves of strands, therefore to go from the definition of Lax operators $\CL$ to their higher coproducts, we simply replace a single strand by a sheaf of $n$ strands:
\begin{align}
	\Delta^{(n)}\CL_{\Kappa',\Kappa}(u):=\begin{array}{c}
		\begin{tikzpicture}
			\draw[ultra thick] (0.8,-0.2) -- (-0.8,-0.2) (0.8,0.2) -- (-0.8,0.2);
			\draw[dashed] (0.8,0) -- (-0.8,0);
			\draw[thick] (0.8,0.6) to[out=180, in=0] (-0.8,-0.6);
			\node[right] at (0.8,0.6) {$\scriptstyle (\Kappa,u)$};
			\node[left] at (-0.8,-0.6) {$\scriptstyle (\Kappa',u)$};
		\end{tikzpicture}
	\end{array}\,.
\end{align}

The coproduct structure in terms of Lax operators acquires an exceptionally simple form:
\begin{align}\label{Lax_co}
	\Delta\CL_{\Kappa',\Kappa}(u)=\sum\lm_{\Kappa''}\CL_{\Kappa'',\Kappa}(u)\otimes \CL_{\Kappa',\Kappa''}(u)\,.
\end{align}

We can re-express the action of the Lax operators in terms of the quiver BPS algebra generators using relation \eqref{main_R}.

To do so let us introduce some simplifying notations.
We will denote a crystal with a fixed number $\alpha\in\IN$ of atoms as $Q_{\alpha}$.
Thus an additional grading for matrix elements of $R$ can be introduced:
\begin{align}
	\begin{split}
		R=\sum\lm_{n\in\IZ} R_n,\quad
		\langle \Kappa',Q_{\beta}'|R_n|Q_{\alpha},\Kappa\rangle\sim\delta_{n,\alpha-\beta}\,.
	\end{split}
\end{align}
We find from \eqref{main_R} (where we use the short hand notation $r=R^{(0)}$, $s_k=S_k(z)$, $s_{-k}=S_k(-z)$):
\begin{align}\label{R-mat}
	\begin{split}
		R_{2}&=r s_{1}^2-r s_{2}\,,\\
		R_{1}&=s_{-1}r s_{1}^2-r s_{1}-s_{-1}r s_{2}\,,\\
		R_{0}&=r-s_{-1}rs_{1}-s_{-2}rs_{2}+s_{-2}rs_{1}^2\,,\\
		R_{-1}&=s_{-1}r-s_{-2}rs_{1}\,,\\
		R_{-2}&=s_{-2}r\,.
	\end{split}
\end{align}

In particular, for the matrix elements we have:
\begin{align}
	\begin{split}
		\langle\Kappa'|\CL_{\varnothing,\varnothing}(z)|\Kappa\rangle&=\rho_{\varnothing,\Kappa}(z)\,\delta_{\Kappa,\Kappa'}\,,\\ \langle\Kappa'|\CL_{\varnothing,\Box}(z)|\Kappa\rangle&=-\rho_{\varnothing,\Kappa'}(z)\,\langle\varnothing,\Kappa'|S_1(z)|\Box,\Kappa\rangle\,,\\
		\langle\Kappa'|\CL_{\Box,\varnothing}(z)|\Kappa\rangle&=\rho_{\varnothing,\Kappa}(z)\,\langle\Kappa',\Box|S_1(-z)|\Kappa,\varnothing\rangle\,,\\
		\langle\Kappa'|\CL_{Q_1',Q_1}(z)|\Kappa\rangle&=\rho_{Q_1,\Kappa}(z)\,\delta_{\Kappa,\Kappa'}\delta_{Q_1,Q_1'}-\\
		&\quad \quad-\langle\Kappa',Q_1'|S_1(-z)R^{(0)}(z)S_1(z)|Q_1,\Kappa\rangle\,.
	\end{split}
\end{align}

Using relations \eqref{1-solit} and \eqref{recurrence} one finds:
\begin{align}\label{Lax_BPS}
	\begin{split}
		\CL_{\varnothing,\varnothing}&=\myt_{\ff}(z)\,,\\
		\CL_{\varnothing,\sqbox{$a$}}&= \left[\sqbox{$a$}\to\varnothing\right]\times \myt_{\ff}(z)\,e^{(a)}\left(z+h_{\sqbox{$a$}}\right)\,,\\
		\CL_{\sqbox{$a$},\varnothing}&=- \left[\varnothing\to \sqbox{$a$}\right]\times f^{(a)}\left(z+h_{\sqbox{$a$}}\right)\,\myt_{\ff}(z)\,,\\
		\CL_{\sqbox{$a$}',\sqbox{$a$}}&=\myt_{\ff}(z)\,\psi^{(a)}(z+h_{\sqbox{$a$}})+\\
		&+\left[\varnothing\to \sqbox{$a$}'\right]\left[\sqbox{$a$}\to \varnothing\right]\times f^{(a)}\left(z+h_{\sqbox{$a$}'}\right)\,\myt_{\ff}(z)\,e^{(a)}\left(z+h_{\sqbox{$a$}}\right)\,.
	\end{split}
\end{align}
Here the operator $t_{\ff}(z)$ (the operator $h$ in the notation of \cite{Litvinov:2020zeq}) depends on the framing and hence on the subcrystal ${}^{\sharp}\CC$ associated with the fixed crystal site that is braided in the construction of Lax operators \eqref{Lax_graph}.
In the crystal basis it acquires the expectation value:
\begin{align}
	t_{\ff}(z)|\Kappa\rangle=\rho_{\varnothing,\Kappa}(z)|\Kappa\rangle=\left(\prod\lm_{a\in Q_0}\prod\lm_{\sqbox{$a$}\in\Kappa}\varphi^{a\Leftarrow \ff}\left(h_{\sqbox{$a$}}-z\right)\right)^{-1}|\Kappa\rangle\,.
\end{align}

The operator $\myt_{\ff}(z)$ can be considered as a set of even Cartan operators in addition to $\psi^{(a)}(z)$; and it has the following commutation relations with the raising/lowering operators (cf. \eqref{QiuvYangian}):
\begin{align}
\begin{split}
    \myt_{\ff}(z)\,\mye^{(a)}(w) &\simeq \varphi^{a\Leftarrow \ff}\left(w-z\right)^{-1} \mye^{(b)}(w)\, \myt_{\ff}(z)\,,\\
    \myt_{\ff}(z)\,\myf^{(a)}(w) &\simeq \varphi^{a\Leftarrow \ff}\left(w-z\right) \myf^{(b)}(w)\, \myt_{\ff}(z)\,.
\end{split}
\end{align}

To conclude this section, let us consider in a similar fashion higher operators of type $\CL_{\Kappa,\varnothing}(z)$.
First we note:
\begin{align}
	S_{k \geq 1}\,|\varnothing\rangle\otimes|\Kappa\rangle=0\quad\Rightarrow \quad U(z)^{-1}|\varnothing\rangle\otimes|\Kappa\rangle=|\varnothing\rangle\otimes|\Kappa\rangle\,.
\end{align}	

Applying \eqref{main_R} one derives:
\begin{align}
	\langle\Kappa'|\CL_{Q_n,\varnothing}|\Kappa\rangle=\langle \Kappa',Q_n|S_n(-z)|\Kappa,\varnothing\rangle\cdot\rho_{\varnothing,\Kappa}(z)\,.
\end{align}	

An expansion similar to \eqref{1-solit} is expected for $S_k$ as well (see Section~\ref{sec:corrections}):
\begin{align}\label{S_k_form}
	S_k\sim \underbrace{f\cdot f\cdot f\cdot \ldots}_{k\;{\rm times}}\otimes \underbrace{e\cdot e\cdot e\cdot \ldots}_{k\;{\rm times}}\,.
\end{align}
This expansion is compatible with the degree assignment \eqref{degrees}.

Using the form \eqref{S_k_form} of $S_n$, its matrix element can be represented in the following way:
\begin{align}
	\begin{split}
	\langle \Kappa',Q_n|S_n(-z)|\Kappa,\varnothing\rangle&=\oint d{\vec y}\; F_{Q_n}(\vec y,z)\times\\
	&\langle \Kappa'|\prod\lm_{b\in Q_0}f^{(b)}\left(y_1^{(b)}\right)f^{(b)}\left(y_2^{(b)}\right)\ldots f^{(b)}\left(y_{|Q_n^{(b)}|}^{(b)}\right)|\Kappa\rangle\,,
	\end{split}
\end{align}
where $F$ is some function depending on the test crystal $Q_n$ and all the other spectral parameters.

Using this representation for $\CL_{\Kappa,\varnothing}(z)$ one finds:
\begin{align}\label{L-rep}
	\CL_{\Kappa,\varnothing}(u)=\oint d{\vec y}\; F_{\Kappa}(\vec y,u)\;\left[\prod\lm_{b\in Q_0}f^{(b)}\left(y_1^{(b)}\right)f^{(b)}\left(y_2^{(b)}\right)\ldots f^{(b)}\left(y_{|\Kappa^{(b)}|}^{(b)}\right)\right]\myt_{\ff}(u)\,.
\end{align}

It is not complicated to generalize these relations to tensor powers of representations. 
For brevity we will denote the higher coproducts in bold font:
\begin{align}
	(\Delta^{(m-1)}\CL)_{\Kappa,\Kappa'}(u)=:\coL_{\Kappa,\Kappa'}(u), \quad 	(\Delta^{(m-1)}e)^{(a)}(u)=:\coe^{(a)}(u),\mbox{ and so on,}
\end{align}
so that we have:
\begin{align}
	\coL_{\Kappa,\varnothing}(u)=\oint d{\vec y}\; F_{\Kappa,a}(\vec y,u)\;\left[\prod\lm_{b\in Q_0}\cof^{(b)}\left(y_1^{(b)}\right)\cof^{(b)}\left(y_2^{(b)}\right)\ldots \cof^{(b)}\left(y_{|\Kappa^{(b)}|}^{(b)}\right)\right]{\bf \myt}_{\ff}(u)\,,
\end{align}
Note that the dependence of the bold font variables on this $m$ is omitted and it can always be read off from the context.

\section{Gauge theory derivation of Gauge/Bethe correspondence} \label{sec:gauge_deriv}

In this section, we discuss the gauge theory derivation of the Gauge/Bethe correspondence. 
In addition to having its own interest, this section serves as a motivation for the assumptions made in 
Section~\ref{sec:coproduct}.

\subsection{Disk partition functions and Higgs-Coulomb duality} \label{sec:Higgs-Coulomb}

As we will discuss soon in Section~\ref{sec:interface}, the discussion of the Gauge/Bethe correspondence requires 
studying interfaces between gauge theories, i.e.\ gauge theories on manifolds with boundaries.

For this purpose we discuss disk (hemisphere) partition function \cite{Hori:2013ika,Honda:2013uca,Sugishita:2013jca}
for the two-dimensional $\mathcal{N}=(2,2)$ gauged linear sigma model (GLSM) \cite{Witten:1993yc} given by the quiver data.
We label the boundary condition at the equator $\p D$ as $\fB$.
The localization technique allows us to reduce the path integral to a finite integral over the Cartan elements
$\sigma_i^{(a)}$, $i=1,\ldots,N_a$ of the complexified gauge group:
\begin{align}
    \prod\lm_{a\in Q_0} GL(N_a,\IC)\,.
\end{align}
The resulting partition function reads \cite{Hori:2013ika,Honda:2013uca,Sugishita:2013jca}:
\begin{align}\label{D_2_pf}
	\begin{split}
		Z_{D^2}(\fB)=\int\lm_{-\I \infty}^{\I\infty} d\vec\sigma\;\prod\lm_{a\in Q_0}\Delta_V^{(a)}(\vec \sigma)\; e^{\I\sum\lm_{a,i}t_a\sigma_i^{(a)}}\prod\lm_{I\in Q_1}\mathscr{G}_I(\vec \sigma)\times \CO_{\fB}(\vec\sigma)\,,
	\end{split}
\end{align}
where $t_a$ are the complexified FI parameters, $\Delta_V^{(a)}$ are variants of Vandermonde determinants:
\begin{align}
    \Delta_V^{(a)}(\vec \sigma)=\prod\lm_{i<j}\left(\sigma^{(a)}_i-\sigma^{(a)}_j\right)\,\sin\left(\sigma^{(a)}_i-\sigma^{(a)}_j\right)\,,
\end{align}
and $\mathscr{G}_I$ is the one-loop contribution of the chiral field associated with the arrow $I: a\rightarrow b$:
\begin{align}
    \mathscr{G}_{I:a\to b}(\vec \sigma)=\prod\lm_{i=1}^{N_a}\prod\lm_{j=1}^{N_b}\Gamma\left(\sigma_j^{(b)}-\sigma_i^{(a)}-h_I\right)\,.
\end{align}
The effect of the boundary condition can be captured by the corresponding observable $\CO_{\fB}$.

To make contact with the Bethe Ansatz equations and integrable models,
we recall that the GLSM for toric Calabi-Yau manifolds has a mirror dual given by the Landau-Ginzburg model with superpotential $\myW$ \cite{Hori:2000kt}.
In the gauge theory language (cf.\ \cite{Galakhov:2020vyb}), we can call the GLSM description the ``Higgs branch'' description, and the LG description as the ``Coulomb branch'' description,
so that the mirror symmetry can be formulated as the Higgs-Coulomb duality.

In the disk partition function, this is reflected in the asymptotic behavior of the partition function \eqref{D_2_pf}
where we apply Stirling's approximation to the gamma functions (see \cite{Gomis:2012wy} for a related discussion for the two-sphere (i.e.\ without boundary)):
\begin{align}\label{LG}
	\begin{split}
	&Z_{D^2}(\CL)\sim\int\lm_{\CL}d\vec\sigma\; e^{\myW(\vec\sigma)} \,.\\
	\end{split}
\end{align}
Here the expression $\myW(\vec\sigma)$ coincides with the effective twisted superpotential of the GLSM introduced in \eqref{W_LG},
whose extremization gives the vacuum equation \eqref{BAE_factor} (i.e.\ the would-be BAE).

In the expression \eqref{LG} we have substituted the effect of the boundary operator $\CO_{\fB}$
by the choice of the integration cycle $\CL$. While a detailed analysis of this effect is not needed for this paper,
let us comment that the substitution of $\fB$ by $\CL$ is the statement anticipated by the mirror
symmetry \cite{Hori:2000ck}: in the GLSM description, $\fB$ is given by coherent sheaves on the toric CY$_{3}$,
whereas in the dual LG description, the integration cycle $\CL$ is defined by a brane wrapping the special Lagrangian submanifold that defines the boundary conditions on the boundary of the disk.
The statement of mirror symmetry should then be that the two different sets of partition functions, $Z_{D^2}(\fB)$ and $Z_{D^2}(\CL)$, should span the same ``vector space of partition functions''.
Indeed, the choice of the basis for the vector space is far from unique, and there exist several natural choices of the basis, each with its own motivation. We will discuss this in the next section.


\subsection{From disk boundary conditions to crystal states} \label{sec:BCtoCrystal}

That we can choose a linear basis for the vector space of the partition functions is expected from general considerations.
The choice of the boundary condition at the boundary of the disk $D$ in either the Higgs phase \eqref{D_2_pf} or the Coulomb phase \eqref{LG} has an internal structure of a triangulated category \cite{D_book_2}. 
From this viewpoint, the partition function provides a functor to a vector space -- the Grothendieck group of the corresponding category.
Under this map, the boundary conditions inherit the linear structure of the space of the partition functions.  

For the system in question, there are five possible choices, summarized by the following pentagram:
\begin{align}
	\begin{array}{c}
			\begin{tikzpicture}
				\draw[fill=white!97!blue] let \n1={1.4}, \n2={0.5} 
				in (\n1*0.,\n1*1.) to[out=180,in=36] (\n2*-0.587785,\n2*0.809017) to[out=216,in=72] (\n1*-0.951057,\n1*0.309017) to[out=252,in=108] (\n2*-0.951057,\n2*-0.309017) to[out=288,in=144] (\n1*-0.587785,\n1*-0.809017) to[out=324,in=180] (\n2*0.,\n2*-1.) to[out=360,in=216] (\n1*0.587785,\n1*-0.809017) to[out=396,in=252] (\n2*0.951057,\n2*-0.309017) to[out=432,in=288] (\n1*0.951057,\n1*0.309017) to[out=468,in=324] (\n2*0.587785,\n2*0.809017) to[out=504,in=360] (\n1*0.,\n1*1.);
				\begin{scope}[scale=1.4]
					\node[above] at (0., 1.)  {(a) Bethe roots};
					\node[left] at (-0.951057, 0.309017)  {$\begin{array}{c}
							\mbox{(b) Jacobian}\\ \mbox{ring }\CO \end{array}$};
					\node[below left] at (-0.587785, -0.809017) {$\begin{array}{c}
							\mbox{(c) Fixed points $\Kappa$ on}\\ \mbox{equivariant moduli space} \end{array}$};
					\node[below right] at (0.587785, -0.809017)  {$\begin{array}{c}
							\mbox{(d) Coherent}\\ \mbox{sheaves $\fB$} \end{array}$};
					\node[right] at (0.951057, 0.309017)  {$\begin{array}{c}
							\mbox{(e) Lagrangian}\\ \mbox{submanifold }\CL \end{array}$};
				\end{scope}
			\node {$\vec Z$};
			\end{tikzpicture}
		\end{array}
\end{align}

Since these choices are simply different ways to choose a basis, they should be related by simple linear maps acting on $\vec Z$.
In the physical picture of IR dynamics, these maps correspond to dualities between different effective descriptions of the same system.
Let us list a few of these dualities:
\begin{itemize}
	\myitem For the gauged linear sigma model with a target space given by the Higgs branch $X_{\rm Higgs}$, the basis corresponds to a set of exceptional objects in the derived category of coherent sheaves (d) on $X_{\rm Higgs}$ \cite{D-book_1,D_book_2}.
	\myitem Mirror symmetry maps the partition function to the partition function of the Landau-Ginzburg model with the effective superpotential $\myW$, and coherent sheaves (d) to Lagrangian submanifolds (e).
	A choice of exceptional objects among the Lagrangian submanifolds is given by a basis of Lefschetz thimbles $\CL$ (e).
	\myitem A Lefschetz thimble (e) can be defined as a union of all Hamiltonian flow trajectories starting from the classical vacua---defined as the solutions to the vacuum equations \eqref{BAE_factor},
	which we hope to view as an analogue of the Bethe ansatz equations (a).\footnote{As we will explain in detail below, although the Bethe ansatz equations \eqref{BAE_factor} are well-defined for any toric CY${}_3$, there might be no apparent underlying integrable model.}
	\myitem Since we consider a GLSM with twisted masses (i.e.\ flavor fugacities), there is an equivariant action of the flavor group on the moduli space, forcing the geometric description to localize the geometry computations on the Higgs branch (d) to a set of fixed points (c).
	\myitem The set of insertions of operators $\CO$ (b) in the Landau-Ginzburg partition function is given by the Jacobian ring $\CJ=\IC[{\rm fileds}]/d\myW$ for the superpotential $\myW$. The basis in this ring corresponds naturally to classical vacua (a) since two operators $\CO_1$ and $\CO_2$ are equivalent for all vacua values if and only if they are equivalent in the ring $\CJ$.
	\myitem On the other hand, a natural choice of the insertion operators $\CO$ (b) for GLSM is given by the cohomolgy ring generated by the Chern classes (brane factors in the terminology of \cite{Hori:2013ika}) of coherent sheaves (d).
	\myitem Eventually, both the fixed points on $X_{\rm Higgs}$ (c) and the classical vacua in the Landau-Ginzburg model (a) are vacua in the dual IR descriptions of the same system.
	These descriptions are related by the 2D mirror symmetry, therefore the descriptions of vacua are related accordingly.
	Projecting crystal atoms to the weight plane, we derive expectation values of scalars $\langle\sigma_i^{(a)}\rangle$ in the gauge multiplet.
	These VEVs have to solve the Bethe ansatz equations as well when the mirror symmetry is applicable, i.e. in the large volume limit ${\rm Re}\,t_a\to +\infty$.
\end{itemize} 

This is a long list of dualities and we will use only some of them.
First we note that a choice of basis in the partition function vector space could be fixed as Lefschetz thimble integration cycles for \eqref{LG}.
These integration cycles are labeled by solutions to \eqref{BAE_factor}, and therefore they are dual to the choices of fixed points \eqref{fp}---molten crystals.
In other words a crystal $\Kappa$ (or a crystal chain for multiple framings) with numbers of atoms $\sqbox{$a$}$ given by $N_a$ defines a valid boundary condition for our theory on $D$, and, therefore the partition function:
\begin{align}
	\Kappa\to Z_{\Kappa}\,.
\end{align}
We will discuss issues with this identification and their consequences in Section~\ref{sec:breakdown}. 
In this section we assume nevertheless that molten crystals give a complete classification of boundary conditions for \eqref{D_2_pf} and \eqref{LG}.

\subsection{\texorpdfstring{From Janus interfaces to twisted $R$-matrices}{From Janus interfaces to twisted R-matrices}} \label{sec:interface}

The $R$-matrix involves an adiabatic continuation of the parameters of the theory.
In general, a natural physical realization of the parallel transport of a system along some parameter space is related to the notion of a Janus interface.

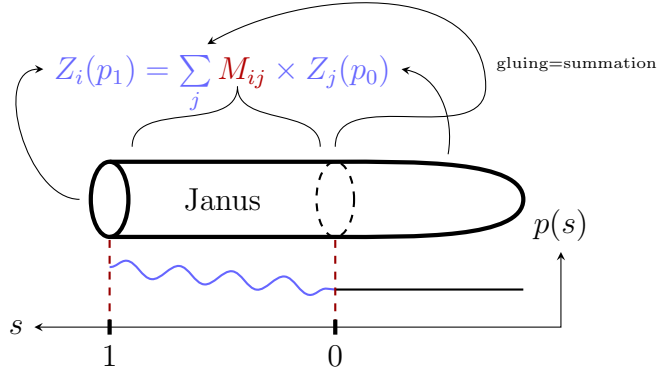
\begin{figure}[htbp]
	\begin{center}
		\begin{tikzpicture}
			\draw[stealth-stealth] (-1,0) -- (6,0) -- (6,1);
			\node[left] at (-1,0) {$s$};
			\node[above] at (6,1) {$p(s)$};
			\begin{scope}[shift={(0,1.7)}]
				\begin{scope}[yscale=0.5]
					\draw[ultra thick] (0,-1) -- (3,-1) to[out=0,in=270] (5.5,0) to[out=90,in=0] (3,1) -- (0,1);
				\end{scope}
			\begin{scope}
				\begin{scope}[xscale=0.5]
				\draw[ultra thick] (0,0) circle (0.5);
				\end{scope}
			\begin{scope}[shift={(3,0)}]
				\begin{scope}[xscale=0.5]
					\draw[thick,dashed] (0,0) circle (0.5);
				\end{scope}
			\end{scope}
			\end{scope}
				\node at (1.5,0) {Janus};
			\end{scope}
			\draw[thick] (3,0.5) -- (5.5,0.5);
			\draw[thick,\myblue,decorate,decoration={snake,amplitude=3pt, segment length=20pt}] (0,0.8) -- (3,0.5);
			\draw[thick, dashed,black!40!red] (0,0) -- (0,1.2) (3,0) -- (3,1.2);
			\draw[ultra thick] (0,-0.1) -- (0,0.1) (3,-0.1) -- (3,0.1);
			\node[below] at (0,-0.1) {$1$};
			\node[below] at (3,-0.1) {$0$};
			\node(A) [above,\myblue] at (1.5, 2.7) {$Z_i(p_1)=\sum\lm_j{\color{black!30!red}M_{ij}} \times Z_j(p_0)$};
			\draw (0.3,2.4) to[out=90,in=270] ([shift={(0.2,0.5)}]A.south) to[out=270,in=90] (2.8,2.4);
			\draw[-stealth] (-0.4,1.7) to[out=180,in=180] ([shift={(0,0.2)}]A.west);
			\draw[-stealth] (3,2.4) to[out=90,in=270] (5,3.2) to[out=90,in=30] ([shift={(-0.2,0)}]A.north);
			\node[right] at (5,3.5) {\tiny gluing$=$summation};
			\draw[-stealth] (4.5,2.3) to[out=80,in=350] ([shift={(0,0.2)}]A.east);
		\end{tikzpicture}
		\caption{Janus interface. $s$ --- ``adiabatic time" along the cylinder}\label{fig:Janus}
	\end{center}
\end{figure}

Let us consider the following construction.
Suppose we extended our disk $D$ into a vial with a long neck and allow some physical parameter $p$ to vary from a value $p(0)$ to a value $p(1)$ along this long neck (see Figure~\ref{fig:Janus}) parameterized by a coordinate $s$.
In what follows we will explain how to modify the theory accordingly.
Suppose we are able to preserve enough supersymmetry to localize the path integral.
If the cylinder is long enough then the theory at each value $p(s)$ can be approximated by the theory at a constant value of the parameter $p(s)$.
As a result, the cylinder neck $[0,1]\times S^1$ can have simultaneously two effective descriptions at $p_0=p(0)$ and $p_1=p(1)$, hence the name \emph{Janus interface} for such a theory in the literature (see \cite{Gaiotto:2015aoa,Dedushenko:2021mds,Bullimore:2021rnr} for recent discussions).

The Janus interface allows one to parallel transport the theory from theory $p_0$ to theory $p_1$ along some path $\wp$, so that for the partition functions we have:
\begin{align}
	Z_i(p_1)=\sum\lm_jM_{ij}(p_1,p_0;\wp)\;Z_j(p_0)\,,
\end{align}
where $M_{ij}(p_1,p_0;\wp)$ is a supersymmetric index of the theory on cylinder $[0,1]\times S^1$ with boundary conditions defined by indices $i$ and $j$.
$M_{ij}(p_1,p_0;\wp)$ provides a connection on the bundle of partition functions $\vec Z$ over the parameter space.\footnote{In principle, one could construct a differential form of this connection analogously to the tt${}^*$-connection \cite{Cecotti:1992rm} (see also \cite[Appendix~A]{Galakhov:2016cji}). 
However we will not need a differential form of this connection in this paper.
Also a rather wide class of problems is solved by constructing connections as Ward identities on spaces of holomorphic QFT quantities like correlators and conformal blocks. 
In these cases the resulting connection can be spotted under the names ``opers'', ``Knizhnik-Zamolodchikov connection'', ``Berry connection'' etc.
We are in no position to list even relevant sources due to popularity of this topic, see e.g. \cite{Okounkov:2016sya,Awata:2017lqa,Frenkel:2020iqq} to have a glimpse of developments and applications of these ideas.
}

To construct an interface that preserves some specific supercharge $\mathsf{Q}$, we first note that the index only counts the contributions of the BPS states that are cohomologies of $\mathsf{Q}$.
The standard localization strategy \cite{Witten:1982im} suggests that the $\mathsf{Q}$-cohomology subspace of the Hilbert space is invariant under conjugation of $\mathsf{Q}$ by some other operators:
\begin{align}
	\mathsf{Q}\to e^{\CO}\;\mathsf{Q}\;e^{-\CO}\,.
\end{align}
If the dependence on the parameter $p$ can be concentrated in the operator $\CO$, then substituting $p$ by the function $p(s)$ does not modify the localization properties of $\mathsf{Q}$.
Although somewhat abstract, this setting is rather universal and could be applied to generic massive 2D $\CN=(2,2)$ theories \cite{Gaiotto:2015aoa} and beyond \cite{Dedushenko:2021mds}.
However, in the actual implementation, one needs to be careful with the divergencies in the field theory.

For our concrete model of the 2D $\CN=(2,2)$ gauged linear sigma model with the gauge/matter content encoded in the quiver-superpotential pair $(Q,W)$, we can simply derive the desired corrections to the Lagrangian of the theory so that a part of the initial supersymmetry parameterized by phase $\vartheta$ is preserved, see Appendix~\ref{app:flavor}.
In particular, we have shown in Appendix~\ref{app:flavor} that the small deformations of the path $\wp$ of the parallel transport are generated by $\mathsf{Q}$-closed operators, leaving the localization result invariant.
Thus we conclude that $M_{ij}(p_1,p_0;\wp)$ provides a \emph{flat} parallel transport, in particular:
\begin{align}\label{Berry_flat}
    M_{ij}(p_1,p_0;\wp)=M_{ij}(p_1,p_0;\wp')\,,
\end{align}
if $\wp$ and $\wp'$ are homotopic.

The localization with respect to the supercharge $\mathsf{Q}$ allows us to compute the parallel transport in the semi-classical approximation:
\begin{align}\label{parallel_expan}
	M_{ij}(p_1,p_0;\wp)=\sum\lm_{\omega\in {\rm solitons}\left(j\longrightarrow i\right)}\left({\rm 1-loop}\right)e^{-\beta \,\CZ(\omega)}\,,
\end{align}
where the summation runs over the BPS soliton solutions $\omega$ that interpolate between the boundary conditions $i$ and $j$, $\CZ(\omega)$ is the central charge of the soliton configuration $\omega$, and $\beta$ is a circumference of the $S^1$.

The BPS soliton equations correspond to the stationary field configurations that annihilate the $\mathsf{Q}$-transformations of the fermionic fields.
In the GLSM model in question, the soliton equations have the form of a flow equation
\cite{Galakhov:2021omc,Dedushenko:2021mds}:
\begin{align}\label{soliton_eq}
	\begin{split}
		&D_s{\rm Im}(\sigma^{(a)})=\mu_{\IR}^{(a)},\qquad a\in Q_0\,,\\
		&D_s\phi_I=\hat G\left({\rm Im}\,e^{-\I\vartheta}\sigma\right)\cdot\phi^I,\qquad I\in Q_1\,,
	\end{split}
\end{align}
where $s$ is the adiabatic time along the interface, and the r.h.s.\ coincides with \eqref{fp}, which means that the crystal states are fixed points of the flow.
Similarly, in the dual LG model the BPS soliton equation reads \cite{Hori:2000ck,Gaiotto:2015aoa}:
\begin{align}\label{flow_LG}
	\p_s\sigma_A=\I e^{-\I\vartheta} g_{A\bar B}\overline{\p_{\sigma_B}\myW}\,,
\end{align}
where $A$, $B$ are coordinates of the field space and $g$ is the metric on the field space.

Although the BPS soliton equations are simpler than the complete equations of motion, their exact solutions are unknown in many cases.
Nevertheless, a combination of the expansion \eqref{parallel_expan} and the flatness condition \eqref{Berry_flat} is rather restrictive, so that the parallel transport $M$ can be computed analytically in certain setups without computing the actual forms of the solitons \cite{Gaiotto:2012rg}.

In this terminology, it is natural to define the $R$-matrix \eqref{R-matrix} as the parallel transport process from one arrangement of the flavor charges $u_i$ to another braiding crystals \eqref{cry_chain} in the weight complex  plane.
A similar approach was exploited in \cite{Bullimore:2017lwu,Dedushenko:2021mds,Aganagic:2017gsx,Aganagic:2017smx}.
Keeping in mind the picture of a crystal chain arrangement and assuming that $u_i$ are ordered as:
\begin{align}\label{ordering}
	{\rm Re}\,u_i\,<\,{\rm Re}\,u_{i+1}\,,
\end{align}
it is natural to assign to the $R$-matrix $R_{i,i+1}$ a process that exchanges the ordering \eqref{ordering} of flavor charges $u_i$ and $u_{i+1}$ in the complex plane.
The strands in the graphical language \eqref{Rm_graph} then become depictions of the crystal center-of-mass world-lines.

Because the parallel transport is flat (see \eqref{Berry_flat}), the $R$-matrix derived in this way automatically satisfies the Yang-Baxter equation \eqref{YBE} and the unitarity constraint \eqref{unitarity}, since the paths in the parameter space defining interfaces in the l.h.s.\ and the r.h.s.\ of these relations are homotopic.

\subsection{From solitonic flows to non-trivial coproducts}\label{sec:solitons}

\subsubsection{Solitonic flows}

The crystal basis we defined in Section~\eqref{sec:cry_chain} is a basis of stable fixed points when the spatial support of the theory is compact (see the examples in \cite{Galakhov:2021vbo}). 
However, if the space-time dimension is D$\geq 2$ the l.h.s.\ of the fixed point equations \eqref{fp} is promoted to a dynamical equation; the resulting equations describing a flow in the field space are given by \eqref{soliton_eq}.

Since the space-time in question is non-compact, the disk soliton equations \eqref{soliton_eq} may have a non-trivial solution interpolating between fixed points of the flow \eqref{soliton_eq}---crystal vacua.

Physically, the appearance of the solitons implies that the semi-classical wave-functions constructed as a description of fluctuations around classical vacua are not actual eigenstates of the Hamiltonian.
They have to be corrected by non-perturbative soliton contributions that arise from the overlaps of wave-functions between crystals located at different $u_i$.
This in turn implies that the crystal basis is not an actual stable basis\footnote{A stable basis is a basis of \emph{leaves} following Maulik-Okounkov \cite{MaulikOkounkov}. Leaves are Higgs duals to Coulomb Lefschetz thimbles and are unions of all soliton trajectories flowing from a fixed point.} of true eigenstates, instead they are related by a linear transformation (which we denote as $U$) constructed as a contribution of a dilute soliton gas analogously to \eqref{parallel_expan}.
This transformation makes the parallel transport \eqref{parallel_expan} non-trivial, which are given by the solitonic contributions above.

The flow \eqref{soliton_eq} has the property that during the flow, the real part of the central charge can only increase along the ``time" $s$.
The central charge in this model can be approximated by the following expression:
\begin{align}
	Z\sim \sum\lm_at_a\,\sum\lm_{\sqbox{$a$}\in{\rm crystals}}x\left(\sqbox{$a$}\right)+O(e^{-t})\,,
\end{align}
where $t$ is the complexified FI parameter and $x(\Box)$ is the spectral parameter---the projection of the position of the atom $\Box$ onto the weight plane.
Let us choose $t_a=t\in \IR_{>0}$. If crystals are ordered according to \eqref{ordering} then the only possible $k$-solitonic processes are those when a group of atoms with colors $a_1$, $a_2$, $\ldots$, $a_k$ are carried from a crystal with lower ${\rm Re}\,u_1$ to a crystal with greater ${\rm Re}\, u_2$.

From the dual point of view, the union of all trajectories for the dual flow equation \eqref{flow_LG} forms a Lefschetz thimble.

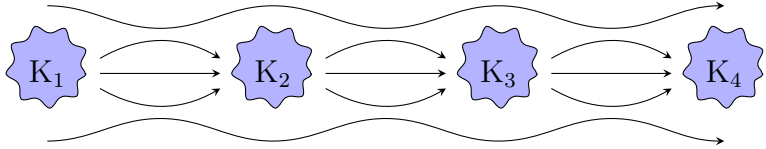
\begin{figure}[htbp]
	\begin{center}
		\begin{tikzpicture}
			\draw[fill=white!70!blue,decorate,decoration={snake,amplitude=1.5pt, segment length=10.35pt}] (0,0) circle (0.5);
			\draw[fill=white!70!blue,decorate,decoration={snake,amplitude=1.5pt, segment length=10.35pt}] (3,0) circle (0.5);
			\draw[fill=white!70!blue,decorate,decoration={snake,amplitude=1.5pt, segment length=10.35pt}] (6,0) circle (0.5);
			\draw[fill=white!70!blue,decorate,decoration={snake,amplitude=1.5pt, segment length=10.35pt}] (9,0) circle (0.5);
			\node at (0,0) {$\Kappa_1$};
			\node at (3,0) {$\Kappa_2$};
			\node at (6,0) {$\Kappa_3$};
			\node at (9,0) {$\Kappa_4$};
			\draw[-stealth] (0.7,0) -- (2.3,0);
			\draw[-stealth] (0.7,0.2) to[out=30,in=150] (2.3,0.2);
			\draw[-stealth] (0.7,-0.2) to[out=-30,in=210] (2.3,-0.2);
			\begin{scope}[shift={(3,0)}]
				\draw[-stealth] (0.7,0) -- (2.3,0);
				\draw[-stealth] (0.7,0.2) to[out=30,in=150] (2.3,0.2);
				\draw[-stealth] (0.7,-0.2) to[out=-30,in=210] (2.3,-0.2);
			\end{scope}
		\begin{scope}[shift={(6,0)}]
			\draw[-stealth] (0.7,0) -- (2.3,0);
			\draw[-stealth] (0.7,0.2) to[out=30,in=150] (2.3,0.2);
			\draw[-stealth] (0.7,-0.2) to[out=-30,in=210] (2.3,-0.2);
		\end{scope}
			\draw[-stealth] (0,0.9) to[out=0,in=180] (1.5,0.6) to[out=0,in=180] (3,0.9) to[out=0,in=180] (4.5,0.6) to[out=0,in=180] (6,0.9) to[out=0,in=180] (7.5,0.6) to[out=0,in=180] (9,0.9);
			\draw[-stealth] (0,-0.9) to[out=0,in=180] (1.5,-0.6) to[out=0,in=180] (3,-0.9) to[out=0,in=180] (4.5,-0.6) to[out=0,in=180] (6,-0.9) to[out=0,in=180] (7.5,-0.6) to[out=0,in=180] (9,-0.9);
		\end{tikzpicture}
	\end{center}
	\caption{Solitonic flow direction in the weight plane.}\label{fig:flow}
\end{figure}

The derivative of the superpotential \eqref{W_LG} consists of a linear term proportional to $t$ and a logarithmic term.
Away from the crystal bodies the logarithmic term can be approximated by $\log\Delta u$, where $\Delta u$ is the distance between crystals and is much smaller than $t$ if one is pursuing the connection between the Higgs and Coulomb branch descriptions in the limit ${\rm Re}\,t_a\to+\infty$.
So the flow in the equivariant weight plane is dominated by the $t_a$ values and is directed along the real axis.
In the dual description we see the same picture: atoms appearing in the gaps between crystals in the crystal chain are carried by the flow from a crystal located at lower ${\rm Re}\,u$ to one with a greater value (see Figure~\ref{fig:flow}).

\subsubsection{From crystal basis to stable basis}\label{sec:stab_bas}

For a generic choice of the integration cycle $\CL$, to study the asymptotic behavior of the partition function \eqref{fp}, one can expand \eqref{fp} over all possible asymptotics:
\begin{align}
	Z=\sum\lm_{\Kappa}c_{\Kappa}(\CC)e^{\myW\left(\langle\sigma\rangle_{\Kappa}\right)}\,,
\end{align}
where $\Kappa$ runs over all crystal fixed points, $\langle\sigma\rangle_{\Kappa}$ is the expectation value that corresponds to the fixed point $\Kappa$, and $c_{\Kappa}(\CC)$ are the expansion coefficients.
Let us call the basis of the cycles $\CC_{\Kappa}$ that are labeled by the fixed points $\Kappa$, which satisfy
$c_{\Kappa}\left(\CC_{\Kappa'}\right)=\delta_{\Kappa,\Kappa'}$, the \emph{vacuum basis}.
We will also use this name (the vacuum basis) for the basis of the corresponding
operators $\CO_{\mathscr{B}}(\Kappa)$.

On the other hand, we will call the basis of Lefschetz thimbles (and also the basis of their corresponding brane operators) the \emph{stable basis}.
In the usual WKB analysis of the asymptotic behavior of the integrals \eqref{fp}, the stable basis (namely the Lefschetz thimble basis) $\CL_{\Kappa}$  coincides with the vacuum basis. 
However, when a Lefshetz thimble corresponding to a fixed point $\Kappa_1$ intersects that of another fixed point $\Kappa_2$, the transition between those bases is controlled by the Stokes matrix $U$:
\begin{align}\label{vacuum_stable}
	\begin{array}{c}
		\begin{tikzpicture}
			\node (A) at (0,0) {Vacuum basis};
			\node (B) at (6,0) {Stable basis};
			\path (A) edge[->] node[above] {$U$} (B);
		\end{tikzpicture}
	\end{array}
\end{align}
The coefficients $c_{\Kappa}(\CL_{\Kappa'})$ are defined by an expression analogous to \eqref{parallel_expan}, where the summation runs over solitonic trajectories that flow from $\Kappa'$ to $\Kappa$.\footnote{Similar coefficients from the point of view of B-type boundary conditions on the Higgs branch are known in the mathematical literature (see e.g. \cite{shenfeld2013abelianization,Smirnov:2014npa,Dedushenko:2021mds,MaulikOkounkov,Aganagic:2016jmx}) as values of stable envelopes in fixed points:
$$
{\rm Stab}_{\Kappa'}\big|_{\Kappa}\,.
$$
} As we will see below, this matrix $U$ is the same matrix as in \textbf{Assumption 1} in Section~\ref{sec:coproduct}.

Let us introduce a degree on crystal two-chain fixed points in the following way:
\begin{align}
	{\rm deg}\,|\Kappa_1\rangle_{u_1}\otimes|\Kappa_2\rangle_{u_2}:=|\Kappa_2|-|\Kappa_1|\,,
\end{align}
where $|\Kappa|$ is the number of atoms in crystal $\Kappa$.
Since all the soliton trajectories are directed from $\Kappa_1$ to $\Kappa_2$ in the basis of two-chains ordered by this grading,
the Stokes matrix $U$ is lower-triangular: equivalently, we can decompose:
\begin{align}\label{U}
	U=1\otimes 1+\sum\lm_{k\geq 1}S_k,\quad {\rm deg}\,S_k=2k\,,
\end{align}
where an operator of degree $n$ raises the degree of a vector by $n$. This explains \textbf{Assumption 2} in Section~\ref{sec:coproduct}.

The explicit construction of $U$ is rather involved: one has to solve the soliton equations \eqref{soliton_eq} or \eqref{flow_LG} first, and then construct a 1-loop determinant in the soliton background.
To circumvent these difficulties we use the relations of these solutions to algebraic structures.

\subsubsection{Soliton algebras vs.\ quiver BPS algebras}\label{sssec:kernel}

A soliton carrying an atom from one crystal to another may be considered as a Heisenberg raising/lowering operator acting on the number of atoms.
Mixing this action with other quantum numbers of states we could derive a rather non-trivial resulting algebra of BPS solitons.

In the given context it is natural to identify such operators with processes of an atom, or a group of atoms carried from/to infinity or a rather distant test crystal to/from the crystal in question.
In \cite{Galakhov:2020vyb} a similar amplitude interpretation was given for the matrix coefficients in the quiver BPS algebra we discussed in Section~\ref{sec:cry_rep}:
\begin{align}\label{processes}
	\begin{array}{c}
		\begin{tikzpicture}
			\draw[fill=white!70!blue,decorate,decoration={snake,amplitude=1.5pt, segment length=10.35pt}] (0,0) circle (0.5);
			\node {$\Kappa$};
			\draw[-stealth] (1,0) -- (3,0) node[pos=0.5,above] {lowering} node[pos=0.5,below] {$\scriptstyle \left[\Kappa\to\Kappa-\Box\right]$};
			\draw[-stealth] (-3,0) -- (-1,0) node[pos=0.5,above] {raising} node[pos=0.5,below] {$\scriptstyle \left[\Kappa\to\Kappa+\Box\right]$};
			\draw[fill=white!70!blue] (-3.2,0) circle (0.08) (3.2,0) circle (0.08);
		\end{tikzpicture}
	\end{array}
\end{align}

A solution to the second set of equations in \eqref{soliton_eq} can be integrated:
\begin{align}\label{homomorphism}
	\begin{split}
	&\phi_{I:a\to b}(s\to-\infty)\cdot \tau_a=\tau_b\cdot\phi_{I:a\to b}(s\to+\infty)\,,\\
	& \tau_a={\rm Pexp}\int\lm_{-\infty}^{+\infty}\left(\I A_s^{(a)}+{\rm Im}\,\sigma^{(a)}\right)\,ds\,.
	\end{split}
\end{align}
The set of transformations $\tau_a$, $a\in Q_0$ is called a \emph{homomorphism} of quiver representations $\phi_I(s\to\pm\infty)$.
The action of the soliton carrying atoms around increases some of the eigenvalues $\sigma$ and $\phi$-values, therefore homomorphisms $\tau_a$ describing solitons are singular in general.
In principle, a projector to a finite sub-representation should be used, so that homomorphisms are between representations of quivers with different dimension vectors $N_a$.

The homomorphism \eqref{homomorphism} can be localized to fixed points, implying that a crystal $\Kappa$ in one representation is a subcrystal of another $\Kappa'$.
The locus $\CI_{\Kappa,\Kappa'}$ in the product space of crystal representations is called the incidence locus.

The soliton partition functions form a Hilbert space of morphisms in the corresponding triangulated category of boundary conditions \cite{D_book_2}.
On the Higgs branch, the triangulated category is a derived category of coherent sheaves and the morphisms are the derived functors of Fourier-Mukai transform.
Localization of the soliton action to \eqref{homomorphism} implies that the Fourier-Mukai kernel also localizes to a structure sheaf of $\CI_{\Kappa,\Kappa'}$, in other words the kernel is supported on $\CI_{\Kappa,\Kappa'}$:
\begin{align}\label{kernel}
	{\rm kernel}\sim \CO_{\CI_{\Kappa,\Kappa'}}\otimes \ldots\,.
\end{align}
Transformations on cohomologies of quiver moduli spaces induced by Fourier-Mukai transforms with kernels given by simply structure sheaves of $\CI_{\Kappa,\Kappa\pm \Box}$ are identified in \cite{Galakhov:2020vyb} with matrix coefficients of the quiver BPS algebra representations \eqref{ef_matrix}.

Unfortunately, in practice the estimate we performed in this section is imprecise.
The reason is that a one-site chain consisting of a single $\Kappa$ is not a true eigen BPS state of the Hamiltonian.
Physically, it is easy to imagine that for a crystal getting long enough in the real axis direction processes driven by the flow depicted in Figure~\ref{fig:flow}  when an atom is dispatched from the left corner and re-captured by the right corner are present.
From the point of view of stable leaf bases even for simple stable envelope bases \cite{Smirnov:2014npa} in one-site crystal chains it is easy to compute that the matrix of stable envelopes computed at fixed points is not diagonal:\footnote{We would like to thank Andrey Smirnov for pointing out this peculiarity.}
\begin{align}
	{\rm Stab}_{\Kappa}\big|_{\Kappa'}\not\sim \delta_{\Kappa,\Kappa'}\,.
\end{align}
A physical consequence of this observation is that soliton flow equations \eqref{soliton_eq} have solutions leading to a self-reshuffling of crystal atoms.
We will \emph{neglect} these fine self-reshuffling processes assuming that they simply redefine the wave functions corresponding to single crystals $\Kappa$ as Sudakov factors.

\subsubsection{Gauge theory origin of coproducts}

The structure of the simple soliton amplitudes we discussed so far is rather suggestive to extend naturally the action of $\myY$ to tensor products.
The only step one should do is to substitute a single crystal in \eqref{processes} by the crystal chain \eqref{cry_chain}.

This extension, however, comes at a price.
In the computations of the soliton action on a crystal chain, one has to take into account the contribution of a gas of solitons that can jump between the crystals in the chain.
The solitons ending on a crystal $\Kappa_i$ may be glued with a soliton starting at $\Kappa_i$ into a longer amplitude.
The result of such a summation is again a Stokes matrix $U$ \eqref{vacuum_stable}.

In the vacuum basis we neglect solitons jumping between crystals $\Kappa_i$ inside the chain, so the representation of the algebra $\myY$ generators takes into account only a distribution of an external atom between $\Kappa_i$ and is given by $\bDelta_0$ in \eqref{chain_naiv_co}.
However the true coproduct $\bDelta$ for the stable basis is non-trivial and captures soliton contribution from conjugation by $U$:
\begin{align}
	\bDelta=U\cdot\bDelta_0\cdot U^{-1}\,.
\end{align}
This is equation \eqref{iso}
for  \textbf{Assumption 1} in Section~\ref{sec:coproduct}.

To define the interface partition function associated to the $R$-matrix,
we can appeal to the algebraic approach instead of directly summing up solitons in \eqref{parallel_expan}.
The homotopy implies that the $R$-matrix can be computed from the commutation relation with the coproduct:
\begin{align}
	R\cdot\Delta=\Delta\cdot R\,.
\end{align}
We discussed an explicit construction of the $R$-matrix from the algebraic constraints in Section~\ref{sec:coproduct}.

\subsubsection{Computing non-trivial coproducts from solitonic corrections}\label{sec:corrections}

To conclude this section, let us estimate the algebraic expressions for the expansion elements $S_k$ in \eqref{U}.

The one-soliton process can be described as follows. 
An atom is removed from the crystal $\Kappa_1$ located at $u_1$, is then carried by the flow \eqref{soliton_eq} to the neighboring crystal $\Kappa_2$ located at $u_2$, and finally is added to $\Kappa_2$.
This is an IR t-channel type process that can be described by the following diagram:\footnote{The process should preserve the total number of atoms,
forbidding the $s$-channel diagram.}
\begin{align}\label{amplitude}
	\CS_1\to\left[\begin{array}{c}
		\begin{tikzpicture}
			\draw[ultra thick,\myblue,decorate,decoration={snake,amplitude=2pt, segment length=10.5pt}] (-1.8,0) -- (1.8,0);
			\draw[ultra thick] (-2.3,-1) -- (-1.8,0) -- (-2.3,1) (2.3,-1) -- (1.8,0) -- (2.3,1);
			\draw[fill=orange] (-1.8,0) circle (0.08) (1.8,0) circle (0.08);
			\node[left] at (-2,0) {$\scriptstyle\left[\Kappa_1\to\Kappa_1-{\color{\myblue}\Box_1}\right]$};
			\node[right] at (2,0) {$\scriptstyle\left[\Kappa_2\to\Kappa_2+{\color{black!40!red}\Box_2}\right]$};
			\node[below] at (0,-0.1) {$\scriptstyle g_1\left(u_1+h_{\color{\myblue}\Box_1}|u_2+h_{\color{black!40!red}\Box_2}\right)$};
			\node[above] at (0,0.1) {$\begin{array}{c}
					\begin{tikzpicture}
						\draw[-stealth] (-1,0) -- (1,0);
						\draw[-stealth] (0,-0.2) -- (0,0.7);
						\node[above] at (0,0.7) {$\scriptstyle \Psi$};
						\begin{scope}[shift={(-0.3,0)}]
							\draw[\myblue] (-0.5,0.1) to[out=0,in=180] (0,0.5) to[out=0,in=180] (0.5,0.1) -- (0.8,0.1);
						\end{scope}
						\begin{scope}[shift={(0.3,0)}]
							\draw[black!40!red] (-0.8,0.1) -- (-0.5,0.1) to[out=0,in=180] (0,0.5) to[out=0,in=180] (0.5,0.1);
						\end{scope}
					\end{tikzpicture}
				\end{array}$};
			\node[below left] at (-2.3,-1) {$\scriptstyle\Kappa_1\;\;\begin{array}{c}
					\begin{tikzpicture}
						\tikzset{style1/.style={thin, fill=white}}
						\tikzset{style2/.style={thin, fill=\myblue}}
						\begin{scope}[scale=0.17]
							\draw[style1] (0.707107,1.22474) -- (0.707107,2.04124) -- (0.,1.63299) -- (0.,0.816497) -- cycle;
							\draw[style1] (-0.707107,-1.22474) -- (-0.707107,-0.408248) -- (-1.41421,-0.816497) -- (-1.41421,-1.63299) -- cycle;
							\draw[style1] (0.707107,-0.408248) -- (0.707107,0.408248) -- (0.,0.) -- (0.,-0.816497) -- cycle;
							\draw[style1] (2.12132,-0.408248) -- (2.12132,0.408248) -- (1.41421,0.) -- (1.41421,-0.816497) -- cycle;
							\draw[style2] (1.41421,-1.63299) -- (1.41421,-0.816497) -- (0.707107,-1.22474) -- (0.707107,-2.04124) -- cycle;
							\draw[style1] (2.82843,-1.63299) -- (2.82843,-0.816497) -- (2.12132,-1.22474) -- (2.12132,-2.04124) -- cycle;
							\draw[style1] (-0.707107,1.22474) -- (-0.707107,2.04124) -- (0.,1.63299) -- (0.,0.816497) -- cycle;
							\draw[style1] (-1.41421,0.) -- (-1.41421,0.816497) -- (-0.707107,0.408248) -- (-0.707107,-0.408248) -- cycle;
							\draw[style1] (-2.12132,-1.22474) -- (-2.12132,-0.408248) -- (-1.41421,-0.816497) -- (-1.41421,-1.63299) -- cycle;
							\draw[style1] (-0.707107,-1.22474) -- (-0.707107,-0.408248) -- (0.,-0.816497) -- (0.,-1.63299) -- cycle;
							\draw[style1] (-0.707107,-0.408248) -- (-0.707107,0.408248) -- (0.,0.) -- (0.,-0.816497) -- cycle;
							\draw[style1] (0.707107,-0.408248) -- (0.707107,0.408248) -- (1.41421,0.) -- (1.41421,-0.816497) -- cycle;
							\draw[style2] (0.,-1.63299) -- (0.,-0.816497) -- (0.707107,-1.22474) -- (0.707107,-2.04124) -- cycle;
							\draw[style1] (1.41421,-1.63299) -- (1.41421,-0.816497) -- (2.12132,-1.22474) -- (2.12132,-2.04124) -- cycle;
							\draw[style1] (0.,2.44949) -- (-0.707107,2.04124) -- (0.,1.63299) -- (0.707107,2.04124) -- cycle;
							\draw[style1] (-0.707107,1.22474) -- (-1.41421,0.816497) -- (-0.707107,0.408248) -- (0.,0.816497) -- cycle;
							\draw[style1] (-1.41421,0.) -- (-2.12132,-0.408248) -- (-1.41421,-0.816497) -- (-0.707107,-0.408248) -- cycle;
							\draw[style1] (0.707107,1.22474) -- (0.,0.816497) -- (0.707107,0.408248) -- (1.41421,0.816497) -- cycle;
							\draw[style1] (0.,0.816497) -- (-0.707107,0.408248) -- (0.,0.) -- (0.707107,0.408248) -- cycle;
							\draw[style1] (1.41421,0.816497) -- (0.707107,0.408248) -- (1.41421,0.) -- (2.12132,0.408248) -- cycle;
							\draw[style2] (0.707107,-0.408248) -- (0.,-0.816497) -- (0.707107,-1.22474) -- (1.41421,-0.816497) -- cycle;
							\draw[style1] (2.12132,-0.408248) -- (1.41421,-0.816497) -- (2.12132,-1.22474) -- (2.82843,-0.816497) -- cycle;
						\end{scope}
					\end{tikzpicture}
				\end{array}$};
			\node[above left] at (-2.3,1) {$\scriptstyle\Kappa_1-{\color{\myblue}\Box_1}\;\;\begin{array}{c}
					\begin{tikzpicture}
						\tikzset{style1/.style={thin, fill=white}}
						\begin{scope}[scale=0.17]
							\draw[style1] (0.707107,1.22474) -- (0.707107,2.04124) -- (0.,1.63299) -- (0.,0.816497) -- cycle;
							\draw[style1] (-0.707107,-1.22474) -- (-0.707107,-0.408248) -- (-1.41421,-0.816497) -- (-1.41421,-1.63299) -- cycle;
							\draw[style1] (0.707107,-1.22474) -- (0.707107,-0.408248) -- (0.,-0.816497) -- (0.,-1.63299) -- cycle;
							\draw[style1] (0.707107,-0.408248) -- (0.707107,0.408248) -- (0.,0.) -- (0.,-0.816497) -- cycle;
							\draw[style1] (2.12132,-0.408248) -- (2.12132,0.408248) -- (1.41421,0.) -- (1.41421,-0.816497) -- cycle;
							\draw[style1] (2.82843,-1.63299) -- (2.82843,-0.816497) -- (2.12132,-1.22474) -- (2.12132,-2.04124) -- cycle;
							\draw[style1] (-0.707107,1.22474) -- (-0.707107,2.04124) -- (0.,1.63299) -- (0.,0.816497) -- cycle;
							\draw[style1] (-1.41421,0.) -- (-1.41421,0.816497) -- (-0.707107,0.408248) -- (-0.707107,-0.408248) -- cycle;
							\draw[style1] (-2.12132,-1.22474) -- (-2.12132,-0.408248) -- (-1.41421,-0.816497) -- (-1.41421,-1.63299) -- cycle;
							\draw[style1] (-0.707107,-1.22474) -- (-0.707107,-0.408248) -- (0.,-0.816497) -- (0.,-1.63299) -- cycle;
							\draw[style1] (-0.707107,-0.408248) -- (-0.707107,0.408248) -- (0.,0.) -- (0.,-0.816497) -- cycle;
							\draw[style1] (0.707107,-1.22474) -- (0.707107,-0.408248) -- (1.41421,-0.816497) -- (1.41421,-1.63299) -- cycle;
							\draw[style1] (0.707107,-0.408248) -- (0.707107,0.408248) -- (1.41421,0.) -- (1.41421,-0.816497) -- cycle;
							\draw[style1] (1.41421,-1.63299) -- (1.41421,-0.816497) -- (2.12132,-1.22474) -- (2.12132,-2.04124) -- cycle;
							\draw[style1] (0.,2.44949) -- (-0.707107,2.04124) -- (0.,1.63299) -- (0.707107,2.04124) -- cycle;
							\draw[style1] (-0.707107,1.22474) -- (-1.41421,0.816497) -- (-0.707107,0.408248) -- (0.,0.816497) -- cycle;
							\draw[style1] (-1.41421,0.) -- (-2.12132,-0.408248) -- (-1.41421,-0.816497) -- (-0.707107,-0.408248) -- cycle;
							\draw[style1] (0.707107,1.22474) -- (0.,0.816497) -- (0.707107,0.408248) -- (1.41421,0.816497) -- cycle;
							\draw[style1] (0.,0.816497) -- (-0.707107,0.408248) -- (0.,0.) -- (0.707107,0.408248) -- cycle;
							\draw[style1] (1.41421,0.816497) -- (0.707107,0.408248) -- (1.41421,0.) -- (2.12132,0.408248) -- cycle;
							\draw[style1] (2.12132,-0.408248) -- (1.41421,-0.816497) -- (2.12132,-1.22474) -- (2.82843,-0.816497) -- cycle;
						\end{scope}
					\end{tikzpicture}
				\end{array}$};
			\node[below right] at (2.3,-1) {$\scriptstyle\begin{array}{c}
					\begin{tikzpicture}
						\tikzset{style1/.style={thin, fill=white}}
						\begin{scope}[scale=0.17]
							\draw[style1] (1.41421,-0.816497) -- (1.41421,0.) -- (0.707107,-0.408248) -- (0.707107,-1.22474) -- cycle;
                    \draw[style1] (1.41421,0.) -- (1.41421,0.816497) -- (0.707107,0.408248) -- (0.707107,-0.408248) -- cycle;
                    \draw[style1] (0.707107,-1.22474) -- (0.707107,-0.408248) -- (0.,-0.816497) -- (0.,-1.63299) -- cycle;
                    \draw[style1] (0.707107,-0.408248) -- (0.707107,0.408248) -- (0.,0.) -- (0.,-0.816497) -- cycle;
                    \draw[style1] (-1.41421,-0.816497) -- (-1.41421,0.) -- (-0.707107,-0.408248) -- (-0.707107,-1.22474) -- cycle;
                    \draw[style1] (-1.41421,0.) -- (-1.41421,0.816497) -- (-0.707107,0.408248) -- (-0.707107,-0.408248) -- cycle;
                    \draw[style1] (-0.707107,-1.22474) -- (-0.707107,-0.408248) -- (0.,-0.816497) -- (0.,-1.63299) -- cycle;
                    \draw[style1] (-0.707107,-0.408248) -- (-0.707107,0.408248) -- (0.,0.) -- (0.,-0.816497) -- cycle;
                    \draw[style1] (0.,1.63299) -- (-0.707107,1.22474) -- (0.,0.816497) -- (0.707107,1.22474) -- cycle;
                    \draw[style1] (-0.707107,1.22474) -- (-1.41421,0.816497) -- (-0.707107,0.408248) -- (0.,0.816497) -- cycle;
                    \draw[style1] (0.707107,1.22474) -- (0.,0.816497) -- (0.707107,0.408248) -- (1.41421,0.816497) -- cycle;
                    \draw[style1] (0.,0.816497) -- (-0.707107,0.408248) -- (0.,0.) -- (0.707107,0.408248) -- cycle;
						\end{scope}
					\end{tikzpicture}
				\end{array}\;\;\Kappa_2$};
			\node[above right] at (2.3,1) {$\scriptstyle\begin{array}{c}
					\begin{tikzpicture}
						\tikzset{style1/.style={thin, fill=white}}
						\tikzset{style2/.style={thin, fill=black!40!red}}
						\begin{scope}[scale=0.17]
						\draw[style2] (0.707107,1.22474) -- (0.707107,2.04124) -- (0.,1.63299) -- (0.,0.816497) -- cycle;
                        \draw[style1] (1.41421,-0.816497) -- (1.41421,0.) -- (0.707107,-0.408248) -- (0.707107,-1.22474) -- cycle;
                        \draw[style1] (1.41421,0.) -- (1.41421,0.816497) -- (0.707107,0.408248) -- (0.707107,-0.408248) -- cycle;
                        \draw[style1] (0.707107,-1.22474) -- (0.707107,-0.408248) -- (0.,-0.816497) -- (0.,-1.63299) -- cycle;
                        \draw[style1] (0.707107,-0.408248) -- (0.707107,0.408248) -- (0.,0.) -- (0.,-0.816497) -- cycle;
                        \draw[style2] (-0.707107,1.22474) -- (-0.707107,2.04124) -- (0.,1.63299) -- (0.,0.816497) -- cycle;
                        \draw[style1] (-1.41421,-0.816497) -- (-1.41421,0.) -- (-0.707107,-0.408248) -- (-0.707107,-1.22474) -- cycle;
                        \draw[style1] (-1.41421,0.) -- (-1.41421,0.816497) -- (-0.707107,0.408248) -- (-0.707107,-0.408248) -- cycle;
                        \draw[style1] (-0.707107,-1.22474) -- (-0.707107,-0.408248) -- (0.,-0.816497) -- (0.,-1.63299) -- cycle;
                        \draw[style1] (-0.707107,-0.408248) -- (-0.707107,0.408248) -- (0.,0.) -- (0.,-0.816497) -- cycle;
                        \draw[style2] (0.,2.44949) -- (-0.707107,2.04124) -- (0.,1.63299) -- (0.707107,2.04124) -- cycle;
                        \draw[style1] (-0.707107,1.22474) -- (-1.41421,0.816497) -- (-0.707107,0.408248) -- (0.,0.816497) -- cycle;
                        \draw[style1] (0.707107,1.22474) -- (0.,0.816497) -- (0.707107,0.408248) -- (1.41421,0.816497) -- cycle;
                        \draw[style1] (0.,0.816497) -- (-0.707107,0.408248) -- (0.,0.) -- (0.707107,0.408248) -- cycle;
						\end{scope}
					\end{tikzpicture}
				\end{array}\;\;\Kappa_2+{\color{black!40!red}\Box_2}$};
		\end{tikzpicture}
	\end{array}\right]\,,
\end{align}
where the two vertices describe the atom adding/removing processes, with coefficients $[\Kappa_1\rightarrow \Kappa_1-\square_1]$ and $[\Kappa_2\rightarrow \Kappa_2+\square_2]$, respectively; and the propagator $g_1$ is defined by the flow.

To estimate the propagator $g_1(x|y)$ one should incorporate in fact the one-loop determinant in the soliton background.
The soliton has a zero-mode corresponding to the center of the soliton mass with equivariant weight $x-y$, where $x$ and $y$ are the initial and final positions of the transferred atom in the weight space.
And the propagator reads:
\begin{align}\label{propag}
	g_1(x|y)\sim\frac{1}{x-y}\,.
\end{align}
There are alternative ways to compute \eqref{propag} explicitly.
One way is to use the compatibility constraints from the algebraic structures, as done in Section~\ref{sec:coproduct}.
Another way is to apply the dual theory where the loop corrections are taken into account in the form of the effective superpotential, see Appendix~\ref{sec:stable}.

We can now see that the combination of \eqref{amplitude} and \eqref{propag} justifies \textbf{Assumption 3} in Section~\ref{sec:coproduct}.

It is harder to describe higher soliton processes, however one could expect the same three stages:
dissolution of an atom group in crystal $\Kappa_1$, transport along the flow, then recombination with $\Kappa_2$.

If we assume that the Fourier-Mukai kernel for the dissolution/recombination amplitudes is given by \eqref{kernel} then the resulting pair of crystals $\Kappa_1'$ and $\Kappa_2'$ are in the following relation with the initial one:
\begin{align}
	\Kappa_1\supset \Kappa_1',\quad \Kappa_2\subset \Kappa_2'\,.
\end{align}
Then $\Kappa_1'$ can be represented as a sequence of actions by lowering operators $f^{(a)}$ on crystal $\Kappa_1$.
Correspondingly, crystal $\Kappa_2'$ is a result of acting by $e^{(a)}$ on $\Kappa_2$.
Eventually, we have the following representation:
\begin{eqnarray}
	S_k\sim f^{a_1}f^{a_2}\ldots f^{a_k}\otimes e^{b_1}e^{b_2}\ldots e^{b_k}\,,
\end{eqnarray}
where the sets of colors $\{a_i\}$ and $\{b_i\}$ are equivalent upon a permutation since the solitons are unable to change atom colors.

\section{No-go against shifts and chiral quivers} \label{sec:no-go}

In this section, we present in sections \ref{sec:obstruction} and \ref{sec:LYcc} two arguments against the Gauge/Bethe correspondence for certain representations (that cause a negative shift in the ground state charge function) and for chiral quivers, respectively.
Our arguments apply both to shifted Yangians associated with non-chiral quivers
and in particular to general quiver Yangians with \emph{chiral} quivers.
We will then discuss the gauge-theoretic origin of the problem in Section~\ref{sec:breakdown}.

\subsection{Constraints from Yang-Baxter equations}\label{sec:obstruction}

For the first no-go argument, 
consider the Yang-Baxter equation with the following initial/final states:
\begin{align}\label{eq:nogo-YBE}
	\begin{array}{c}
	\begin{tikzpicture}[scale=1.2]
		\draw[thick] (0.984808, 0.173648) -- (-0.642788, -0.766044) node[pos=0.5, below]{$\scriptstyle \varnothing$};
		\draw[thick] (-0.34202, 0.939693) -- (-0.34202, -0.939693) node[pos=0.5, left]{$\scriptstyle Q_1$};
		\draw[thick] (-0.642788, 0.766044) -- (0.984808, -0.173648) node[pos=0.5, above]{$\scriptstyle Q_{|\Kappa|}$};
		\node[above right] at (0.984808, 0.173648) {$\scriptstyle\varnothing$};
		\node[below right] at (0.984808, -0.173648) {$\scriptstyle\Kappa$};
		\node[above] at (-0.34202, 0.939693) {$\scriptstyle\varnothing$};
		\node[above left] at (-0.642788, 0.766044) {$\scriptstyle\Kappa'$};
		\node[below] at (-0.34202, -0.939693) {$\scriptstyle\Box$};
		\node[below left] at (-0.642788, -0.766044) {$\scriptstyle\varnothing$};
	\end{tikzpicture}
	\end{array}+\begin{array}{c}
	\begin{tikzpicture}[scale=1.2]
		\draw[thick] (0.984808, 0.173648) -- (-0.642788, -0.766044) node[pos=0.5, below]{$\scriptstyle Q_1$};
		\draw[thick] (-0.34202, 0.939693) -- (-0.34202, -0.939693) node[pos=0.5, left]{$\scriptstyle \varnothing$};
		\draw[thick] (-0.642788, 0.766044) -- (0.984808, -0.173648) node[pos=0.5, above]{$\scriptstyle Q_{|\Kappa|+1}$};
		\node[above right] at (0.984808, 0.173648) {$\scriptstyle\varnothing$};
		\node[below right] at (0.984808, -0.173648) {$\scriptstyle\Kappa$};
		\node[above] at (-0.34202, 0.939693) {$\scriptstyle\varnothing$};
		\node[above left] at (-0.642788, 0.766044) {$\scriptstyle\Kappa'$};
		\node[below] at (-0.34202, -0.939693) {$\scriptstyle\Box$};
		\node[below left] at (-0.642788, -0.766044) {$\scriptstyle\varnothing$};
	\end{tikzpicture}
\end{array}=\begin{array}{c}
	\begin{tikzpicture}[scale=1.2,xscale=-1]
		\draw[thick] (0.984808, 0.173648) -- (-0.642788, -0.766044) node[pos=0.5, below]{$\scriptstyle Q_{|\Kappa|+1}$};
		\draw[thick] (-0.34202, 0.939693) -- (-0.34202, -0.939693) node[pos=0.5, right]{$\scriptstyle \varnothing$};
		\draw[thick] (-0.642788, 0.766044) -- (0.984808, -0.173648) node[pos=0.5, above]{$\scriptstyle \varnothing$};
		\node[above left] at (0.984808, 0.173648) {$\scriptstyle\Kappa'$};
		\node[below left] at (0.984808, -0.173648) {$\scriptstyle\varnothing$};
		\node[above] at (-0.34202, 0.939693) {$\scriptstyle\varnothing$};
		\node[above right] at (-0.642788, 0.766044) {$\scriptstyle\varnothing$};
		\node[below] at (-0.34202, -0.939693) {$\scriptstyle\Box$};
		\node[below right] at (-0.642788, -0.766044) {$\scriptstyle\Kappa$};
	\end{tikzpicture}
\end{array}\,,
\end{align}
where $\Kappa'$ has one more atom than $\Kappa$, and can be written as $\Kappa+\Box'$, and $Q_n$ denotes a summation over all crystals of dimension $n$ (i.e.\ with $n$ atoms) that can appear in the relevant intermediate channel.\footnote{In this subsection, we drop the color $a$ labels on the operators and the atoms to reduce clutter; they are not essential to the argument and  can be easily reinstated.}

Plugging the expansions of $R$-matrices \eqref{R-mat} into the  YBE \eqref{eq:nogo-YBE}, we obtain the following constraint:
\begin{align}\label{nogoYBE1}
\begin{split}
	&\langle\varnothing, Q_1|S_1(u_{21})|\Box,\varnothing\rangle\langle\varnothing,\Kappa'|S_1(u_{13})|Q_1,\Kappa\rangle+\\
	&\frac{\rho_{\varnothing,\Kappa}(u_{13})}{\rho_{\varnothing,\Kappa'}(u_{13})}\langle\varnothing,\Kappa'|S_1(u_{23})|\Box,\Kappa\rangle=\frac{\rho_{\varnothing,\varnothing}(u_{12})}{\rho_{\varnothing,\Box}(u_{12})}\langle\varnothing,\Kappa'|S_1(u_{23})|\Box,\Kappa\rangle\,,
\end{split}
\end{align}
where $u_{ij}\equiv u_i-u_j$.
Then substituting the expression of  $S_1$ from \eqref{1-solit} and the recurrence relation of $\rho$ from \eqref{recurrence}, we rewrite the constraint \eqref{nogoYBE1} into:
\begin{align}\label{E3}
	\begin{split}
&(-1)^{|a|+1}\Psi_{\varnothing}(h_{\Box'}+u_{3}-u_1)\frac{\left[\Box\to\varnothing\right]\left[\Kappa\to\Kappa+\Box'\right]}{(h_{\Box}+u_2)-(h_{\Box'}+u_3)}+\\
& \qquad+\sum_{\tilde{\Box}\in \textrm{Add}(\varnothing)}
\frac{\left[\Box\to\varnothing\right]\left[\varnothing\to \tilde{\Box}\to\varnothing\right]\left[\Kappa\to\Kappa+\Box'\right]}{\left((h_{\Box}+u_2)-(h_{\tilde{\Box}}+u_1)\right)\left((h_{\Box}+u_1)-(h_{\Box'}+u_3)\right)}\\
&=(-1)^{|a|+1}\Psi_{\varnothing}(h_{\Box}+u_2-u_1)\frac{\left[\Box\to\varnothing\right]\left[\Kappa\to\Kappa+\Box'\right]}{(h_{\Box}+u_2)-(h_{\Box'}+u_3)}\,,
	\end{split}
\end{align}
where in the second line we have summed over all possible choices of $Q_1$, which corresponds to the set $\textrm{Add}(\varnothing)$, i.e.\ all the atoms that can be added to the ground state of this crystal representation (namely the set of all its starters).

On the other hand, since $\Psi_{\varnothing}(z)$ is the ground state contribution of the charge function, it can be written as 
\begin{align}\label{pole_exp}
	\Psi_{\varnothing}(z)=\sum\lm_{k\geq 0}\Psi_{\varnothing,-k}z^k+\sum\lm_{\tilde{\Box}\in \textrm{Add}(\varnothing)}\frac{\mathop{\rm Res}\lm_{w=h_{\tilde{\Box}}}\Psi_{\varnothing}(w)}{z-h_{\tilde{\Box}}}\,.
\end{align}
Substituting this expression in \eqref{E3} we derive a consistency requirement:
\begin{align}\label{ostruction_0}
	\sum\lm_{k\geq 1}\Psi_{\varnothing,-k}\frac{\left(h_{\Box'}+u_3-u_1\right)^k-\left(h_{\Box}+u_2-u_1\right)^k}{(h_{\Box}+u_2)-(h_{\Box'}+u_3)}=0\,,
\end{align}
where we have used the fact $[\varnothing\to \tilde{\Box}\to \varnothing]=(-1)^{|a|+1}\mathop{\rm Res}\lm_{w=h_{\tilde{\Box}}}\Psi_{\varnothing}(w)$, see the last equation of \eqref{ef_matrix}.

The condition \eqref{ostruction_0} means that the $R$-matrix given by \eqref{main_R} solves the YBE only if there is no negative shift in the charge function of the ground state $|\varnothing\rangle$, namely
\begin{equation}\label{no-go-1final}
    \Psi_{\varnothing,-k}=0\,, \qquad \textrm{for } k\geq 1\,.
\end{equation}
Note that we are allowed to have a non-zero $\Psi_{\varnothing,0}$.

The implication of the constraint \eqref{no-go-1final} is that in the ground state charge function (the $\varphi^{a\Leftarrow \ff}(z)$ defined in \eqref{psi_vac_0} or \eqref{psi_vac}), the degree of the denominator cannot be less than that of the numerator, namely, 
\begin{equation}\label{constraint-s}
    \mathfrak{s}_a \geq 0, \quad \forall\,a\in Q_0\,.
\end{equation}
This translates into the shape of the crystal ${}^{\sharp}\mathcal{C}$ that defines the representation via the positive-negative crystal decomposition of \cite{Galakhov:2021xum}: we are only allowed to consider crystal representations in which\footnote{In the presence of the truncation factors, this equation becomes $(\# \textrm{  starters}) - (\# \textrm{  stoppers})- (\# \textrm{ pausers} ) \geq (\# \textrm{  truncation factors})$, within the same color $a$, see Footnote~7 on page~16.} 
\begin{equation}\label{constraint-ss}
 \textrm{for the same color $a$}: \qquad  (\# \textrm{  starters}) - (\# \textrm{  stoppers})- (\# \textrm{ pausers} ) \geq 0\,.
\end{equation}

Finally, we mention that if we restrict to the 2D crystal representations, as will be necessary when we consider the BAE in Section \ref{sec:BAE}, then all the shifts have to vanish:
\begin{equation}\label{s2Dcrystal}
    \textrm{for 2D crystal:}\qquad   \mathfrak{s}_a=0\,, 
    \quad \forall\,a\in Q_0\,.
\end{equation}
The reason is the following. 
For 2D crystals, since  we need stoppers to stop the crystal from growing in the third direction,  the number of the stoppers cannot be smaller than the number of the starters (in the 2D plane), i.e.\ $\sum_{a}\mathfrak{s}_a \leq 0$.\footnote{Note that the colors of starters and stoppers are not constrained to match in general, therefore at this step we could impose this relation only on the total shift.}
Then together with the constraint \eqref{constraint-s} from the YBE, we arrive at \eqref{s2Dcrystal}.


\subsection{Consistency between coproducts of Lax operators and those of quiver BPS algebras}\label{sec:LYcc}

The previous no-go argument disallows negative shift (more precisely, we require $\mathfrak{s}_{a}\geq 0$ in the mode expansion of the \emph{ground state charge function}), and hence restricts the type of 2D crystal representations that we are allowed to consider. 
Next we will give a stronger no-go argument, which will rule out all the chiral quivers.

The second no-go argument uses the consistency of the coproduct originating from two different sources.
As we have seen in Section~\ref{sec:coproduct}, Lax operators have a natural coproduct structure \eqref{Lax_co}.
Using relations \eqref{Lax_BPS} we could derive a Lax-induced coproduct on $\myY$.
Now we are in a position to pose the question if the Lax-induced coproduct on $\myY$ and the original $\myY$ coproduct constructed in Section~\ref{sec:co-mul} are compatible:
\begin{align}
    \Delta^{({\rm Lax})}\;\overset{???}{=}\;\Delta^{(\myY)}\,.
\end{align}
As we will see in this section the compatibility imposes a constraint on the negative modes of the Cartan operators $\psi^{(a)}(z)$.

For the coproduct \eqref{Lax_co} of Lax operators, we have:
\begin{align}
	\begin{split}
		\Delta\CL_{\varnothing,\varnothing}=\CL_{\varnothing,\varnothing}\otimes\CL_{\varnothing,\varnothing}+\CL_{\Box,\varnothing}\otimes\CL_{\varnothing,\Box}\quad {\rm mod}\; \myY_{4}^{\otimes 2}\,,\\
		\Delta\CL_{\varnothing,\Box}=\CL_{\varnothing,\Box}\otimes\CL_{\varnothing,\varnothing}+\CL_{\Box,\Box}\otimes\CL_{\varnothing,\Box}\quad {\rm mod}\; \myY_{3}^{\otimes 2}\,,\\
		\Delta\CL_{\Box,\varnothing}=\CL_{\varnothing,\varnothing}\otimes\CL_{\Box,\varnothing}+\CL_{\Box,\varnothing}\otimes\CL_{\Box,\Box}\quad {\rm mod}\; \myY_{3}^{\otimes 2}\,.
	\end{split}
\end{align}
We then find:
\begin{align}
	\begin{split}
		\left(\Delta\CL_{\varnothing,\varnothing}\right)^{-1}\Delta\CL_{\varnothing,\Box}=&\left(\CL_{\varnothing,\varnothing}^{-1}\CL_{\varnothing,\Box}\right)\otimes 1+\\&+\left[\left(\CL_{\varnothing,\varnothing}^{-1}\CL_{\Box,\Box}\right)-\left(\CL_{\varnothing,\varnothing}^{-1}\CL_{\Box,\varnothing}\CL_{\varnothing,\varnothing}^{-1}\CL_{\varnothing,\Box}\right)\right]\otimes\left(\CL_{\varnothing,\varnothing}^{-1}\CL_{\varnothing,\Box}\right) \\
		&\hspace{5cm} {\rm mod}\;\myY_3^{\otimes 2}\,.
	\end{split}
\end{align}
Substituting \eqref{Lax_BPS} one derives:
\begin{align}\label{obstr_1}
    \Delta e(z)=\Delta_1 e(z)\quad {\rm mod}\;\myY_3^{\otimes 2}\,.
\end{align}

Comparing \eqref{obstr_1} and \eqref{coprod},  we see that these two expressions are compatible only if\footnote{Note that for simplicity, here we will not consider the presence of the truncation factors, which will shift the mode by the number of the truncation factor but will not change the result of this subsection, namely, the inconsistency for the chiral quivers.}
\begin{tcolorbox}[ams equation]\label{obstruction}
    \psi_{-k}=0,\quad k\geq 1\,.
\end{tcolorbox}

This constraint means that in the algebra no negative shifts (as defined in \eqref{shift_psi_exp}) are allowed. 
Namely, we require
\begin{align}
    \mathscr{S}_a\geq0,\quad \forall\, a\in Q_0\,.
\end{align}
Since the quiver Yangians for chiral quivers always requires negative shifts, even infinitely negative shifts when one considers infinite representations (as one generally does), see Section~\ref{sec:cry_rep}, this second no-go argument rules out all the chiral quivers.

\subsection{\texorpdfstring{Gauge-theoretical argument against chiral quivers: \\ breakdown of Higgs-Coulomb duality}{Gauge-theoretical argument against chiral quivers: breakdown of Higgs-Coulomb duality}}
\label{sec:breakdown}

Let us next explain the gauge-theory origin of the no-go results for chiral quivers.

For 2D $\CN=(2,2)$ gauged linear sigma models with a target space given by a chiral quiver $Q$,
there are two phenomena modifying the quantum description of the moduli space: the chiral anomaly and the RG running of the FI parameters.
As a result, the exponentiated complex FI parameters $e^{\I t_a}$, $a\in Q_0$ have anomalous dimensions:
\begin{align}
	\delta_a=\sum\lm_{b} N_a N_{b}\chi_{a b}\,,
\end{align}
where index $b$ runs over all quiver nodes---both gauge and framing ones.
The parameters $\chi_{ab}$ are quiver chirality parameters defined in \eqref{chirality},
and hence $\delta_a$ vanish for non-chiral quivers.

The vacuum equations (the would-be BAE) \eqref{BAE_factor} are equations for the complex scalars $\sigma_{\alpha}^{(a)}$, $a\in Q_0$, $\alpha=1,\ldots,N_a$ and take the following form:
\begin{align}\label{BAE_simp}
	\frac{P_{(a,\alpha)}\left(\sigma_{\alpha}^{(a)},\sigma_{\beta}^{(b)}\right)}{Q_{(a,\alpha)}\left(\sigma_{\alpha}^{(a)},\sigma_{\beta}^{(b)}\right)}=e^{\I t_a}\,,
\end{align}
where we imply that index pairs $(b,\beta)\neq(a,\alpha)$, and the scalars $\sigma_{(a,\alpha)}$ have dimension 1.
Comparing dimensions of the left and right hand sides we conclude:
\begin{align}
	{\rm deg}\,P_{(a,\alpha)}-{\rm deg}\,Q_{(a,\alpha)}=\delta_a\,.
\end{align}

As a consequence of the Higgs-Coulomb duality,
we expect a direct map between the crystal vacua on the Higgs branch, and the Bethe roots---solutions to \eqref{BAE_simp}---on the Coulomb branch in the large K\"ahler volume limit $\left|e^{\I t_a}\right|\to \infty$ in the cyclic chamber.
This identification goes as follows.
In the limit $\left|e^{\I t_a}\right|\to \infty$, the equations \eqref{BAE_simp} reduce to equations for denominator zeroes:
\begin{align}
	Q_{(a,\alpha)}\left(\sigma_{\alpha}^{(a)},\sigma_{\beta}^{(b)}\right)=0\,.
\end{align}
Roots of these equations correspond to vacuum expectation values of fields $\sigma_{\alpha}^{(a)}$ on the Coulomb branch that coincide with corresponding values on the Higgs branch.
On the Higgs branch $\sigma_{\alpha}^{(a)}$ acquire values in projections of crystal atoms to the complex weight plane (see Section~\ref{sec:cry_rep}).
In this way we derive a set of solutions to \eqref{BAE_simp} with the following behavior:
\begin{align}\label{roots}
	\sigma_{\alpha}^{(a)}=u_r+O\left(h_i,\left|e^{-\I t_a}\right|\right)\,,
\end{align}
for some choice of moduli $u_r$ so that $\sigma_{\alpha}^{(a)}$ belongs to crystal $\Kappa_r$ in the crystal chain.
If the quiver is chiral,\footnote{This issue is present even when only the framing part is chiral and $\chi_{\ff a}<0$ for some node $a$, so that the corresponding quiver Yangian has a negative shift $\fs_a < 0$, see Section~ \ref{sec:obstruction}.} however, there are choices of quiver dimensions $N_a$ such that $\delta_a>0$ for some $a\in Q_0$.
In this case, in addition to the roots \eqref{roots} associated with a configuration of a crystal chain, \eqref{BAE_simp} will have extra roots with the following behavior:
\begin{align}\label{roots_extra}
	\sigma_{\alpha}^{(a)}\sim \left|e^{\I\, \upsilon\, t_a}\right|,\quad \upsilon >0\,.
\end{align}
In the large volume limit, these roots run to infinity in the weight plane and do not correspond to vacua on the Higgs branch.

The $R$-matrix for quiver $Q$ is constructed as an interface in the 2D $\CN=(2,2)$ theory with a target space given by the quiver $Q$ moduli space.
As we discussed in Section~\ref{sec:gauge_deriv}, the set of Yang-Baxter equations follows from two properties:
\begin{enumerate}
	\item Flatness of parallel transport leads to an equality of parallel transport along homotopic paths on the parameter space as in \eqref{YBE}.
	\item Separability of the crystals in crystal chains allows one to assign one-site crystal modules to strands in \eqref{YBE}.
\end{enumerate}
That the quiver $Q$ is chiral does not spoil the flatness property of the parallel transport and of an associated interface.
The unitarity is spoiled, however, by an appearance of vacua \eqref{roots_extra} not associated with crystal chains: in the splitting process of the l.h.s.\ and the r.h.s.\ of \eqref{YBE} in a product of three $R$-matrices we have to sum over \emph{all} BPS vacua appearing in the theory, including \eqref{roots_extra}, for the relation \eqref{YBE} to hold.
Therefore if we restrict our construction of $R$-matrices to the crystal bases and associated coproduct structure, then the resulting $R$-matrix is not a solution of the YBE.

\section{Deriving BAE from quiver BPS algebras} \label{sec:BAE}

\subsection{Preliminaries}

In this section we extend the construction of \cite{Litvinov:2020zeq,Chistyakova:2021yyd} (see also recent developments in \cite{Kolyaskin:2022tqi,Bao:2022fpk}) for Bethe ansatz equations to BPS algebras that are not disallowed by the no-go arguments of Section~\ref{sec:no-go}, namely, the unshifted version of the BPS algebras that correspond to non-chiral quivers coming from toric CY$_3$.
The only possibilities are then the unshifted affine Yangian of $\mathfrak{g}$ where $\mathfrak{g}$ is $\mathfrak{gl}_n$, $\mathfrak{gl}_{m|n}$, or $D(2,1;\alpha)$.\footnote{For the explicit algebraic relations of $\myY(\widehat{\fg\fl}_{n})$ and $\myY(\widehat{\fg\fl}_{m|n})$, see Section~8\ of \cite{Li:2020rij}; for the $D(2,1;\alpha)$ quiver, see Section~6\ of \cite{Noshita:2021ldl}, which gave the trigonometric version of the algebra. }

This construction is rather technical and relies on the properties of algebra $\myY$ and the algebra of Lax operators.

To proceed we first assume that the generators of the algebra $\myY$ satisfy algebraic relations \eqref{QiuvYangian}, and the Lax operators can be re-expressed in terms of $\myY$ according to \eqref{Lax_BPS}, \eqref{L-rep}.

As for the $R$-matrix, it is expected to satisfy the YBE \eqref{YBE} and the unitarity constraint \eqref{unitarity}.
In addition to those standard relations we expect the $R$-matrix to have a specific relation in the limit for the spectral parameter $u_{12}\to 0$.\footnote{For the cases of $\myY(\widehat{\fg\fl}_1)$ and $\myY(\widehat{\fg\fl}_2)$, these properties were proven in \cite{Litvinov:2020zeq} and \cite{Chistyakova:2021yyd}, respectively, using the relation to the corresponding CFT. However, a CFT description is unknown for generic quiver BPS algebras in question.}
If the representations at $u_1$ and $u_2$ are isomorphic we require:
\begin{align}
    R_{12}(0)=1\otimes 1\,.
\end{align}
This property follows naturally from the existence of the coproduct $\Delta$ for $\myY$.
The coproduct behaves smoothly in the limit $u_{12}\to 0$, and there is a well-defined limit $\Delta(u_{12}=0)$. 
Moreover the $R$-matrix maps $\Delta(u_{12})$ to $\Delta(-u_{12})$, which are identical at the point $u_{12}=0$, therefore the $R$-matrix at $u_{12}=0$ acts as an identity operator.

We can graphically depict this property as:
\begin{align}\label{id}
    \begin{array}{c}
        \begin{tikzpicture}
            \draw[thick] (-0.5,0.3) to[out=0,in=180] (0.5,-0.3) (-0.5,-0.3) to[out=0,in=180] (0.5,0.3);
            \node[left] at (-0.5,0.3) {$\scriptstyle u$};
            \node[left] at (-0.5,-0.3) {$\scriptstyle u$};
            \node[right] at (0.5,0.3) {$\scriptstyle u$};
            \node[right] at (0.5,-0.3) {$\scriptstyle u$};
        \end{tikzpicture}
    \end{array}=
    \begin{array}{c}
        \begin{tikzpicture}
            \draw[thick] (-0.5,0.3) -- (0.5,0.3) (-0.5,-0.3) -- (0.5,-0.3);
            \node[left] at (-0.5,0.3) {$\scriptstyle u$};
            \node[left] at (-0.5,-0.3) {$\scriptstyle u$};
            \node[right] at (0.5,0.3) {$\scriptstyle u$};
            \node[right] at (0.5,-0.3) {$\scriptstyle u$};
        \end{tikzpicture}
    \end{array}\,.
\end{align}

We remind the reader that here  we will only consider \emph{representations corresponding to 2D crystals}, see the discussion at the end of Section~\ref{sec:cry_dim}.

\subsection{Off-shell Bethe vectors}

Let us assign some choices of colors $b\in Q_0$ to crystals with centers-of-mass located in different positions $x_i^{(b)}$.
Let us call this set of parameters:
\begin{align}
	\vec x:=\left\{\left\{x_i^{(b)}\right\}_{i=1}^{N_b}\right\}_{b\in Q_0}\,,
\end{align}
and denote the corresponding empty crystal chain state as:
\begin{align}
	|\tilde{\pmb{\varnothing}}\rangle=|\varnothing\rangle_{x_1^{(b_1)}}\otimes  \cdots \otimes |\varnothing\rangle_{x_{d_{b_m}}^{(b_m)}}\,.
\end{align}
 Following \cite{Litvinov:2020zeq} we define a state $|\chi(\vec x)\rangle$ as the following contour integral:
 \begin{align}\label{chi}
 	|\chi(\vec x)\rangle:=\oint\lm_{x_1^{(b_1)}}dz_1^{(b_1)}\, \coe^{(b_1)}\left(z_1^{(b_1)}\right)\cdots \oint\lm_{x_{d_{b_m}}^{(b_m)}}dz_1^{(b_m)}\, \coe^{(b_m)}\left(z_{d_{b_m}}^{(b_m)}\right)\;|\tilde{\pmb{\varnothing}}\rangle\,,
 \end{align}
where all the integrals are computed along small circle contours around the corresponding poles $\vec x$.\footnote{Let us note that some authors \cite{Litvinov:2020zeq,Bao:2022fpk} make an unjustified step and identify the resulting state $|\chi\rangle$ with a simple crystal chain having a single atom at each site. This would be a correct observation if the coproduct for raising operators had the simple structure of \eqref{chain_naiv_co}. However, due to the non-trivial coproduct structure---a conjugation by the matrix $U$ bringing in non-trivial poles in $z$, the actual state $|\chi\rangle$ is a mixed state in the space of crystal chains in the general case:
$$
	|\chi(\vec x)\rangle \neq |\Box\rangle_{x_1^{(b_1)}}\otimes  \cdots \otimes |\Box\rangle_{x_{d_{b_m}}^{(b_m)}}\,.
$$}
	
We will consider a chain of modules, all corresponding to the same canonical framing $\ff$.
We denote the empty crystal in this chain module in the following way:
\begin{align}
	|\pmb{\varnothing}\rangle=|\varnothing\rangle_{u_1}\otimes |\varnothing\rangle_{u_2}\otimes \cdots \otimes |\varnothing\rangle_{u_n}\,.
\end{align}

We construct an off-shell Bethe vector via a scattering process where $|\chi(\vec{x})\rangle$ goes to $|\tilde{\pmb{\varnothing}}\rangle$, whereas vacuum $|\pmb{\varnothing}\rangle$ scatters to the off-shell Bethe vector $|B(\vec x)\rangle$:
	\begin{align}\label{diag:off-shell-B}
		|B(\vec x)\rangle:=\begin{array}{c}
			\begin{tikzpicture}[xscale=0.8]
				\draw[dashed] (4,0) to[out=180,in=0] (0,1.5);
				\draw[thick] (4,0.3) to[out=180,in=0] (0,1.8);
				\draw[dashed] (4,0.6) to[out=180,in=0] (0,2.1);
				\draw[thick] (4,0.9) to[out=180,in=0] (0,2.4);
				\draw[dashed] (4,1.2) to[out=180,in=0] (0,2.7);
				\draw[thick] (4,2.1) to[out=180,in=0] (0,0);
				\draw[dashed] (4,2.4) to[out=180,in=0] (0,0.3);
				\draw[thick] (4,2.7) to[out=180,in=0] (0,0.6);
				%
				\node[right] at (4,0.6) {$\scriptstyle |\chi(\vec x)\rangle$};
				\node[right] at (4,2.1) {$\scriptstyle (\varnothing,u_1)$};
				\node[right] at (4,2.7) {$\scriptstyle (\varnothing,u_n)$};
				\node[left] at (0,1.8) {$\scriptstyle (\varnothing,x_1^{(b)})$};
				\node[left] at (0,2.4) {$\scriptstyle (\varnothing,x_{d_b}^{(b)})$};
			\end{tikzpicture}
		\end{array}
	\end{align}
	with unspecified states on the external legs on the lower right.
This pictorial notation should be translated into:
\begin{align}\label{eq:off-shell-B}
    |B(\vec x)\rangle=\sum\lm_{\vec K}\coL_{\varnothing,\Kappa_1^{(b_1)}}^{(b_1)}\left(x_1^{(b_1)}\right)\cdots \coL_{\varnothing,\Kappa_{d_m}^{(b_m)}}^{(b_m)}\left(x_{d_m}^{(b_m)}\right)|\pmb{\varnothing}\rangle \cdot \langle\vec \Kappa|\chi\rangle\,.
\end{align}

\subsection{Derivation of Bethe ansatz equations}

In this subsection,  we will show that for an off-shell Bethe vector $|B(\vec x)\rangle$ to satisfy the eigenvalue equation \eqref{eigen_value}, the Bethe variables $\vec x$ have to solve the Bethe ansatz equations.

Let us first project equation \eqref{eigen_value}, with $|B\rangle$ given by the off-shell Bethe vector \eqref{eq:off-shell-B}, to the state:
\begin{align}
\langle\vec \Kappa|=\bigotimes\lm^{n}_{i=1} {}_{u_i}\langle \Kappa_i|\,.
\end{align}
Such a projection means that we assign the state ${}_{ u_i}\langle \Kappa_i|$ to the $i$-th external leg at the lower right part of the diagram \eqref{diag:off-shell-B}, and can be represented by 
\begin{align}
    \begin{array}{c}
        \begin{tikzpicture}[yscale=0.8]
            \draw[dashed] (0,0.5) -- (1,0.5) -- (1,-2) -- (0,-2) -- cycle (0,-0.5) -- (-1,-0.5);
            \draw[thick] (0,0) -- (-1,0) (0,-1) -- (-1,-1) (0,-1.5) -- (-1,-1.5);
        \end{tikzpicture}
    \end{array}\mathop{\longrightarrow}\lm^{\langle\vec \Kappa|}
    \begin{array}{c}
        \begin{tikzpicture}[yscale=0.8]
            \draw[dashed] (0,0.5) -- (1,0.5) -- (1,-2) -- (0,-2) -- cycle (0,-0.5) -- (-1,-0.5);
            \draw[thick] (0,0) -- (-1,0) (0,-1) -- (-1,-1) (0,-1.5) -- (-1,-1.5);
            \node[left] at (-1,0) {$\scriptstyle (\Kappa_n,a,u_n)$};
            \node[left] at (-1,-1) {$\scriptstyle (\Kappa_2,a,u_2)$};
            \node[left] at (-1,-1.5) {$\scriptstyle (\Kappa_1,a,u_1)$};
        \end{tikzpicture}
    \end{array}\,.
\end{align}
Note that for each color $b\in Q_0$, the total number of atoms with color $b$ in this collection of crystal states $\{\Kappa_i\}$ is constrained to be:
\begin{align}\label{const1}
\sum\lm_{i=1}^n|\Kappa_i^{(b)}|=N_b,\quad\forall b\in Q_0\,.
\end{align}

After applying this projection, we notice that the eigenvalue problem \eqref{eigen_value} can be transformed into the following set of graphic relations for arbitrary  $\vec \Kappa$ that satisfies the constraint \eqref{const1}:
\begin{align}\label{Bethe}
	{\color{\myblue}\sum\lm_{\Kappa}}\begin{array}{c}
		\begin{tikzpicture}[xscale=0.8]
			\draw[dashed] (4,0.3) to[out=180,in=0] (0,1.8);
			\draw[thick] (4,0.6) to[out=180,in=0] (0,2.1);
			\draw[dashed] (4,0.9) to[out=180,in=0] (0,2.4);
			\draw[thick] (4,1.2) to[out=180,in=0] (0,2.7);
			\draw[dashed] (4,1.5) to[out=180,in=0] (0,3.0);
			\draw[thick] (4,2.4) to[out=180,in=0] (0,0);
			\draw[dashed] (4,2.7) to[out=180,in=0] (0,0.3);
			\draw[thick] (4,3.0) to[out=180,in=0] (0,0.6);
			\node[right] at (4,0.9) {$\scriptstyle |\chi(\vec x)\rangle$};
			\node[right] at (4,2.4) {$\scriptstyle (\varnothing,u_1)$};
			\node[right] at (4,3.0) {$\scriptstyle (\varnothing,u_n)$};
			\node[left] at (0,2.1) {$\scriptstyle (\varnothing,x_1^{(b)})$};
			\node[left] at (0,2.7) {$\scriptstyle (\varnothing,x_{d_b}^{(b)})$};
			\node[left] at (0,0) {$\scriptstyle (\Kappa_1,u_1)$};
			\node[left] at (0,0.6) {$\scriptstyle (\Kappa_n,u_n)$};
			\draw[thick,\myblue] (4,0) to[out=180,in=0] node[pos=0.9] {
					\begin{tikzpicture}
						\draw[fill=white] (0,0) circle (0.08);
					\end{tikzpicture}
				} (0,1.5);
			\node[right,black!20!blue] at (4,0) {$\scriptstyle (\Kappa,z)$};
			\node[left,black!20!blue] at (0,1.5) {$\scriptstyle (\Kappa,z)$};
		\end{tikzpicture}
	\end{array}=\Lambda(z)\begin{array}{c}
	\begin{tikzpicture}[xscale=0.8]
		\draw[dashed] (4,0) to[out=180,in=0] (0,1.5);
		\draw[thick] (4,0.3) to[out=180,in=0] (0,1.8);
		\draw[dashed] (4,0.6) to[out=180,in=0] (0,2.1);
		\draw[thick] (4,0.9) to[out=180,in=0] (0,2.4);
		\draw[dashed] (4,1.2) to[out=180,in=0] (0,2.7);
		\draw[thick] (4,2.1) to[out=180,in=0] (0,0);
		\draw[dashed] (4,2.4) to[out=180,in=0] (0,0.3);
		\draw[thick] (4,2.7) to[out=180,in=0] (0,0.6);
		\node[right] at (4,0.6) {$\scriptstyle |\chi(\vec x)\rangle$};
		\node[right] at (4,2.1) {$\scriptstyle (\varnothing,u_1)$};
		\node[right] at (4,2.7) {$\scriptstyle (\varnothing,u_n)$};
		\node[left] at (0,1.8) {$\scriptstyle (\varnothing,x_1^{(b)})$};
		\node[left] at (0,2.4) {$\scriptstyle (\varnothing,x_{d_b}^{(b)})$};
		\node[left] at (0,0) {$\scriptstyle (\Kappa_1,u_1)$};
		\node[left] at (0,0.6) {$\scriptstyle (\Kappa_n,u_n)$};
	\end{tikzpicture}
\end{array}
\end{align}
where the eigenvalue $\Lambda(z)$ is the same for all $\vec \Kappa$.

To simplify these relations, we apply the trick from \cite{Litvinov:2020zeq} and set $z=u_1$.
The l.h.s.\ of \eqref{Bethe} simplifies after applying \eqref{id}:
\begin{align}\label{trick}
	\begin{array}{c}
		\begin{tikzpicture}[xscale=0.8]
			\draw[dashed] (4,0.3) to[out=180,in=0] (0,1.8);
			\draw[thick] (4,0.6) to[out=180,in=0] (0,2.1);
			\draw[dashed] (4,0.9) to[out=180,in=0] (0,2.4);
			\draw[thick] (4,1.2) to[out=180,in=0] (0,2.7);
			\draw[dashed] (4,1.5) to[out=180,in=0] (0,3.0);
			\draw[dashed] (4,2.7) to[out=180,in=0] node(A) [pos=0.6] {} (0,0.3);
			\draw[thick] (4,3.0) to[out=180,in=0] (0,0.6);
			\draw[thick] (4,0) to[out=180,in=0] ([shift={(0,-0.4)}]A.south) to[out=180,in=0] (0,0);
			\node[right] at (4,0.9) {$\scriptstyle |\chi(\vec x)\rangle$};
			\node[right] at (4,0) {$\scriptstyle (\Kappa_1,u_1)$};
			\node[right] at (4,3.0) {$\scriptstyle (\varnothing,u_n)$};
			\node[left] at (0,2.1) {$\scriptstyle (\varnothing,x_1^{(b)})$};
			\node[left] at (0,2.7) {$\scriptstyle (\varnothing,x_{d_b}^{(b)})$};
			\node[left] at (0,0) {$\scriptstyle (\Kappa_1,u_1)$};
			\node[left] at (0,0.6) {$\scriptstyle (\Kappa_n,u_n)$};
			\draw[thick] (4,2.4) to[out=180,in=0] ([shift={(0,-0.2)}]A.south) to [out=180,in=0] node[pos=0.7] {
				\begin{tikzpicture}
					\draw[fill=white] (0,0) circle (0.08);
				\end{tikzpicture}
			} (0,1.2);
			\node[right] at (4,2.4) {$\scriptstyle (\varnothing,u_1)$};
			\node[left] at (0,1.2) {$\scriptstyle (\Kappa_1,u_1)$};
		\end{tikzpicture}
	\end{array}=\begin{array}{c}
	\begin{tikzpicture}[xscale=0.8]
		\draw[dashed] (4,0) to[out=180,in=0] (0,1.5);
		\draw[thick] (4,0.3) to[out=180,in=0] (0,1.8);
		\draw[dashed] (4,0.6) to[out=180,in=0] (0,2.1);
		\draw[thick] (4,0.9) to[out=180,in=0] (0,2.4);
		\draw[dashed] (4,1.2) to[out=180,in=0] (0,2.7);
		\draw[dashed] (4,2.1) to[out=180,in=0] (0,0);
		\draw[thick] (4,2.4) to[out=180,in=0] (0,0.3);
		\draw[thick] (4,2.7) to[out=180,in=0] node[pos=0.9] {
			\begin{tikzpicture}
				\draw[fill=white] (0,0) circle (0.08);
			\end{tikzpicture}
		} (0,0.6);
		\node[right] at (4,0.6) {$\scriptstyle |\chi(\vec x)\rangle$};
		\node[right] at (4,2.4) {$\scriptstyle (\varnothing,u_n)$};
		\node[right] at (4,2.7) {$\scriptstyle (\varnothing,u_1)$};
		\node[left] at (0,1.8) {$\scriptstyle (\varnothing,x_1^{(b)})$};
		\node[left] at (0,2.4) {$\scriptstyle (\varnothing,x_{d_b}^{(b)})$};
		\node[left] at (0,0.3) {$\scriptstyle (\Kappa_n,u_n)$};
		\node[left] at (0,0.6) {$\scriptstyle (\Kappa_1,u_1)$};
	\end{tikzpicture}
\end{array}\,,
\end{align}
where we have used the fact that the $R$-matrix act trivially on empty crystals.

Now we would like to compare the right hand sides of \eqref{Bethe} and \eqref{trick}. 
One looks at those diagrams from a different angle and treats now the braiding operations as actions of $\CL$ operators with spectral parameters $u_1,\ldots,u_n$ on a tensor power of Fock spaces specified by the spectral parameters $\vec x$.
In these terms we are able to rewrite these equations as:
\begin{align}\label{main}
	\begin{split}
	&\langle\tilde{\pmb{\varnothing}}|{\bf s}_{\Kappa_1}\coL_{\Kappa_1,\varnothing}(u_1)\coL_{\Kappa_n,\varnothing}(u_n)\ldots \coL_{\Kappa_2,\varnothing}(u_2)\;\left|\chi(\vec x)\right\rangle=\\
	&=\Lambda(u_1)\langle\tilde{\pmb{\varnothing}}|\coL_{\Kappa_n,\varnothing}(u_n)\ldots \coL_{\Kappa_2,\varnothing}(u_2)\coL_{\Kappa_1,\varnothing}(u_1)\;\left|\chi(\vec x)\right\rangle
	\end{split}
\end{align}

To simplify this equation further we apply the representation \eqref{L-rep} for the $\CL$ operators:
\begin{align}\label{main_expand}
	\begin{split}
	\oint d\vec y\;F_{\vec \Kappa}\left(\vec y\right)\left[\prod\lm_{b\in Q_0}\fq_b^{|\Kappa_1^{(b)}|}\right]\Bigg\langle\prod\lm_{b\in Q_0}\prod\lm_{\alpha=1}^{|\Kappa_1^{(b)}|}\cof^{(b)}\left(y_{(\alpha|1)}^{(b)}\right){\bf \myt}_{\ff}(u_1)\times\\
	\times\prod\lm_{i=2}^n \prod\lm_{b\in Q_0}\prod\lm_{\alpha=1}^{|\Kappa_i^{(b)}|}\cof^{(b)}\left(y_{(\alpha|i)}^{(b)}\right){\bf \myt}_{\ff}(u_i)\Bigg\rangle=\\
	=\Lambda(u_1)\oint d\vec y\;F_{\vec \Kappa}\left(\vec y\right)\Bigg\langle\prod\lm_{i=2}^n \prod\lm_{b\in Q_0}\prod\lm_{\alpha=1}^{|\Kappa_i^{(b)}|}\cof^{(b)}\left(y_{(\alpha|i)}^{(b)}\right){\bf \myt}_{\ff}(u_i)\times\\
	\times\prod\lm_{b\in Q_0}\prod\lm_{\alpha=1}^{|\Kappa_1^{(b)}|}\cof^{(b)}\left(y_{(\alpha|1)}^{(b)}\right){\bf \myt}_{\ff}(u_1)\Bigg\rangle\,,
	\end{split}
\end{align}
where
$$
F_{\vec \Kappa}\left(\vec y\right)=\prod\lm_{i=1}^nF_{\Kappa_i}\left(y_{(*|i)}^{(*)}\right)\,,
$$
and the matrix elements $\langle\ldots\rangle$ are computed between the same states as in \eqref{main}.

The eigenvalue $\Lambda(u_1)$ in \eqref{main} can be easily fixed from the choice $\Kappa_1=\varnothing$. 
Since it is independent from the set of the crystal states $\vec \Kappa$, it is clear that its eigenvalue corresponds to a factor  appearing from a permutation of $h(u_1)$ from the end of the operator expression to the front, thus we can eliminate $\Lambda(u_1)$ from \eqref{main_expand}:
\begin{align}\label{main_mod2}
	\begin{split}
		\oint d\vec y\;F_{\vec \Kappa}\left(\vec y\right)\left[\prod\lm_{b\in Q_0}\fq_b^{|\Kappa_1^{(b)}|}\right]\Bigg\langle\prod\lm_{b\in Q_0}\prod\lm_{\alpha=1}^{|\Kappa_1^{(b)}|}\cof^{(b)}\left(y_{(\alpha|1)}^{(b)}\right){\bf \myt}_{\ff}(u_1)\times\\
		\times\prod\lm_{i=2}^n \prod\lm_{b\in Q_0}\prod\lm_{\alpha=1}^{|\Kappa_i^{(b)}|}\cof^{(b)}\left(y_{(\alpha|i)}^{(b)}\right){\bf \myt}_{\ff}(u_i)\Bigg\rangle=\\
		=\oint d\vec y\;F_{\vec \Kappa}\left(\vec y\right)\Bigg\langle {\bf \myt}_{\ff}(u_1)\prod\lm_{i=2}^n \prod\lm_{b\in Q_0}\prod\lm_{\alpha=1}^{|\Kappa_i^{(b)}|}\cof^{(b)}\left(y_{(\alpha|i)}^{(b)}\right){\bf \myt}_{\ff}(u_i)\times \\ \times\prod\lm_{b\in Q_0}\prod\lm_{\alpha=1}^{|\Kappa_1^{(b)}|}\cof^{(b)}\left(y_{(\alpha|1)}^{(b)}\right)\Bigg\rangle\,.
	\end{split}
\end{align}
In this relation, the expressions in brackets differ only by the ordering of the lowering operators.
Then using the commutation relations of the $\cof$-operators and the
${\bf \myt}$-operators, we can move the lowering operators one by one from the rightmost to the leftmost in the r.h.s.\ of \eqref{main_mod2}.
Each such movement of the lowering operator of $y^{(b)}_A$ (where $A=(i|1)$ is some double index) will produce a factor:
\begin{align}\label{factor}
\BAE_A^{(b)}(\vec y,\vec u,\vec q)\,,
\end{align}
where the phase $\BAE$ is defined in \eqref{BAE_factor}.

When computing  both sides of \eqref{main_mod2}, we have to pair the $\cof$ operators with the $\coe$ operators that appear in the definition of the state $|\chi(\vec x)\rangle$. The elementary pairing takes the following form:
\begin{align}
    \langle\varnothing|\cof^{(a)}(y)\coe^{(b)}(z)|\varnothing\rangle= \sum\lm_p\frac{\delta_{ab}\;\mathop{\rm res}\lm_{u=p}\langle\varnothing|\copsi^{(a)}(u)|\varnothing\rangle}{(y-p)(z-p)}\,,
\end{align}
where $p$ runs over all the possible poles of the operator $\copsi^{(a)}$. 
The following integration over $\vec z$ along small circles around $\vec x$ cuts out in \eqref{main_mod2} a contribution from only those poles $c$ located at some of $\vec x$.
Similarly, the integration over $\vec y$ forces the support of the integrand to shrink to $\vec x$ only.
Therefore different pairings between $\cof$ and $\coe$ operators induce different permutations of maps of the integrand support $\vec y\to \vec x$.

Using an identification between $\vec y$ and $\vec x$ we see that \eqref{main_mod2} and \eqref{main} are satisfied trivially and sufficiently if the points $\vec x$ are roots of the Bethe ansatz equations:
\begin{tcolorbox}[ams equation]\label{BAE}
\BAE_i^{(b)}\left(\vec x,\vec u,\vec \fq\right)=1,\quad \forall\; b\in Q_0, \;i=1,\ldots,N_b\,,
\end{tcolorbox}
\noindent where the $\BAE$ is defined in \eqref{BAE_factor}.

Finally, we mention that since the BAE of a physical system has net degree zero, the BAE \eqref{BAE} as defined by \eqref{BAE_factor} provides another argument against the shift and chiral quivers. 

\section{\texorpdfstring{Rational vs.\ trigonometric quiver algebras: \\ comparative analysis}{Rational vs. trigonometric quiver algebras: comparative analysis}} \label{sec:trig}

The quiver BPS algebras admit generalizations to trigonometric \cite{Galakhov:2021vbo,Noshita:2021dgj,Noshita:2021ldl} and elliptic \cite{Galakhov:2021vbo} versions.
For a rational BPS algebra $\myY$ we will denote its trigonometric version by $\ddot{\myY}$, which is also referred to as the \emph{quantum toroidal} algebra.\footnote{In this paper we will set the central elements (denoted by $c$ in \cite{Galakhov:2021vbo}) to be $c=0$.}
$\ddot{\myY}$ plays the same role for the Gauge/Bethe correspondence for 3D $\mathcal{N}=2$ quiver gauge theories as the one $\myY$ plays for the Gauge/Bethe correspondence for 2D $\mathcal{N}=(2,2)$ quiver gauge theories.\footnote{Similarly, the quiver elliptic algebras defined in \cite{Galakhov:2021vbo} are the relevant algebras for the Gauge/Bethe correspondence for 4D $\mathcal{N}=1$ quiver gauge theories, but we will leave them to future study.}

Generalization from $\myY$ to $\ddot\myY$ is straightforward as far as the generators and relations are concerned, yet it has its own peculiarities.
We devote this section to a comparative analysis of the coproduct, the $R$-matrix and BAE in the cases of $\myY$ and $\ddot\myY$. We will not give a thorough review of $\myY$ here: for the definition of $\ddot\myY$, see Section~2.2 (in particular (2.7)) of \cite{Galakhov:2021vbo}, and for their crystal representations, see Section~3.2 (in particular (3.21)) of \cite{Galakhov:2021vbo}. 

The basic step in this generalization $\myY\to \ddot\myY$ is the promotion of the basic building block---the rational bond factor \eqref{bond_factor}---to a trigonometric function:
\begin{align}\label{trig_bf}
	\ddot\varphi^{a\Leftarrow b} (u_1,u_2)\myequiv (-1)^{|b\rightarrow a| \chi_{ab}} e^{\chi_{ab}\beta(u_1+u_2)} \frac{\prod\lm_{I\in \{a\rightarrow b\}} \sinh \beta\left(u_1-u_2+h_{I}\right) }{\prod\lm_{J\in \{b\rightarrow a\}} \sinh \beta\left( u_1-u_2-h_{J} \right)} \;,
\end{align}
where $\beta$ parameterizes the trigonometric deformation, so that in the limit $\beta=0$ \eqref{trig_bf} reduces to \eqref{bond_factor} with $u=u_1-u_2$.
As observed in \cite{Galakhov:2021vbo}, there is a non-trivial dependence of the bond factor on the center of mass variable $u_1+u_2$ in \eqref{bond_factor} that is lifted if $\chi_{ab}=0$.

\subsection{Spectral parameters}\label{ssec:spec_param}

The first discrepancy in the structures of $\myY$ and $\ddot\myY$ arises in the properties of the spectral parameter:
in the case of $\myY$ the spectral parameter $u\in \IC$, whereas in the case of $\ddot\myY$ the parameter $u$ belongs to a cylinder with periodicity $2\pi\I\beta^{-1}$.
It is useful to map the latter cylinder to $\IC^{\times}$ with coordinate $U=e^{\beta u}$.

For the field description of the generators of the algebra, the notion of the delta function is relevant.
We have 
\begin{align}
	\delta^{\rm rat.}(u)=\frac{1}{u},\quad \delta^{\rm trig.}(U)=\sum\lm_{k\in\IZ}U^k\,,
\end{align}
for $u\in\IC$ and $U\in\IC^{\times}$, respectively.
The localizing properties of these functions are slightly different.
For any polynomial $f(x)$ in $x\in\IC$ we have:
\begin{align}\label{del_rat}
	\delta^{\rm rat.}(u-x)f(u)=\delta^{\rm rat.}(u-x)f(x)+\sum\lm_{k\geq 0}t_k (u-x)^k\,,
\end{align}
and for any Laurent series $f(X)$ in $X\in\IC^{\times}$ we have:
\begin{align}\label{del_trig}
	\delta^{\rm trig.}(U/X)f(U)=\delta^{\rm trig.}(U/X)f(X)\,.
\end{align}

Apparently, raising/lowering generators in the crystal representation \eqref{crystal_rep} contain delta-functions.
Due to the ``non-exactness" in the localizing relation for $\delta^{\rm rat.}$, or exactness up to $u^{k\geq 0}$ corrections, the algebraic relations \eqref{QiuvYangian} contain equivalence signs $\simeq$ and $\sim$ up to positive modes in the spectral parameters rather than equality signs.
In the case of $\ddot\myY$, analogous $ee$- and $ff$-relations can be promoted to equalities.

\subsection{Coproduct, \texorpdfstring{$R$}{R}-matrices and obstruction}

In the case of $\ddot\myY$, the mode expansion for the generators is performed on $\IC^{\times}$:
\begin{align}
	\ddot e^{(a)}(U)=\sum\lm_{k\in\IZ}\ddot e^{(a)}_k U^{-k},\quad \ddot f^{(a)}(U)=\sum\lm_{k\in\IZ}\ddot f^{(a)}_k U^{-k},\quad \ddot \psi_{\pm}^{(a)}(U)=\sum\lm_{k\in\IZ}\ddot{\psi}^{(a)}_{\pm,k} U^{-k}\,,
\end{align} 
where the Cartan $\psi$-generators are doubled.

Another drastic difference between $\myY$ and $\ddot\myY$ also has its origin in the exactness \eqref{del_trig} of $\delta^{\rm trig}$.
A substitution of the denominator factors in \eqref{chain_naiv_co} representing $\delta^{\rm rat.}$ by $\delta^{\rm trig.}$ allows the naive tensor representation ${\ddot\bDelta}_0$ to factorize:
\begin{align}
	{\ddot\bDelta}_0=\ddot{\rm Rep}\circ {\ddot\Delta}_1\,,
\end{align}
where $\ddot\Delta_1$ is a coproduct on $\ddot\myY$ (compare with $\Delta_1$ \eqref{Delta_1} that is not a coproduct on $\myY$):
\begin{align}\label{dotDelta_1}
	\begin{split}
		&\ddot\Delta_1:\quad \ddot\myY \longrightarrow \ddot\myY\otimes \ddot\myY\,,\\
		&\ddot\Delta_1\ddot e^{(a)}(U)=\ddot e^{(a)}(U)\otimes 1+\ddot\psi^{(a)}_+(U)\otimes\ddot e^{(a)}(U)\,,\\
		&\ddot\Delta_1\ddot f^{(a)}(U)=\ddot f^{(a)}(U)\otimes \ddot\psi^{(a)}_-(U)+1\otimes\ddot f^{(a)}(U)\,,\\
		&\ddot\Delta_1\ddot \psi_{\pm}^{(a)}(U)=\ddot \psi_{\pm}^{(a)}(U)\otimes \ddot \psi_{\pm}^{(a)}(U)\,.
	\end{split}
\end{align}

The resulting $R$-matrix for $\ddot\Delta_1$ is equivalent to an $R$-matrix for $\ddot\bDelta_0$ and is a straightforward trigonometric generalization of the diagonal matrix \eqref{R-diag}.
The trigonometric $R$-matrix generalizing the non-trivial rational $R$-matrix \eqref{main_R} intertwines another coproduct structure $\ddot\Delta$, so that $\ddot\myY$ has \emph{two} nonequivalent coproduct structures: $\ddot\Delta$ and $\ddot\Delta_1$.

Unfortunately, in the case of $\ddot\myY$ the strategy of Section~\ref{sec:co-mul} to find a representation homomorphism $\ddot U$ (cf. \eqref{iso}) conjugating $\ddot\Delta_1$ into $\ddot\Delta$ is inapplicable, and we have to determine $\ddot U$ from other principles.
In the case of the toroidal algebra $U_{q_1,q_2}(\widehat{\widehat{\fg\fl}}_1)$ (a trigonometric counterpart of $\myY(\widehat{\fg\fl}_1)$) $\ddot U$ results from the Miki automorphism (see Appendix~\ref{app:Miki}).

For a generic $\myY$ we  \emph{assume} that $\ddot\Delta_1$ and $\ddot\Delta$  are conjugated by a  homomorphism $\ddot U$ that has an expansion similar to \eqref{U_op} with $S_k$ of the form \eqref{S_k_form}. Moreover for $S_1$ we simply generalize expression \eqref{tor_ff} derived for the quantum toroidal algebra $U_{q_1,q_2}(\widehat{\widehat{\fg\fl}}_1)$ to multiple colors.

Then for the Lax operators we can derive expressions analogous to \eqref{Lax_BPS} and \eqref{L-rep}:
\begin{align}\label{qLax}
	\scalebox{1.0}{$
	\begin{array}{l}
		\ddot{\CL}_{\varnothing,\varnothing}\left(u_1/u_2\right)|\Kappa\rangle_{u_2}=\left(\prod\lm_{\Box\in\Kappa}{}_{u_1}\langle\varnothing|\Psi_q\left(\frac{u_2}{u_1}H_{\Box}\right)|\varnothing\rangle_{u_1}^{-1}\right)\times |\Kappa\rangle_{u_2}\,,\\
		\ddot{\CL}_{\varnothing,\varnothing}\left(u_1/u_2\right)^{-1}\ddot{\CL}_{\varnothing,\Box}\left(u_1/u_2\right)|\Kappa\rangle_{u_2}=\sum\lm_{\Box\in{\rm Add}(\Kappa)}\frac{\kappa\left[\Kappa\to\Kappa+\Box\right]_q}{1-\frac{u_2H_{\Box}}{u_1}}|\Kappa+\Box\rangle_{u_2}\,,\\
		\ddot{\CL}_{\Box,\varnothing}\left(u_1/u_2\right)\ddot{\CL}_{\varnothing,\varnothing}\left(u_1/u_2\right)^{-1}|\Kappa\rangle_{u_2}=\sum\lm_{\Box\in{\rm Rem}(\Kappa)}\frac{\kappa\left[\Kappa\to\Kappa-\Box\right]_q}{1-\frac{u_2H_{\Box}}{u_1}}|\Kappa-\Box\rangle_{u_2}\,,\\
		\ddot{\CL}_{\Kappa,\varnothing}\left(u_1/u_2\right)\ddot{\CL}_{\varnothing,\varnothing}\left(u_1/u_2\right)^{-1}|\Kappa'\rangle_{u_2}=\\
		=\sum\lm_{\Box_1\in{\rm Rem}(\Kappa')}\sum\lm_{\Box_2\in{\rm Rem}(\Kappa'-\Box_1)}\sum\lm_{\Box_3\in{\rm Rem}(\Kappa'-\Box_1-\Box_2)}\ldots\\
		G_{\Kappa}\left(\frac{u_2 H_{\Box_1}}{u_1},\frac{u_2 H_{\Box_2}}{u_1},\frac{u_2 H_{\Box_3}}{u_1},\ldots\right)\left[\Kappa'\to\Kappa'-\Box_1\to\Kappa'-\Box_1-\Box_2\to\ldots\right]_q\times\\
		|\Kappa-\Box_1-\Box_2-\Box_3-\ldots\rangle_{u_2}\,,
	\end{array}$}
\end{align}
where the subscript $q$ in $\Psi_q$ and $[\Kappa\to\Kappa']_q$ indicates that those are \emph{quantum} toroidal analogues of similar quantities in $\myY$, see equation (3.19) in \cite{Galakhov:2021vbo}.
These relations have transparent generalizations to multiple colors.

For the rational algebra $\myY$ we have derived in Sections \ref{sec:obstruction} and \ref{sec:LYcc}  obstructions for BPS algebras corresponding to toric CY${}_3$ with compact 4-cycles (represented by chiral quivers) and to the BPS algebras with negative shifts.  For the trigonometric versions $\ddot\myY$ we can follow the same derivation step-by-step. 
The key ingredient of the obstacle---the appearance of negative modes in an expansion of rational trigonometric functions:
\begin{align}
	f(u)=\frac{\prod\lm_{i\in \CI} \sinh \beta(u-p_i)}{\prod\lm_{j\in \CJ} \sinh \beta(u-q_j)}\,,
\end{align}
where $\CI$ and $\CJ$ are some index sets,
 over pole contributions---follows from the  expansion (the analogue of \eqref{pole_exp}):
\begin{align}
	f(u)=2\sum\lm_{j\in\CJ}e^{-\beta\left(|\CI|-|\CJ|\right)(u-q_j)}\times\left(\frac{\lim\lm_{w\to 0}\left(\sinh\beta w\right) f(w+q_k)}{e^{2\beta(u-q_k)}-1}+\sum\lm_{r\geq 0}\alpha_re^{r\beta u}\right)\,.
\end{align}

\subsection{Bethe ansatz equations}

The major difficulty in generalizing the derivation of the BAE to the trigonometric case along the lines of Section~\ref{sec:BAE} is that the canonical derivation \cite{2015arXiv150207194F} for quantum toroidal $U_{q_1,q_2}(\widehat{\fg\fl}_1)$ implements a computation in terms of shuffle modules.
And the map back to the crystal description is an involved problem on its own.
A review of shuffle algebras associated with quiver BPS algebras can be found in \cite[Section~5.2]{Galakhov:2021vbo}.

In this section we will try to mimic the basic elements of the computation in Section~\ref{sec:BAE} in terms of $\ddot\myY$.

Let us strip the weight lattice $\mathscr{L}$ \eqref{lattice} of edges and consider only points on the complex plane.
These points can be further mapped to $\IC^{\times}$ by an exponential map $H=e^{\beta h}$.
We can split the group of lattice points into subgroups $\mathscr{P}\!\!\mathscr{L}^{(a)}$ of atoms of definite color $a\in Q_0$.
Let us denote an atom corresponding to a point $p\in\mathscr{P}\!\!\mathscr{L}^{(a)}$ as $\sqbox{$a$}(p)$, and construct a natural raising (lowering) operator $E_p^{(a)}$ ($F_p^{(a)}$) according to the following rule:
\begin{align}
	\begin{split}
	E_p^{(a)}|\Kappa\rangle&:=\left\{\begin{array}{ll}
		[\Kappa\to \Kappa+\sqbox{$a$}(p)]_q|\Kappa+\sqbox{$a$}(p)\rangle,&\mbox{ if }\sqbox{$a$}(p)\in{\rm Add}(\Kappa)\,;\\
		0,& \mbox{ otherwise}\,;
	\end{array}\right.\\
	F_p^{(a)}|\Kappa\rangle&:=\left\{\begin{array}{ll}
		[\Kappa\to \Kappa-\sqbox{$a$}(p)]_q|\Kappa-\sqbox{$a$}(p)\rangle,&\mbox{ if }\sqbox{$a$}(p)\in{\rm Rem}(\Kappa)\,;\\
		0,& \mbox{ otherwise}\,.
	\end{array}\right.
	\end{split}
\end{align}

These operators satisfy the following commutation relations (cf. \eqref{QiuvYangian}):
\begin{align}
	\begin{split}
	&E^{(a)}_xE^{(b)}_y=(-1)^{|a||b|}\ddot\varphi^{a\Leftarrow b}(x,y)\,E^{(b)}_yE^{(a)}_x\,,\\ &F^{(a)}_xF^{(b)}_y=(-1)^{|a||b|}\ddot\varphi^{a\Leftarrow b}(x,y)^{-1}\,F^{(b)}_yF^{(a)}_x\,,\\
	&[E^{(a)}_x,F^{(b)}_y\}\sim-\delta_{ab}\delta_{xy}\left\{\begin{array}{ll}
		1,&\mbox{ if }\sqbox{$a$}(x)\in {\rm Rem}(\Kappa)\cup{\rm Add}(\Kappa)\,,\\
		0, &\mbox{ otherwise}\,.
	\end{array}\right.\\
	\end{split}
\end{align}

It is a simple task to rewrite raising/lowering operators of $\ddot\myY$ in terms of these operators:
\begin{align}
    {\ddot e}^{(a)}(U)=\sum\lm_{k\in\IZ}\sum\lm_{x \in\mathscr{P}\!\!\mathscr{L}^{(a)}} E_x^{(a)}\frac{x^k}{U^k},\quad {\ddot f}^{(a)}(U)=\sum\lm_{k\in\IZ}\sum\lm_{x \in\mathscr{P}\!\!\mathscr{L}^{(a)}} F_x^{(a)}\frac{x^k}{U^k}\,.
\end{align}
What is not so trivial is to construct an inverse map from fields $\ddot e^{(a)}(U)$ and $\ddot f^{(a)}(U)$ back to lattice point generators $E^{(a)}_x$ and $F^{(a)}_x$.
Instead we use the nice pole structure of the Lax operators \eqref{qLax}:
\begin{align}\label{point_gen}
    E^{(a)}_x=\oint\lm_x\frac{dU}{U}\,\CL^{(a)}_{\varnothing,\varnothing}(U)^{-1}\CL^{(a)}_{\varnothing,\sqbox{$a$}}(U)\,.
\end{align}

The higher Lax operators also acquire a form reminiscent of the rational $\myY$ (cf.\ \eqref{L-rep}):
\begin{align}
	\ddot\CL_{\Kappa,\varnothing}(U)=\sum\lm_{\vec y}\; G_{\Kappa}(\vec y,U)\;\left[\prod\lm_{b\in Q_0}F^{(b)}_{y_1^{(b)}}F^{(b)}_{y_2^{(b)}}\ldots F^{(b)}_{y_{|\Kappa^{(b)}|}^{(b)}}\right]\ddot\myt_{\ff}(U)\,.
\end{align}
where the summation runs over points of lattices $\mathscr{P}\!\!\mathscr{L}^{(a)}$.

It is straightforward to extend the coproduct structure $\ddot\Delta$ on lattice point generators using the coproduct in the Lax representation \eqref{point_gen}.

Eventually, we have to define the test state $|\chi\rangle$ \eqref{chi} as with similar localizing properties delivered by the contour integrals in the case of $\myY$.
This is simply done by inserting operators $\ddot\Delta^{(*)} E_x^{(a)}$ in points $x$ located at each crystal site, so that combining all the ingredients together we derive (cf. \eqref{chi}):
\begin{align}
	|\chi(\vec x)\rangle:= \ddot\Delta^{(N)} E^{(b_1)}_{x_1^{(b_1)}}\;\cdots \; \ddot\Delta^{(N)} E^{(b_m)}_{x_{d_{b_m}}^{(b_m)}}\;|\tilde{\pmb{\varnothing}}\rangle\,, 
\end{align}
where $N$ is the total number of particles $\vec x$.

Eventually, by following all the steps of Section~\ref{sec:BAE} we arrive at a set of trigonometric Bethe ansatz equations \eqref{BAE} with the rational bond-factors substituted by the trigonometric bond-factors \eqref{trig_bf}.

\section{Summary, discussions, and future directions}\label{sec:summary}
Let us first summarize the main \emph{positive} result of this paper.
For non-chiral quivers associated with the toric Calabi-Yau three-folds without compact four-cycles, we have derived the Gauge/Bethe correspondence for the 2D $\mathcal{N}=(2,2)$ supersymmetric quiver gauge theories using the quiver BPS algebras, in this case the unshifted quiver Yangians, and their crystal-chain representations, where on each site is a 2D crystal representation with zero shift. 
In particular, we have reproduced the BAE that correspond to the vacua equations of these quiver gauge theories. 

We have also described the algebraic aspects of this derivation for the Gauge/Bethe correspondence for the 3D $\mathcal{N}=2$ supersymmetric quiver gauge theories, whose quiver BPS algebras are the trigonometric generalizations of the unshifted quiver Yangians, namely the unshifted quiver toroidal algebras. 

More intriguing is the main 
\emph{negative} result of this paper. 
We have found that there are  obstructions to the Gauge/Bethe correspondence, when the quiver is chiral or when the 2D crystal representations considered have non-zero shifts (which force the quiver BPS algebras to have shifts as well). 

To be more precise, in the standard discussion of the Gauge/Bethe correspondence,
we expect that there exists a coproduct of the BPS algebra
such that we can find a consistent $R$-matrix satisfying the YBE and the unitary constraint;
moreover, by choosing a good coproduct, one can reproduce the vacuum equation of the 
supersymmetric gauge theory as the BAE of the integrable model
as determined by the coproduct. However, in this paper, We find that (under some well-motivated assumptions listed in 
Section~\ref{sec:coproduct} and motivated further in Section~\ref{sec:gauge_deriv}) 
this expectation \emph{does not hold} whenever we have negative modes (namely the modes multiplying the positive powers of the spectral parameter) in the Cartan elements, which happens either to chiral quivers (for all representations) or to non-chiral quivers when the representations used has non-zero shifts.

Since it is easy to find 2D representations with zero-shifts, the restriction of our no-go arguments is most significant for the chiral quivers. 
Namely, our no-go arguments have ruled out Gauge/Bethe correspondence for all the examples involving chiral quivers, and hence
all the examples arising from toric Calabi-Yau three-folds with compact four-cycles (i.e.\ for generic toric Calabi-Yau manifolds).

Let us quickly point out that not everything is lost, despite our no-go arguments.
Even for chiral quivers, we still have well-defined quiver BPS algebras (quiver Yangians and trigonometric/elliptic counterparts),
which have perfectly legitimate representations in terms of crystal melting.
The problem happens only when we construct representations associated with crystal chains (with non-trivial interactions between neighboring sites), and 
consider coproducts, $R$-matrices and YBE of the integrable models,
all of which are needed for the final match between the gauge-theory vacuum equations and the BAE:

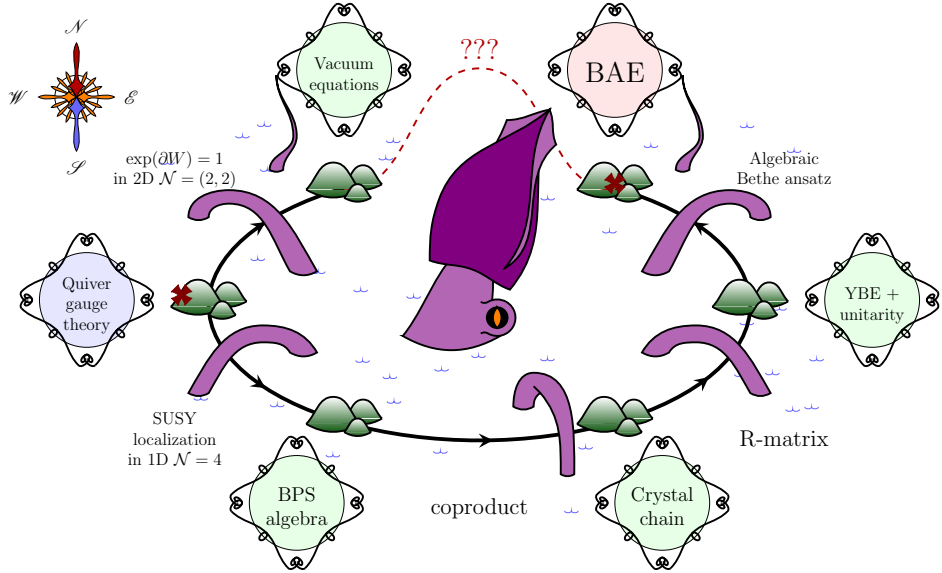
\def\bckclr{white}
\begin{align}\nonumber
\scalebox{0.9}{$
\begin{array}{c}
	\begin{tikzpicture}
		\foreach \x/\y in {0.990563930248332/0.2586602719722503,0.44399712575644323/0.3517767990834775,0.8951112520943217/0.9514470103922079,0.172434996336304/0.6240071880352236,0.9979974734103211/0.37678873536373014,0.6003718400724132/0.37713214162855513,0.7531666838846893/0.9024789780272383,0.40274015661797247/0.5310266899137746,0.0882712620320949/0.27998467346061606,0.3615513864123181/0.8719479558345539,0.635759500426785/0.0025982787694591725,0.8949070055284931/0.4659028379767004,0.2793725601405379/0.2431754620600146,0.18200025344534276/0.9543112362962636,0.9326740491630308/0.3425305608727124,0.5986162117663887/0.29802545991590534,0.9432561058440279/0.5016357480831934,0.6567718654347532/0.15968136546823453,0.26180868558692794/0.5920481341448122,0.7830834692940267/0.41511095708507095,0.32446044697631504/0.5471515991689114,0.08990468069089852/0.3248005978412998,0.4985129603121349/0.3851288151913538,0.03914673799675761/0.8610148155150661,0.32734316493588167/0.3731471627327295,0.13283170275734457/0.5498776821375131,0.9626295692365495/0.25250355488933274,0.5398043771898127/0.6351719626808294,0.14988737962797394/0.9262188587881925,0.6007477261361934/0.7732993325448833,0.37374520807853984/0.2723946478883428,0.28341305461022237/0.14016381068086503,0.6919514633397602/0.6919262227305609,0.44222552169180096/0.14922248937103466,0.36748811908070356/0.3330981684573968,0.3726184604710475/0.9146208802851589,0.8869139439869038/0.3384726695909288,0.9646995772222321/0.8902953468854865,0.20120386807796176/0.2637505678888966,0.18649907844561586/0.8437080354363191}
		{
			\node at (10*\x - 5, 6*\y - 3) {$\begin{array}{c}
					\begin{tikzpicture}
						\draw[thin,\myblue] (-0.1,0) to[out=270,in=270] (0,0) to[out=270,in=270] (0.1,0);
					\end{tikzpicture}
				\end{array}$};		
		}
		\tikzset{sty1/.style={ultra thick,postaction={decorate},decoration={markings, 
					mark= at position 0.55 with {\arrow{stealth}}}}}
		\foreach \r  in {4.0}
		{
			\begin{scope}[yscale=0.5]
			\draw[sty1] ([shift={(180:\r)}]0,0) arc (180:240:\r);
			\draw[sty1] ([shift={(240:\r)}]0,0) arc (240:300:\r);
			\draw[sty1] ([shift={(300:\r)}]0,0) arc (300:360:\r);
			\draw[sty1] ([shift={(0:\r)}]0,0) arc (0:60:\r);
			\draw[sty1] ([shift={(180:\r)}]0,0) arc (180:120:\r);
			\end{scope}
		}
		\draw[thick, dashed, black!30!red] (-2,1.7) to[out=0,in=180] (0, 3.5) to[out=0,in=180] (2,1.7);
		\node[above, black!30!red] at (0,3.5) {???};
		\begin{scope}[yscale=0.5]
		\foreach \a in {0, 60, 120, 180, 240, 300}
		{
		\begin{scope}[rotate=\a]
		\node at (4,0) {$\begin{array}{c}
				\begin{tikzpicture}[scale=0.4]
					\begin{scope}[shift={(1.1,0.2)}]
						\begin{scope}[scale=0.7]
							\draw[thick,bottom color=black!70!green,top color=\bckclr] (-1,0) to[out=330,in=210] (1,0) (-1,0) to[out=45,in=180] (0,1) to[out=0,in=135] (1,0);
						\end{scope}
					\end{scope}
					\draw[thick,bottom color=black!70!green,top color=\bckclr] (-1,0) to[out=330,in=210] (1,0) (-1,0) to[out=45,in=180] (0,1) to[out=0,in=135] (1,0);
					\begin{scope}[shift={(0.9,-0.3)}]
						\begin{scope}[scale=0.5]
							\draw[thick,bottom color=black!70!green,top color=\bckclr] (-1,0) to[out=330,in=210] (1,0) (-1,0) to[out=45,in=180] (0,1) to[out=0,in=135] (1,0);
						\end{scope}
					\end{scope}
				\end{tikzpicture}
			\end{array}$};
		\end{scope}
		}
		\end{scope}
		\node at (0,1) {$\begin{array}{c}
				\begin{tikzpicture}[scale=0.9]
					\draw[thick,fill=white!40!violet] (-0.8,0) to[out=330,in=210] (0.8,0) to[out=130,in=320] (0.5,1.3) to[out=140,in=50] cycle;
					\begin{scope}[shift={(0.75,0.3)}]
						\foreach \s in {0.3}
						{
							\draw[thick,fill=white!40!violet] (-0.9*\s, -0.866025403784439*\s) to[out=30,in=150] ([shift={(240:\s)}]0,0) arc (240:480:\s) to[out=210,in=330] (-0.9 * \s, 0.866025403784439 * \s);
						}
						\draw[thick,fill=black] (0,0) circle (0.17);
						\begin{scope}[scale=0.17]
							\draw[thick,fill=orange] (0,1) to[out=225,in=135] (0,-1) to[out=45,in=315] cycle;
						\end{scope}
					\end{scope}
					\begin{scope}[shift={(0.3,1)}]
						\begin{scope}[rotate=-20]
							\draw[thick,fill=white!40!violet] (-0.5,0.3) to[out=100,in=250] (0.3,3) to[out=210,in=20] (-0.3,2.6) to[out=270,in=110] cycle;
							\draw[thick,fill=violet] (-0.7,0) to[out=0,in=180] (0,0.1) to[out=0,in=180] (0.7,0)  to[out=70,in=270] (0.3,3) to[out=240,in=90] (-0.9,1) to[out=270,in=110] cycle;
							\draw[thick,fill=white!40!violet] (0.5,0.3) to[out=70,in=260] (0.3,3) to[out=315,in=135] (0.6,2.5) to[out=240,in=50] cycle;
						\end{scope}
					\end{scope}
				\end{tikzpicture}
			\end{array}$};
		\draw[thick,fill=white!40!violet] (-2.4,0.5) to[out=120,in=0] (-3.7,1.7)  to[out=180,in=90] (-4.3,1.3) to[out=330,in=210] (-4,1.3) to[out=90,in=180] (-3.7,1.5) to[out=0,in=120] (-2.8,0.5) to[out=330,in=210] cycle;
		\begin{scope}[xscale=-1]
			\draw[thick,fill=white!40!violet] (-2.4,0.5) to[out=120,in=0] (-3.7,1.7)  to[out=180,in=90] (-4.3,1.3) to[out=330,in=210] (-4,1.3) to[out=90,in=180] (-3.7,1.5) to[out=0,in=120] (-2.8,0.5) to[out=330,in=210] cycle;
		\end{scope}
		\draw[thick,fill=white!40!violet] (-2.4,-0.7) to[out=120,in=0] (-3.2,-0.3) to[out=180,in=60] (-4.3,-1.3) to[out=330,in=210] (-4,-1.3) to[out=60,in=180] (-3.2,-0.55) to[out=0,in=120] (-2.8,-0.7) to[out=330,in=210] cycle;
		\begin{scope}[xscale=-1]
			\draw[thick,fill=white!40!violet] (-2.4,-0.7) to[out=120,in=0] (-3.2,-0.3) to[out=180,in=60] (-4.3,-1.3) to[out=330,in=210] (-4,-1.3) to[out=60,in=180] (-3.2,-0.55) to[out=0,in=120] (-2.8,-0.7) to[out=330,in=210] cycle;
		\end{scope}
		\draw[thick,fill=white!40!violet] (0.5,-1.5) to[out=90,in=180] (0.8,-1) to[out=0,in=90] (1.4,-2.5) to[out=210,in=330] (1.2,-2.5) to[out=90,in=310] (0.9,-1.3) to[out=210,in=90] (0.8,-1.5) to[out=210,in=330] cycle;
		\draw[thick] (0.9,-1.3) -- (0.8,-1.2);
		\begin{scope}[shift={(-3,2)}]
			\draw[thick,fill=white!40!violet] (-0.1,0) to[out=90,in=270] (0.25,0.5) to[out=90,in=270] (0,1.4) to[out=270,in=90] (0.3,0.5) to[out=270,in=90] (0.1,0) to[out=210,in=330] cycle; 
		\end{scope}
		\begin{scope}[shift={(3,2)}]
			\draw[thick,fill=white!40!violet] (-0.1,0) to[out=90,in=270] (0.25,0.5) to[out=90,in=270] (0,1.4) to[out=270,in=90] (0.3,0.5) to[out=270,in=90] (0.1,0) to[out=210,in=330] cycle; 
		\end{scope}
		\node at (-6,3) {$\begin{array}{c}
			\begin{tikzpicture}[scale=0.4]
				\tikzset{sty2/.style={fill=\bckclr}}
				\draw (0,0) circle (0.8);
				\foreach \t in {22.5,45,67.5,112.5,135,157.5}
				{
					\begin{scope}[rotate=\t]
						\begin{scope}[scale=0.5]
							\begin{scope}
								\draw[sty2,fill=orange] (0,0) to[out=60,in=210] (0.3,0.4) to[out=150,in=300] (0,2) to[out=240,in=30] (-0.3,0.4) to[out=330,in=120] cycle;
							\end{scope}
							\begin{scope}[rotate=180]
								\draw[sty2,fill=orange] (0,0) to[out=60,in=210] (0.3,0.4) to[out=150,in=300] (0,2) to[out=240,in=30] (-0.3,0.4) to[out=330,in=120] cycle;
							\end{scope}
						\end{scope}
					\end{scope}
				}
				\begin{scope}[rotate=90]
					\begin{scope}[scale=0.7]
						\begin{scope}
							\draw[sty2,fill=orange] (0,0) to[out=60,in=210] (0.3,0.4) to[out=150,in=300] (0,2) to[out=240,in=30] (-0.3,0.4) to[out=330,in=120] cycle;
						\end{scope}
						\begin{scope}[rotate=180]
							\draw[sty2,fill=orange] (0,0) to[out=60,in=210] (0.3,0.4) to[out=150,in=300] (0,2) to[out=240,in=30] (-0.3,0.4) to[out=330,in=120] cycle;
						\end{scope}
						\node[left] at (0,2) {$\scriptstyle\mathscr{W}$};
						\node[right] at (0,-2) {$\scriptstyle\mathscr{E}$};
					\end{scope}
				\end{scope}
				\begin{scope}
					\draw[sty2,fill=black!30!red] (0,0) to[out=60,in=210] (0.3,0.4) to[out=150,in=300] (0,2) to[out=240,in=30] (-0.3,0.4) to[out=330,in=120] cycle;
				\end{scope}
				\begin{scope}[rotate=180]
					\draw[sty2,fill=\myblue] (0,0) to[out=60,in=210] (0.3,0.4) to[out=150,in=300] (0,2) to[out=240,in=30] (-0.3,0.4) to[out=330,in=120] cycle;
				\end{scope}
				\node[above] at (0,2) {$\scriptstyle\mathscr{N}$};
				\node[below] at (0,-2) {$\scriptstyle\mathscr{S}$};
			\end{tikzpicture}
		\end{array}$};
		\foreach \x/\y/\clr in {-5.8/0/white!90!blue,5.8/0/white!90!green,-2/3.4/white!90!green,2/3.4/white!90!red,-2.7/-3/white!90!green,2.7/-3/white!90!green}
		{
		\node at (\x,\y) {$\begin{array}{c}
				\begin{tikzpicture}[scale=0.7]
					\draw[fill=\clr] (0,0) circle (1);
					\foreach \p in {45,135,225,315}
					{
						\begin{scope}[rotate=\p]
							\begin{scope}[shift={(0,0.8)}]
								\begin{scope}[yscale=1]
									\draw[thick] (-0.85,0.1) to[out=180,in=180] (-0.85,0.2) to[out=0,in=0] (-0.85,0) to[out=180,in=180] (-0.85,0.3) to[out=0,in=180] (0,0.1) to[out=0,in=0] (0,0.3) to[out=180,in=180] (0,0.1) to[out=0,in=180] (0.85,0.3);
									\begin{scope}[xscale=-1]
										\draw[thick] (-0.85,0.1) to[out=180,in=180] (-0.85,0.2) to[out=0,in=0] (-0.85,0) to[out=180,in=180] (-0.85,0.3);
									\end{scope}
								\end{scope}
							\end{scope}
						\end{scope}
					}
				\end{tikzpicture}
			\end{array}$};
		}
		\node at(-5.8,0) {\scalebox{0.6}{$\begin{array}{c}
					\mbox{Quiver}\\
					\mbox{gauge}\\
					\mbox{theory}
				\end{array}$}};
		\node at(5.8,0) {\scalebox{0.6}{$\begin{array}{c}
					\mbox{YBE }+\\
					\mbox{unitarity}
				\end{array}$}};
		\node at(-2,3.4) {\scalebox{0.6}{$\begin{array}{c}
					\mbox{Vacuum}\\
					\mbox{equations}
				\end{array}$}};
		\node at(2,3.4) {\scalebox{1}{$\begin{array}{c}
					\mbox{BAE}
				\end{array}$}};
		\node at(-2.7,-3) {\scalebox{0.7}{$\begin{array}{c}
					\mbox{BPS}\\
					\mbox{algebra}
				\end{array}$}};
		\node at(2.7,-3) {\scalebox{0.7}{$\begin{array}{c}
					\mbox{Crystal}\\
					\mbox{chain}
				\end{array}$}};
		\node at (-4.5,2) {\scalebox{0.6}{$\begin{array}{c}
					\exp(\p W)=1\\
					\mbox{in 2D }\CN=(2,2)
				\end{array}$}};
		\node at (4.5,2) {\scalebox{0.6}{$\begin{array}{c}
					\mbox{Algebraic}\\
					\mbox{Bethe ansatz}
				\end{array}$}};
		\node at (-4.5,-2) {\scalebox{0.6}{$\begin{array}{c}
					\mbox{SUSY}\\
					\mbox{localization}\\
					\mbox{in 1D } \CN=4
				\end{array}$}};
		\node at (4.5,-2) {\scalebox{0.8}{$\begin{array}{c}
					\mbox{R-matrix}
				\end{array}$}};
		\node at (0,-3) {\scalebox{0.8}{$\begin{array}{c}
					\mbox{coproduct}
				\end{array}$}};
		\begin{scope}[shift={(-4.4,0.15)}]
		\begin{scope}[scale=0.7]
		\begin{scope}[rotate=45]
			\draw[black!50!red,fill=black!50!red] (-0.1,0.2) -- (0.1,0.2) to[out=240,in=120] (0.1,-0.2) -- (-0.1,-0.2) to[out=60,in=300] cycle;
		\end{scope}
		\begin{scope}[rotate=135]
			\draw[black!50!red,fill=black!50!red] (-0.1,0.2) -- (0.1,0.2) to[out=240,in=120] (0.1,-0.2) -- (-0.1,-0.2) to[out=60,in=300] cycle;
		\end{scope}
		\end{scope}
		\end{scope}
		\begin{scope}[shift={(2,1.8)}]
			\begin{scope}[scale=0.7]
				\begin{scope}[rotate=45]
					\draw[black!50!red,fill=black!50!red] (-0.1,0.2) -- (0.1,0.2) to[out=240,in=120] (0.1,-0.2) -- (-0.1,-0.2) to[out=60,in=300] cycle;
				\end{scope}
				\begin{scope}[rotate=135]
					\draw[black!50!red,fill=black!50!red] (-0.1,0.2) -- (0.1,0.2) to[out=240,in=120] (0.1,-0.2) -- (-0.1,-0.2) to[out=60,in=300] cycle;
				\end{scope}
			\end{scope}
		\end{scope}
	\end{tikzpicture}
\end{array}$}
\end{align}

Of course, our ``no-go'' result relies on several assumptions,
any of which could in principle be violated. It seems fair to say, however, 
that any such possibility requires a deviation from the standard narratives in the Gauge/Bethe correspondence.
For example, any violation of \textbf{Assumption 1} in \eqref{iso} will require serious reconsideration of stable envelops in Section~\ref{sec:stab_bas};
skeptics of \textbf{Assumption 1} are encouraged to come up with a concrete expression for the coproduct not satisfying \textbf{Assumption 1}
yet still reproducing the BAE.
Let us also emphasize that our ``yes-go'' result in section \ref{sec:BAE} means that any such subtlety will arise
only when our no-go results apply, i.e.\ to shifted quiver Yangians for non-chiral quivers or to
general quiver Yangians for chiral quivers.
For this reason, it seems fair to say that there are at least important subtleties
in the Gauge/Bethe correspondence yet to be clarified.

While a complete understanding is still lacking either physically or mathematically, 
we already discussed a gauge-theory origin of the obstruction 
in Section \ref{sec:breakdown}: the obstruction has to do
with the vacua running off to infinity.
One natural possibility then is to introduce a suitable 
regularization where all the vacua are kept in the finite region.

One possible regularization is to embed the chiral $Q$ quiver into 
an extended non-chiral quiver $\hat{Q}$ by doubling the number of arrows: for each arrow we 
add another arrow in the opposite direction.\footnote{We thank Nikita Nekrasov for a brief discussion on this point.}
Since the extended quiver is non-chiral, some complications (e.g.\ the running of the FI parameters)
go away and the bond factor \eqref{bond_factor} has net degree zero as a polynomial in the 
spectral parameter (i.e.\ the polynomials in the numerator and the denominator have the same degree). This makes the vacuum equation \eqref{BAE_factor}
closer to a BAE, which typically involves rational functions of net degree zero.

While such an embedding into a non-chiral quiver might work in principle, 
a satisfactory resolution of our no-go result requires many new ingredients.
First, we will need to identify the quiver BPS algebras (e.g.\ find explicit generators and relations)
for $\hat{Q}$.\footnote{The quiver Yangian in \cite{Li:2020rij} in itself can be defined for any quiver (and a superpotential). 
It is a separate question, however,
if we can identify the algebra as the physical BPS algebra for a non-toric theory.} 
The complication is that in general $\hat{Q}$ is not associated with a toric
CY$_{3}$, even when $Q$ is. 
This likely means that there are not enough equivariant actions,
and hence many of the known results for quiver Yangians, e.g.\ the crystal-melting representations,
need to be revised at least. 
In addition to identifying the BPS algebra for $\hat{Q}$,
one needs to find a suitable coproduct and the associated $R$-matrix. One also needs to 
discuss suitable limits to discuss the BAE for $Q$ from that for $\hat{Q}$.
It would be interesting to explore this direction in future research.

Before concluding this paper, let us 
comment on some more questions for further research:
\begin{itemize}
    \myitem The perfect candidates for 2D crystals producing no shift (therefore no obstruction) in the case of $\myY(\widehat{\fg\fl}_1)$ are integer partitions spanning Fock modules (see Figure~\ref{fig:MacMahon}).
    As we explained in section \ref{sec:cry_dim} the quiver framing that corresponds to the Fock modules is such that the resulting quiver variety is Nakajima-type (i.e.\ the gauge theory acquires a supersymmetry enhancement).
    Therefore it is not difficult to invent canonical analogues for Fock modules in the case of $\myY(\widehat{\fg\fl}_n)$---it suffices to consider Nakajima varieties of $\hat A_n$-type.
    However quivers for other allowed algebras, including $\myY(\widehat{\fg\fl}_{m|n})$ and the affine Yangian of $D(2,1;\alpha)$, can not be made Nakajima-type with any choice of the framing.
    It would be interesting to construct and investigate canonical analogues of Fock modules for those algebras.
    \myitem Similarly to the previous point, the structure of stable envelopes \cite{MaulikOkounkov} incorporates a splitting of the cotangent bundle into two halves with respect to the action of the symplectic structure.
	This construction can be performed for $\myY(\widehat{\fg\fl}_1)$ \cite{Smirnov:2014npa} relatively easily, and it is not surprising that there is a generalization to $\myY(\widehat{\fg\fl}_n)$ \cite{2021arXiv210709569D} (elliptic version) for Nakajima $\hat{A}_n$-varieties.
	In this context, a natural question arises if this construction can be generalized to $\myY(\widehat{\fg})$ (i.e.\ the affine Yangian of $\mathfrak{g}$) when $\mathfrak{g}$ is the super Lie algebra $\mathfrak{gl}_{m|n}$ or $D(2,1;\alpha)$.
	Even if the standard way to construct the stable envelopes is unavailable in this situation, it is natural to expect a concise formula for the stable basis choice in terms of Higgs branch operators analogous to \cite{Bullimore:2017lwu}.
    \myitem In Section~\ref{sssec:kernel} we presented our motivation to use Fourier-Mukai kernel \eqref{kernel} supported on the incidence locus to construct the action of $\myY$ on crystals.
    In the construction of \cite{Galakhov:2020vyb} somewhat distant from 2D $\CN=(2,2)$ GLSM a similar kernel was proposed due to its resemblance with the Hecke modifications caused by adding/subtracting fractional D-branes to/from D-brane system wrapping a toric CY${}_3$.
    A similar kernel was used in the original construction of the action of $\myY(\widehat{\fg\fl}_1)$ on Hilbert schemes \cite{Nakajima_book}.
    It would be interesting to derive \eqref{kernel} from first principles and soliton dynamics discussed in Section~\ref{sec:gauge_deriv}.
    \myitem The rational/trigonometric/elliptic trichotomy of quiver BPS algebras can be treated uniformly as far as their generators and relations are concerned \cite{Galakhov:2021vbo}. As we discussed in Section~\ref{sec:trig}, however, the uniformity can be misleading when we discuss more structures in the algebra, such as coproducts. 
    This is the reason that we have decided to postpone the discussion of the elliptic quiver BPS algebras to future work, since it needs special case.
    We only note here that in the BAE \eqref{BAE}, if we replace the rational bond factors $\varphi^{a\Leftarrow b}(u)$ by the corresponding elliptic versions, namely with each rational factor $u$ replaced by a theta function $\Theta_q(z)$ (see Section 2 of \cite{Galakhov:2021vbo}), then we would reproduce the so-called ``Bethe Ansatz Equations" seen in the computation of the superconformal indices of the corresponding 4D quiver gauge theory \cite{Benini:2018ywd,GonzalezLezcano:2019nca}.\footnote{Note that  \cite{GonzalezLezcano:2019nca} include chiral quivers, which might not correspond to any integrable models in the traditional sense, as shown by our no-go arguments.} 
    In other words, the elliptic version of the current paper might provide an explanation for the appearance of these Bethe Ansatz Equations in the computation of 4D superconformal indices.

    In addition to discussing more general algebras,
    another strategy is to study some special theories in more detail, for example 3d $\mathcal{N}=4$ theories \cite{Dedushenko:2021mds,Bullimore:2021rnr} corresponding to quantum toroidal BPS algebras for non-chiral quivers.
    
    \myitem The tensor product structures $\bDelta_0$ and $\Delta$ we discussed in section \ref{sec:coproduct} have natural higher analogues $\bDelta_0^{(n)}$ and $\Delta^{(n)}$ acting now on tensor powers since both $\bDelta_0^{(n)}$ and $\Delta^{(n)}$ are associative.
    It is natural to define higher homomorphisms
    $U^{(n)}$ conjugating one into the other:
    $$\bDelta_0^{(n)}U^{(n)}=U^{(n)} \bDelta^{(n)}\,.$$
    The field-theoretic consideration of Section~\ref{sec:gauge_deriv} also suggests that such a structure exists as a transform between the vacuum and the stable bases in long crystal chains.
    It would be interesting to study the properties of $U^{(n)}$ and the possibility to reconstruct it from elementary operations $U$.
    \myitem If we can find a map from the a quiver Yangian to some $\mathcal{W}$-algebra, then the information on the tensor representations of this  $\mathcal{W}$ algebra can help us determine the coproduct structure of the corresponding quiver Yangian. 
    For a non-chiral quiver from the toric CY$_3$, the quiver Yangian is the affine Yangian of $\mathfrak{g}$, where $\mathfrak{g}$ is $\mathfrak{gl}_{m|n}$ or $D(2,1;\alpha)$; and it is expected to be isomorphic to (the UEA) of the $\mathfrak{g}$-extended $\mathcal{W}$-algebra.\footnote{For more on the $\mathfrak{g}$-extended $\mathcal{W}$-algebras, see \cite{Creutzig:2018pts,Eberhardt:2020zgt} for the $\mathfrak{g}=\mathfrak{gl}_n$ case and \cite{Creutzig:2019qos,Rapcak:2019wzw} for the $\mathfrak{g}=\mathfrak{gl}_{m|n}$ case.} 
    Indeed, for $\mathfrak{g}=\mathfrak{gl}_{m|n}$,  the map to the corresponding  $\mathcal{W}$ algebra was a useful ingredient for the identification of 
    the coproduct \cite{Prochazka:2015deb,Litvinov:2020zeq,Chistyakova:2021yyd,Kolyaskin:2022tqi,Bao:2022fpk}. 
    
    Conversely, if we succeed in determining the coproduct $\Delta$ completely,
    it will provide very useful information in determining the full dictionary between the affine Yangian of $\mathfrak{g}$ and the $\mathfrak{g}$-extended $\mathcal{W}$ algebra.
    
    More generally, since it is not yet clear whether the quiver Yangian for a chiral quiver can be mapped to a $\mathcal{W}$ algebra, the information from the coproduct can provide invaluable clue as to  
     whether or not there exists an isomorphism from a quiver BPS algebra to a $W$-algebra for general quivers.
     Finally, it can also help us  determine the Serre relations of the BPS algebra.
    
\end{itemize}

\section*{Acknowledgements}
We would like to thank David Hernandez, Alexei Morozov, Hiraku Nakajima, Nikita Nekrasov, Andrei Okounkov, Alexei Sleptsov, Andrey Smirnov and Zijun Zhou for stimulating discussions. WL is grateful for support
from NSFC No.\ 11875064 and 11947302, CAS Grant No.\ XDPB15, the Max-Planck Partnergruppen fund, and the hospitality of ETH Zurich and Albert-Einstein-Institut (Potsdam). The work of MY and DG was supported in part
by WPI Research Center Initiative, MEXT, Japan. MY was also supported by
the JSPS Grant-in-Aid for Scientific Research (17KK0087, 19K03820, 19H00689, 20H05860).
DG would like to thank Moscow Institute of Physics and Technology for their generous hospitality.

\appendix
\section{\texorpdfstring{$R$}{R}-matrices and quiver Yangian from stable envelopes}\label{sec:stable}

In this section we consider an application of the technology of stable envelopes to the construction of quiver BPS algebras.
The stable envelopes were defined in \cite{MaulikOkounkov}, see also \cite{Smirnov:2014npa,Aganagic:2016jmx,2020arXiv201207814R} and references therein for development and practical applications.
In Section~\ref{sec:BCtoCrystal} we treated stable envelopes similarly to \cite{Bullimore:2017lwu,Dedushenko:2021mds,Bullimore:2021rnr} as a transformation between bases in the disk boundary conditions: from a naive basis of classical vacua to a basis of stable objects in the corresponding triangulated category of boundary branes. 

The original construction of stable envelopes relies heavily on the symplectic structure of Nakajima quiver varieties and a subsequent base-fiber duality.
In general, a quiver variety associated with a toric CY${}_3$ is not of Nakajima-type, therefore we have to modify our techniques.
We combine the construction of stable envelopes for low-dimensional quivers when the map between bases can be carried out explicitly, and the technique of \cite{Litvinov:2020zeq} to produce the algebraic structures from low level $R$-matrices and the YBE.

A simple way to take into account the stable envelopes for low-dimensional cases is to consider the Lefschetz thimble basis and compare it with the basis in the ring of operators.

The value of the partition function \eqref{D_2_pf} for the brane given by a Lefschetz thimble can be computed easily.
We should substitute $\CO$ by 1, and all the gamma-functions by the corresponding disk partition function of a single chiral field:
\begin{align}
	\Gamma_0(z)=\int\lm_{\CL}dY\; \exp\left(-zY-e^{-Y}\right)\,,
\end{align}
where $Y$ is the mirror dual to the chiral field \cite{Hori:2000kt}, and $\CL$ is a single Lefschetz thimble integration contour.

However depending if ${\rm Re}\,z>0$ {\color{black!60!green} (a)} or ${\rm Re}\,z<0$ {\color{\myblue} (b)} the topology of the thimble jumps drastically
\begin{align}
	\mbox{in }e^{-Y}\mbox{-plane: }\begin{array}{c}
		\begin{tikzpicture}[scale=0.8]
			\draw[black!60!green, thick] (0,0) -- (3,0);
			\draw[thick, \myblue] (3,0) to[out=160, in=90] (-1,0) to[out=270, in=200] (3,0);
			\draw[fill=white] (0,0) circle (0.08) (3,0) circle (0.08);
			\node[left] at (0,0) {$0$};
			\node[right] at (3,0) {$+\infty$};
			\draw[fill=black!60!green] (1,0) circle (0.05);
			\node[below left, black!60!green] at (1,0) {$\CL_{\rm a}$};
			\draw[fill=\myblue] (-1,0) circle (0.05);
			\node[left, \myblue] at (-1,0) {$\CL_{\rm b}$};
		\end{tikzpicture}
	\end{array}\,,
\end{align}
so that we have:
\begin{align}\label{G0}
	\Gamma_0(z)=\Gamma(z)\times\left(2\I\,\sin\pi z\right)^{\Theta\left(-{\rm Re}\,z\right)}\,,
\end{align}
where $\Theta$ is the Heaviside step function.

\subsection{\texorpdfstring{Warmup: $R$-matrix for $\myY(\fs\fl_2)$}{Warmup: R-matrix for Y(sl2)}}

As a warm-up let us start with the case of $\myY(\fs\fl_2)$ acting on a Heisenberg XXX $\frac{1}{2}$-spin chain of two spins.
The corresponding QFT has the moduli space $T^*\IC\IP^1$, and its matter content can be defined by the following quiver \cite{Bullimore:2017lwu}:
\begin{align}
	\begin{array}{c}
		\begin{tikzpicture}
			\draw[postaction={decorate},decoration={markings, 
				mark= at position 0.65 with {\arrow{stealth}}}] (0,0.06) -- (2,0.06) node[pos=0.5,above] {$0$};
			\draw[postaction={decorate},decoration={markings, 
				mark= at position 0.40 with {\arrow{stealth}}}] (2,-0.06) -- (0,-0.06) node[pos=0.5,below] {$h$};
			\draw[postaction={decorate},decoration={markings, 
				mark= at position 0.35 with {\arrow{stealth}}}] (0,0) to[out=120,in=90] (-1,0) to[out=270,in=240] (0,0);
			\draw[fill=white] (0,0) circle (0.1);
			\begin{scope}[shift={(2,0)}]
				\draw[fill=white] (-0.1,-0.1) -- (-0.1,0.1) -- (0.1,0.1) -- (0.1,-0.1) -- cycle;
			\end{scope}
			\node[above] at (0,0.1) {$\sigma$};
			\node[right] at (2.1,0) {$\left(\begin{array}{c}
					u_1\\ u_2\\
				\end{array}\right)$};
		\end{tikzpicture}
	\end{array}
\end{align}
where $u_1$ and $u_2$ are complex flavor charges corresponding to the framing node, and without loss of generality we choose the weights for the two chiral fields to be $0$ and $h$, respectively.

The disk partition function reads:
\begin{align}
	Z=\int\lm_{-\I\infty}^{+\I\infty} d\sigma\;e^{t\sigma}\,\Gamma_0(\sigma-u_1)  \Gamma_0(\sigma-u_2)\Gamma_0(u_1-\sigma-h)\Gamma_0(u_2-\sigma-h)\,.
\end{align}

If ${\rm Re}\, t>0$ and its absolute value is large, one should close the integration cycle in the right complex half-plane, so there are two integration cycles, encircling poles of the gamma-functions.
Now we also adopt another useful notation:
\begin{align}
\begin{split}
\gamma_1 = \left[\begin{array}{c|c}
	u_1 & u_2\\
	\Box & \varnothing\\
\end{array}\right]: \quad &{\rm poles}:\;\sigma=u_1-\IZ_{\geq 0}\,,\\
\gamma_2 = \left[\begin{array}{c|c}
	u_1 & u_2\\
	\varnothing & \Box\\
\end{array}\right]: \quad &{\rm poles}:\;\sigma=u_2-\IZ_{\geq 0}\,.
\end{split}
\end{align}

On the other hand if we return to the Lefschetz thimble boundary conditions, we have two Bethe roots:
\begin{align}
	\sigma_{*1}=u_1+O\left(e^{-t}\right),\quad \sigma_{*2}=u_2+O\left(e^{-t}\right)\,.
\end{align}

Let us assume ${\rm Re}\,u_1> {\rm Re}\,u_2$. 
Comparing contributions of the Bethe roots to the chiral partition functions \eqref{G0}, we derive the operators corresponding to the thimble boundary conditions as the contributions of Heaviside factors.
Here we adopt the following notation:
\begin{align}
	\left(\begin{array}{cc}
		u_1 & u_2\\
		\Box &  \varnothing
	\end{array}\right)=\sin\pi(u_2-\sigma-h),\quad \left(\begin{array}{ccc}
	u_1 & u_2\\
	\varnothing & \Box 
\end{array}\right)=\sin\pi(\sigma-u_1)\,.
\end{align}
For the opposite situation we have:
\begin{align}
	\left(\begin{array}{cc}
		u_2 & u_1\\
		\Box &  \varnothing
	\end{array}\right)=\sin\pi(u_1-\sigma-h),\quad \left(\begin{array}{ccc}
		u_2 & u_1\\
		\varnothing & \Box
	\end{array}\right)=\sin\pi(\sigma-u_2)\,.
\end{align}
This is to be compared with operator-thimble boundary condition matching in \cite{Bullimore:2017lwu}.

\begin{subequations}
Calculating the integrals we derive the follwoing relation between boundary conditions:
\begin{align}\label{solit_a}
	\begin{split}
	\left(\begin{array}{cc}
		u_1 & u_2\\
		\Box &  \varnothing
	\end{array}\right)&=A_1\overset{{\color{\myblue}\gamma_1}}{\left[\begin{array}{c|c}
	u_1 & u_2\\
	\Box &  \varnothing
\end{array}\right]}+A_2\overset{{\color{\myblue}\gamma_2}}{\left[\begin{array}{c|c}
u_1 & u_2\\
\varnothing & \Box  
\end{array}\right]}\,,\\
\left(\begin{array}{cc}
	u_1 & u_2\\
	\varnothing & \Box 
\end{array}\right)&=A_3\overset{{\color{\myblue}\gamma_2}}{\left[\begin{array}{c|c}
	u_1 & u_2\\
	\varnothing & \Box  
\end{array}\right]}\,;
\end{split}
\end{align}
and similarly,
\begin{align}\label{solit_b}
	\begin{split}
		\left(\begin{array}{cc}
			u_2 & u_1\\
			\Box &  \varnothing
		\end{array}\right)&=B_1\overset{{\color{\myblue}\gamma_2}}{\left[\begin{array}{c|c}
			u_2 & u_1\\
			\Box &  \varnothing
		\end{array}\right]}+B_2\overset{{\color{\myblue}\gamma_1}}{\left[\begin{array}{c|c}
			u_2 & u_1\\
			\varnothing & \Box  
		\end{array}\right]}\,,\\
		\left(\begin{array}{cc}
			u_2 & u_1\\
			\varnothing & \Box
		\end{array}\right)&=B_3\overset{{\color{\myblue}\gamma_1}}{\left[\begin{array}{c|c}
			u_2 & u_1\\
			\varnothing & \Box  
		\end{array}\right]}\,;
	\end{split}
\end{align}
\end{subequations}
where
\begin{eqnarray}
\begin{aligned}
	&A_1=\sin\pi (-u_{12}-h),\quad A_2=\sin\pi(-h),\quad A_3=\sin\pi(-u_{12}),\\
	&B_1=\sin\pi (u_{12}-h),\qquad B_2=\sin\pi(-h),\quad B_3=\sin\pi(u_{12})\,.
\end{aligned}
\end{eqnarray}

Integrals over the cycles $\gamma_1$, $\gamma_2$ in the r.h.s.\ of expressions \eqref{solit_a} and \eqref{solit_b} are analytic in the spectral parameters $u_1$ and $u_2$.
Therefore the result of the corresponding integral does not change if we permute the columns of the tables in the r.h.s.
This manipulation allows us to compare thimble bases for ${\rm Re}\,u_1> {\rm Re}\,u_2$ and ${\rm Re}\,u_2> {\rm Re}\,u_1$.
The resulting parallel transport matrix is given by interface partition functions and corresponds to the twisted $R$-matrix as we have discussed in Section~\ref{sec:interface}.

Hence for the twisted $R$-matrix we have:
\begin{align}\label{R-m_1}
	\begin{array}{cc}
		\begin{array}{c}
			\begin{tikzpicture}
				\draw[thick] (0,0) to[out=0,in=180] (0.7,0.5) (0,0.5) to[out=0,in=180] (0.7,0);
				\node[left] at (0,0) {$\scriptstyle (\varnothing,u_1)$};
				\node[left] at (0,0.5) {$\scriptstyle (\Box,u_2)$};
				\node[right] at (0.7,0) {$\scriptstyle (\varnothing,u_2)$};
				\node[right] at (0.7,0.5) {$\scriptstyle (\Box,u_1)$};
			\end{tikzpicture}
		\end{array}=\dfrac{A_2}{B_1}\,, & \begin{array}{c}
		\begin{tikzpicture}
			\draw[thick] (0,0) to[out=0,in=180] (0.7,0.5) (0,0.5) to[out=0,in=180] (0.7,0);
			\node[left] at (0,0) {$\scriptstyle (\Box,u_1)$};
			\node[left] at (0,0.5) {$\scriptstyle (\varnothing,u_2)$};
			\node[right] at (0.7,0) {$\scriptstyle (\varnothing,u_2)$};
			\node[right] at (0.7,0.5) {$\scriptstyle (\Box,u_1)$};
		\end{tikzpicture}
	\end{array}=\dfrac{A_1}{B_3}-\dfrac{A_2B_2}{B_1B_3}\,,\\
	\begin{array}{c}
		\begin{tikzpicture}
			\draw[thick] (0,0) to[out=0,in=180] (0.7,0.5) (0,0.5) to[out=0,in=180] (0.7,0);
			\node[left] at (0,0) {$\scriptstyle (\varnothing,u_1)$};
			\node[left] at (0,0.5) {$\scriptstyle (\Box,u_2)$};
			\node[right] at (0.7,0) {$\scriptstyle (\Box,u_2)$};
			\node[right] at (0.7,0.5) {$\scriptstyle (\varnothing,u_1)$};
		\end{tikzpicture}
	\end{array}=\dfrac{A_3}{B_1}\,, & \begin{array}{c}
		\begin{tikzpicture}
			\draw[thick] (0,0) to[out=0,in=180] (0.7,0.5) (0,0.5) to[out=0,in=180] (0.7,0);
			\node[left] at (0,0) {$\scriptstyle (\Box,u_1)$};
			\node[left] at (0,0.5) {$\scriptstyle (\varnothing,u_2)$};
			\node[right] at (0.7,0) {$\scriptstyle (\Box,u_2)$};
			\node[right] at (0.7,0.5) {$\scriptstyle (\varnothing,u_1)$};
		\end{tikzpicture}
	\end{array}=-\dfrac{A_3B_2}{B_1B_3}\,.
	\end{array}
\end{align}

For the case of $T^*\IC\IP^1$ discussed this expression reproduces the canonical trigonometric $R$-matrix for the XXZ spin chain:
\begin{align}
	\left(\begin{array}{cc}
		\dfrac{\sin\pi(h)}{\sin\pi(h-u_{12})} & -\dfrac{\sin\pi(u_{12})}{\sin\pi(h-u_{12})} \\
		-\dfrac{\sin\pi(u_{12})}{\sin\pi(h-u_{12})} & \dfrac{\sin\pi(h)}{\sin\pi(h-u_{12})} \\		
	\end{array}\right)\,.
\end{align}

For simplicity in what follows we will work with rational $R$-matrices.
For a generic quiver we have:
\begin{align}
\begin{aligned}
&	A_1=\prod\lm_{J\in\{a\to\ff\}}\left(-u_{12}-h_J\right),\;A_2=\prod\lm_{J\in\{a\to \ff\}}(-h_J),\;\quad A_3=-u_{12}\,,\\
&	B_1=\prod\lm_{J\in\{a\to\ff\}}\left(u_{12}-h_J\right),\;\quad B_2=\prod\lm_{J\in\{a\to \ff\}}(-h_J),\;\quad B_3=u_{12}\,.
\end{aligned}
\end{align}

\subsection{\texorpdfstring{$R$-matrix for general quiver}{R-matrix for general quiver}}

\def\bbox{{\Box\!\Box}}
In the rest of this section we will denote a crystal with two atoms as $\bbox$.
The symbol $\bbox$ is a mnemonic symbol implying the whole variety of 2-atom crystals allowed for a given representation.

Without loss of generality we can construct analogous relations between thimble and cycle bases of the disk partition functions for all states with two atoms. Here we use the fact that the flow of the soliton transports the atoms from the crystal with a higher real part of the spectral parameter to the one with a lower real part:
\begin{subequations}
\begin{align}\label{solit_2_a}
	\begin{split}
		\left(\begin{array}{cc}
			u_1 & u_2\\
			\bbox &  \varnothing
		\end{array}\right)&=P_1\overset{{\color{\myblue}\gamma_1}}{\left[\begin{array}{c|c}
				u_1 & u_2\\
				\bbox &  \varnothing
			\end{array}\right]}+P_2\overset{{\color{\myblue}\gamma_2}}{\left[\begin{array}{c|c}
			u_1 & u_2\\
			\Box &  \Box
		\end{array}\right]}+P_3\overset{{\color{\myblue}\gamma_3}}{\left[\begin{array}{c|c}
		u_1 & u_2\\
		\varnothing &  \bbox
	\end{array}\right]}\,,\\
		\left(\begin{array}{cc}
			u_1 & u_2\\
			\Box &  \Box
		\end{array}\right)&=P_4\overset{{\color{\myblue}\gamma_2}}{\left[\begin{array}{c|c}
				u_1 & u_2\\
				\Box &  \Box
			\end{array}\right]}+P_5\overset{{\color{\myblue}\gamma_3}}{\left[\begin{array}{c|c}
				u_1 & u_2\\
				\varnothing &  \bbox
			\end{array}\right]}\,,\\
		\left(\begin{array}{cc}
			u_1 & u_2\\
			\varnothing &  \bbox
		\end{array}\right)&=P_6\overset{{\color{\myblue}\gamma_3}}{\left[\begin{array}{c|c}
				u_1 & u_2\\
				\varnothing &  \bbox
			\end{array}\right]}\,;\\
	\end{split}
\end{align}
and 
\begin{align}\label{solit_2_b}
	\begin{split}
		\left(\begin{array}{cc}
			u_2 & u_1\\
			\bbox &  \varnothing
		\end{array}\right)&=Q_1\overset{{\color{\myblue}\gamma_3}}{\left[\begin{array}{c|c}
				u_2 & u_1\\
				\bbox &  \varnothing
			\end{array}\right]}+Q_2\overset{{\color{\myblue}\gamma_2}}{\left[\begin{array}{c|c}
				u_2 & u_1\\
				\Box &  \Box
			\end{array}\right]}+Q_3\overset{{\color{\myblue}\gamma_1}}{\left[\begin{array}{c|c}
				u_2 & u_1\\
				\varnothing &  \bbox
			\end{array}\right]}\,,\\
		\left(\begin{array}{cc}
			u_2 & u_1\\
			\Box &  \Box
		\end{array}\right)&=Q_4\overset{{\color{\myblue}\gamma_2}}{\left[\begin{array}{c|c}
				u_2 & u_1\\
				\Box &  \Box
			\end{array}\right]}+Q_5\overset{{\color{\myblue}\gamma_1}}{\left[\begin{array}{c|c}
				u_2 & u_1\\
				\varnothing &  \bbox
			\end{array}\right]}\,,\\
		\left(\begin{array}{cc}
			u_2 & u_1\\
			\varnothing &  \bbox
		\end{array}\right)&=Q_6\overset{{\color{\myblue}\gamma_1}}{\left[\begin{array}{c|c}
				u_2 & u_1\\
				\varnothing &  \bbox
			\end{array}\right]}\,.\\
	\end{split}
\end{align}
\end{subequations}
The expressions for the expansion coefficients and the $R$-matrix elements are rather long, and we will not give them here.
However we stress that a special coefficient $P_4/Q_4$ corresponds to a pure permutation of two single atom crystals located at $u_1$ and $u_2$ in the complex weight plane.
It is not surprising that it can be expressed in terms of the bond factors (see also \cite[Section~2.7]{Galakhov:2020vyb}):
\begin{align}\label{P4/Q4}
	\frac{P_4}{Q_4}=\frac{\prod\lm_{K\in \{a\to\ff\}}(u_{21}-h_K)}{\prod\lm_{L\in \{b\to\ff\}}(u_{12}-h_L)}\times\varphi^{a\Leftarrow b}(u_{12})\,.
\end{align}

\subsection{Quiver Yangian from Lax operators}

In this subsection we follow \cite{Litvinov:2020zeq} and show that Lax operators representing elementary braids form the algebra $\myY$ via the relations \eqref{Lax_BPS}.

As a starting assumption in this derivation, we assume that $R$-matrices satisfy the YBE \eqref{YBE}.
Surely, this derivation works only if the model is not excluded by the no-go arguments of Section~\ref{sec:no-go}.

\subsubsection{\texorpdfstring{$hh$ relations}{hh relations}}
Let us denote:
\begin{align}
	\CL_{\varnothing,\varnothing}(u)=:\tilde \myt_{\ff}(u)\,,
\end{align}
and study properties of the new operators $\tilde \myt_{\ff}(u)$.
The operators $\tilde \myt_{\ff}(u)$ form a commuting algebra.
This simple observation follows from a set of diagram equivalences and the fact that for trivial crystals the $R$-matrix is also trivial:
\begin{align}
	\begin{array}{c}
		\begin{tikzpicture}
			\draw[thick] (-1,0.5) to[out=0,in=180] (1,-0.5) (-1,0) to[out=0,in=180] (1,0.5) (-1,-0.5) to[out=0,in=180] (1,0); 
			\node[left] at (-1,-0.5) {$\scriptstyle \varnothing, u_2$};
			\node[left] at (-1,0) {$\scriptstyle \varnothing, u_1$};
			\node[left] at (-1,0.5) {$\scriptstyle \Kappa'$};
			\node[right] at (1,-0.5) {$\scriptstyle \Kappa$};
			\node[right] at (1,0) {$\scriptstyle \varnothing, u_2$};
			\node[right] at (1,0.5) {$\scriptstyle \varnothing, u_1$};
		\end{tikzpicture}
	\end{array}=
\begin{array}{c}
	\begin{tikzpicture}
		\draw[thick] (-1,0.5) to[out=0,in=180]  (1,-0.5) (-1,0) to[out=0,in=180] (0,-0.5) to[out=0,in=180] (1,0.5) (-1,-0.5) to[out=0,in=180] (0,0.5) to[out=0,in=180] (1,0); 
		\node[left] at (-1,-0.5) {$\scriptstyle \varnothing, u_2$};
		\node[left] at (-1,0) {$\scriptstyle \varnothing, u_1$};
		\node[left] at (-1,0.5) {$\scriptstyle \Kappa'$};
		\node[right] at (1,-0.5) {$\scriptstyle \Kappa$};
		\node[right] at (1,0) {$\scriptstyle \varnothing, u_2$};
		\node[right] at (1,0.5) {$\scriptstyle \varnothing, u_1$};
	\end{tikzpicture}
\end{array}=
\begin{array}{c}
	\begin{tikzpicture}
		\draw[thick] (-1,0.5) to[out=0,in=180] (1,-0.5) (-1,0) to[out=0,in=180] (1,0.5) (-1,-0.5) to[out=0,in=180] (1,0); 
		\node[left] at (-1,-0.5) {$\scriptstyle \varnothing, u_1$};
		\node[left] at (-1,0) {$\scriptstyle \varnothing, u_2$};
		\node[left] at (-1,0.5) {$\scriptstyle \Kappa'$};
		\node[right] at (1,-0.5) {$\scriptstyle \Kappa$};
		\node[right] at (1,0) {$\scriptstyle \varnothing, u_1$};
		\node[right] at (1,0.5) {$\scriptstyle \varnothing, u_2$};
	\end{tikzpicture}
\end{array}\,.
\end{align}
Moreover we can argue that crystal states $|\Kappa\rangle$ form an eigen basis of commuting operators $\tilde \myt_{\ff}(u)$, and we will compute eigenvalues momentarily.

Indeed, if we put an empty crystal at a puncture at $u_1$ and a crystal $\Kappa$ at a puncture at $u_2$ assuming that ${\rm Re}\,u_1>{\rm Re}\,u_2$ there will be no atoms that can be carried by solitons from the left to the right, therefore for the thimble-cycle change of basis one has:
\begin{align}
	\left(\begin{array}{cc}
		u_1 & u_2\\
		\varnothing &  \Kappa\\
	\end{array}\right)=G_1\overset{{\color{\myblue}\gamma_0}}{\left[\begin{array}{c|c}
			u_1 & u_2\\
			\varnothing &  \Kappa
		\end{array}\right]}\,.
\end{align}

The opposite situation is very different and we will get all the possible soliton contributions:
\begin{align}
	\left(\begin{array}{cc}
		u_2 & u_1\\
		\Kappa &  \varnothing\\
	\end{array}\right)=T_1\overset{{\color{\myblue}\gamma_0}}{\left[\begin{array}{c|c}
			u_2 & u_1\\
			\Kappa &  \varnothing
		\end{array}\right]}+\underbrace{\sum\lm_{\Box\in{\rm Rem}(\Kappa)}T_2(\Box)\left[\begin{array}{c|c}
		u_2 & u_1\\
		\Kappa-\Box &  \Box
	\end{array}\right]+\ldots}_{\sim O(e^{-{\rm Re}(u_2-u_1)})}\,.
\end{align}
However the ``solitonic" tail in this expression is suppressed by the exponentiated soliton action.
Using the fact that the $R$-matrix coefficients are rational functions in the spectral parameters we derive:
\begin{align}
	\begin{array}{c}
		\begin{tikzpicture}
			\draw[thick] (0,0) to[out=0,in=180] (0.7,0.5) (0,0.5) to[out=0,in=180] (0.7,0);
			\node[left] at (0,0) {$\scriptstyle (\varnothing,u_1,a)$};
			\node[left] at (0,0.5) {$\scriptstyle (\Kappa,u_2)$};
			\node[right] at (0.7,0) {$\scriptstyle (\Kappa,u_2)$};
			\node[right] at (0.7,0.5) {$\scriptstyle (\varnothing,u_1,a)$};
		\end{tikzpicture}
	\end{array}=\frac{G_1}{T_1}=\prod\lm_{\sqbox{$b$}\in\Kappa}\varphi^{\ff\Leftarrow b}\left(u_{12}-h_{\sqbox{$b$}}\right)\,.
\end{align}
This expression is an eigenvalue of the operator $\myt_{\ff}(u)$ on the crystal vector $|\Kappa\rangle$.
Therefore we conclude:
\begin{align}
	\tilde \myt_{\ff}(u)=\myt_{\ff}(u)\,.
\end{align}

\subsubsection{\texorpdfstring{$\myt e$ relations}{te relations}}

To check the rest of the relations \eqref{Lax_BPS}, we introduce the notation:
\begin{align}\label{Lax}
    \CL_{\varnothing,\Box}^{(a)}(u)=\myt_{\ff}(u)\, \tilde e^{(a)}(u),\quad 
			\CL_{\Box,\varnothing}^{(a)}(u)= \tilde f^{(a)}(u)\,\myt_{\ff}(u)\,.
\end{align}
In what follows we will use the recipes of \cite{Litvinov:2020zeq,Chistyakova:2021yyd,Kolyaskin:2022tqi} and argue that the generators $\tilde e^{(a)}(u)$ and $\tilde f^{(a)}(u)$ satisfy the quiver Yangian relations. The tilded generators thus can be identified with their untilded counterparts, moreover the BPS states form the crystal module defined by the quiver framing.

Consider the following equality:
\begin{align}
	\begin{array}{c}
		\begin{tikzpicture}
			\draw[thick] (0,0) to[out=0,in=180] (1,0.5) to[out=0,in=180] (1.7,1) (0,0.5) to[out=0,in=180] (1,0) to[out=0,in=180] (1.7,0.5) (0,1) -- (1,1) to[out=0,in=180] (1.7,0); 
			\node[left] at (0,0) {$\scriptstyle \varnothing,u_1$};
			\node[left] at (0,0.5) {$\scriptstyle \varnothing,u_2$};
			\node[left] at (0,1) {$\scriptstyle \Kappa'$};
			\node[right] at (1.7,0) {$\scriptstyle \Kappa$};
			\node[right] at (1.7,0.5) {$\scriptstyle \Box,u_2$};
			\node[right] at (1.7,1) {$\scriptstyle \varnothing,u_1$};
			\node[above] at (0.8,0.45) {$\scriptstyle \varnothing$};
			\node[below] at (0.8,0.05) {$\scriptstyle \varnothing$};
		\end{tikzpicture}
	\end{array}=\begin{array}{c}
	\begin{tikzpicture}[scale=-1]
		\draw[thick] (0,0) to[out=0,in=180] (1,0.5) to[out=0,in=180] (1.7,1) (0,0.5) to[out=0,in=180] (1,0) to[out=0,in=180] (1.7,0.5) (0,1) -- (1,1) to[out=0,in=180] (1.7,0); 
		\node[right] at (0,0) {$\scriptstyle \varnothing,u_1$};
		\node[right] at (0,0.5) {$\scriptstyle \Box,u_2$};
		\node[right] at (0,1) {$\scriptstyle \Kappa$};
		\node[left] at (1.7,0) {$\scriptstyle \Kappa'$};
		\node[left] at (1.7,0.5) {$\scriptstyle \varnothing,u_2$};
		\node[left] at (1.7,1) {$\scriptstyle \varnothing,u_1$};
		\node[below] at (0.8,0.45) {$\scriptstyle \Box$};
		\node[above] at (0.8,0.05) {$\scriptstyle \varnothing$};
	\end{tikzpicture}
\end{array}+\begin{array}{c}
\begin{tikzpicture}[scale=-1]
	\draw[thick] (0,0) to[out=0,in=180] (1,0.5) to[out=0,in=180] (1.7,1) (0,0.5) to[out=0,in=180] (1,0) to[out=0,in=180] (1.7,0.5) (0,1) -- (1,1) to[out=0,in=180] (1.7,0); 
	\node[right] at (0,0) {$\scriptstyle \varnothing,u_1$};
	\node[right] at (0,0.5) {$\scriptstyle \Box,u_2$};
	\node[right] at (0,1) {$\scriptstyle \Kappa$};
	\node[left] at (1.7,0) {$\scriptstyle \Kappa'$};
	\node[left] at (1.7,0.5) {$\scriptstyle \varnothing,u_2$};
	\node[left] at (1.7,1) {$\scriptstyle \varnothing,u_1$};
	\node[below] at (0.8,0.45) {$\scriptstyle \varnothing$};
	\node[above] at (0.8,0.05) {$\scriptstyle \Box$};
\end{tikzpicture}
\end{array}\,.
\end{align}
In terms of generators this diagrammatic equality can be rewritten in the following way:
\begin{align}
\begin{split}
	\myt_{\ff_1}(u_1)\myt_{\ff_2}(u_2)\tilde e^{(b)}(u_2)=&-\frac{A_3B_2}{B_1B_3}\myt_{\ff_2}(u_2)\myt_{\ff_1}(u_1)\tilde e^{(a)}(u_1)+\\
	&+\frac{A_3}{B_1}\myt_{\ff_2}(u_2)e^{(b)}(u_2)\myt_{\ff_1}(u_1)\,.
\end{split}
\end{align}
Dividing this equation by $\myt_{\ff_2}(u_2)$ from the left, and noting that the first term in the r.h.s.\ will drop out if we substitute the equality sign by $\simeq$, we can reduce this relation to:
\begin{align}
	\myt_{\ff}(u_1)\,\tilde e^{(b)}(u_2)\simeq \varphi^{\ff\Leftarrow b}(u_{12})\times \tilde e^{(b)}(u_2)\,\myt_{\ff}(u_1)\,.
\end{align}

We can similarly derive the $\myt f$ relations.

\subsubsection{\texorpdfstring{$ee$ and $ff$ relations}{ee and ff relations}}

Consider a set of relations including two atom crystals:
\begin{align}\label{2-atom}
\scalebox{0.8}{$\begin{array}{l}
	\begin{array}{c}
		\begin{tikzpicture}
			\draw[thick] (0,0) to[out=0,in=180] (1,0.5) to[out=0,in=180] (1.7,1) (0,0.5) to[out=0,in=180] (1,0) to[out=0,in=180] (1.7,0.5) (0,1) -- (1,1) to[out=0,in=180] (1.7,0); 
			\node[left] at (0,0) {$\scriptstyle \varnothing,u_1$};
			\node[left] at (0,0.5) {$\scriptstyle \varnothing,u_2$};
			\node[left] at (0,1) {$\scriptstyle \Kappa'$};
			\node[right] at (1.7,0) {$\scriptstyle \Kappa$};
			\node[right] at (1.7,0.5) {$\scriptstyle \Box,u_2$};
			\node[right] at (1.7,1) {$\scriptstyle \Box,u_1$};
			\node[above] at (0.8,0.45) {$\scriptstyle \varnothing$};
			\node[below] at (0.8,0.05) {$\scriptstyle \varnothing$};
		\end{tikzpicture}
	\end{array}=\begin{array}{c}
		\begin{tikzpicture}[scale=-1]
			\draw[thick] (0,0) to[out=0,in=180] (1,0.5) to[out=0,in=180] (1.7,1) (0,0.5) to[out=0,in=180] (1,0) to[out=0,in=180] (1.7,0.5) (0,1) -- (1,1) to[out=0,in=180] (1.7,0); 
			\node[right] at (0,0) {$\scriptstyle \Box,u_1$};
			\node[right] at (0,0.5) {$\scriptstyle \Box,u_2$};
			\node[right] at (0,1) {$\scriptstyle \Kappa$};
			\node[left] at (1.7,0) {$\scriptstyle \Kappa'$};
			\node[left] at (1.7,0.5) {$\scriptstyle \varnothing,u_2$};
			\node[left] at (1.7,1) {$\scriptstyle \varnothing,u_1$};
			\node[below] at (0.8,0.45) {$\scriptstyle \Box$};
			\node[above] at (0.8,0.05) {$\scriptstyle \Box$};
		\end{tikzpicture}
	\end{array}+\begin{array}{c}
	\begin{tikzpicture}[scale=-1]
		\draw[thick] (0,0) to[out=0,in=180] (1,0.5) to[out=0,in=180] (1.7,1) (0,0.5) to[out=0,in=180] (1,0) to[out=0,in=180] (1.7,0.5) (0,1) -- (1,1) to[out=0,in=180] (1.7,0); 
		\node[right] at (0,0) {$\scriptstyle \Box,u_1$};
		\node[right] at (0,0.5) {$\scriptstyle \Box,u_2$};
		\node[right] at (0,1) {$\scriptstyle \Kappa$};
		\node[left] at (1.7,0) {$\scriptstyle \Kappa'$};
		\node[left] at (1.7,0.5) {$\scriptstyle \varnothing,u_2$};
		\node[left] at (1.7,1) {$\scriptstyle \varnothing,u_1$};
		\node[below] at (0.8,0.45) {$\scriptstyle \varnothing$};
		\node[above] at (0.8,0.05) {$\scriptstyle \bbox$};
	\end{tikzpicture}
\end{array}+\begin{array}{c}
\begin{tikzpicture}[scale=-1]
	\draw[thick] (0,0) to[out=0,in=180] (1,0.5) to[out=0,in=180] (1.7,1) (0,0.5) to[out=0,in=180] (1,0) to[out=0,in=180] (1.7,0.5) (0,1) -- (1,1) to[out=0,in=180] (1.7,0); 
	\node[right] at (0,0) {$\scriptstyle \Box,u_1$};
	\node[right] at (0,0.5) {$\scriptstyle \Box,u_2$};
	\node[right] at (0,1) {$\scriptstyle \Kappa$};
	\node[left] at (1.7,0) {$\scriptstyle \Kappa'$};
	\node[left] at (1.7,0.5) {$\scriptstyle \varnothing,u_2$};
	\node[left] at (1.7,1) {$\scriptstyle \varnothing,u_1$};
	\node[below] at (0.8,0.45) {$\scriptstyle \bbox$};
	\node[above] at (0.8,0.05) {$\scriptstyle \varnothing$};
\end{tikzpicture}
\end{array}\,,\\
\begin{array}{c}
	\begin{tikzpicture}
		\draw[thick] (0,0) to[out=0,in=180] (1,0.5) to[out=0,in=180] (1.7,1) (0,0.5) to[out=0,in=180] (1,0) to[out=0,in=180] (1.7,0.5) (0,1) -- (1,1) to[out=0,in=180] (1.7,0); 
		\node[left] at (0,0) {$\scriptstyle \varnothing,u_1$};
		\node[left] at (0,0.5) {$\scriptstyle \varnothing,u_2$};
		\node[left] at (0,1) {$\scriptstyle \Kappa'$};
		\node[right] at (1.7,0) {$\scriptstyle \Kappa$};
		\node[right] at (1.7,0.5) {$\scriptstyle \bbox,u_2$};
		\node[right] at (1.7,1) {$\scriptstyle \varnothing,u_1$};
		\node[above] at (0.8,0.45) {$\scriptstyle \varnothing$};
		\node[below] at (0.8,0.05) {$\scriptstyle \varnothing$};
	\end{tikzpicture}
\end{array}=\begin{array}{c}
	\begin{tikzpicture}[scale=-1]
		\draw[thick] (0,0) to[out=0,in=180] (1,0.5) to[out=0,in=180] (1.7,1) (0,0.5) to[out=0,in=180] (1,0) to[out=0,in=180] (1.7,0.5) (0,1) -- (1,1) to[out=0,in=180] (1.7,0); 
		\node[right] at (0,0) {$\scriptstyle \varnothing,u_1$};
		\node[right] at (0,0.5) {$\scriptstyle \bbox,u_2$};
		\node[right] at (0,1) {$\scriptstyle \Kappa$};
		\node[left] at (1.7,0) {$\scriptstyle \Kappa'$};
		\node[left] at (1.7,0.5) {$\scriptstyle \varnothing,u_2$};
		\node[left] at (1.7,1) {$\scriptstyle \varnothing,u_1$};
		\node[below] at (0.8,0.45) {$\scriptstyle \Box$};
		\node[above] at (0.8,0.05) {$\scriptstyle \Box$};
	\end{tikzpicture}
\end{array}+\begin{array}{c}
	\begin{tikzpicture}[scale=-1]
		\draw[thick] (0,0) to[out=0,in=180] (1,0.5) to[out=0,in=180] (1.7,1) (0,0.5) to[out=0,in=180] (1,0) to[out=0,in=180] (1.7,0.5) (0,1) -- (1,1) to[out=0,in=180] (1.7,0); 
		\node[right] at (0,0) {$\scriptstyle \varnothing,u_1$};
		\node[right] at (0,0.5) {$\scriptstyle \bbox,u_2$};
		\node[right] at (0,1) {$\scriptstyle \Kappa$};
		\node[left] at (1.7,0) {$\scriptstyle \Kappa'$};
		\node[left] at (1.7,0.5) {$\scriptstyle \varnothing,u_2$};
		\node[left] at (1.7,1) {$\scriptstyle \varnothing,u_1$};
		\node[below] at (0.8,0.45) {$\scriptstyle \varnothing$};
		\node[above] at (0.8,0.05) {$\scriptstyle \bbox$};
	\end{tikzpicture}
\end{array}+\begin{array}{c}
	\begin{tikzpicture}[scale=-1]
		\draw[thick] (0,0) to[out=0,in=180] (1,0.5) to[out=0,in=180] (1.7,1) (0,0.5) to[out=0,in=180] (1,0) to[out=0,in=180] (1.7,0.5) (0,1) -- (1,1) to[out=0,in=180] (1.7,0); 
		\node[right] at (0,0) {$\scriptstyle \varnothing,u_1$};
		\node[right] at (0,0.5) {$\scriptstyle \bbox,u_2$};
		\node[right] at (0,1) {$\scriptstyle \Kappa$};
		\node[left] at (1.7,0) {$\scriptstyle \Kappa'$};
		\node[left] at (1.7,0.5) {$\scriptstyle \varnothing,u_2$};
		\node[left] at (1.7,1) {$\scriptstyle \varnothing,u_1$};
		\node[below] at (0.8,0.45) {$\scriptstyle \bbox$};
		\node[above] at (0.8,0.05) {$\scriptstyle \varnothing$};
	\end{tikzpicture}
\end{array}\,,\\
\begin{array}{c}
	\begin{tikzpicture}
		\draw[thick] (0,0) to[out=0,in=180] (1,0.5) to[out=0,in=180] (1.7,1) (0,0.5) to[out=0,in=180] (1,0) to[out=0,in=180] (1.7,0.5) (0,1) -- (1,1) to[out=0,in=180] (1.7,0); 
		\node[left] at (0,0) {$\scriptstyle \varnothing,u_1$};
		\node[left] at (0,0.5) {$\scriptstyle \varnothing,u_2$};
		\node[left] at (0,1) {$\scriptstyle \Kappa'$};
		\node[right] at (1.7,0) {$\scriptstyle \Kappa$};
		\node[right] at (1.7,0.5) {$\scriptstyle \bbox,u_2$};
		\node[right] at (1.7,1) {$\scriptstyle \varnothing,u_1$};
		\node[above] at (0.8,0.45) {$\scriptstyle \varnothing$};
		\node[below] at (0.8,0.05) {$\scriptstyle \varnothing$};
	\end{tikzpicture}
\end{array}=\begin{array}{c}
	\begin{tikzpicture}[scale=-1]
		\draw[thick] (0,0) to[out=0,in=180] (1,0.5) to[out=0,in=180] (1.7,1) (0,0.5) to[out=0,in=180] (1,0) to[out=0,in=180] (1.7,0.5) (0,1) -- (1,1) to[out=0,in=180] (1.7,0); 
		\node[right] at (0,0) {$\scriptstyle \varnothing,u_1$};
		\node[right] at (0,0.5) {$\scriptstyle \bbox,u_2$};
		\node[right] at (0,1) {$\scriptstyle \Kappa$};
		\node[left] at (1.7,0) {$\scriptstyle \Kappa'$};
		\node[left] at (1.7,0.5) {$\scriptstyle \varnothing,u_2$};
		\node[left] at (1.7,1) {$\scriptstyle \varnothing,u_1$};
		\node[below] at (0.8,0.45) {$\scriptstyle \Box$};
		\node[above] at (0.8,0.05) {$\scriptstyle \Box$};
	\end{tikzpicture}
\end{array}+\begin{array}{c}
	\begin{tikzpicture}[scale=-1]
		\draw[thick] (0,0) to[out=0,in=180] (1,0.5) to[out=0,in=180] (1.7,1) (0,0.5) to[out=0,in=180] (1,0) to[out=0,in=180] (1.7,0.5) (0,1) -- (1,1) to[out=0,in=180] (1.7,0); 
		\node[right] at (0,0) {$\scriptstyle \varnothing,u_1$};
		\node[right] at (0,0.5) {$\scriptstyle \bbox,u_2$};
		\node[right] at (0,1) {$\scriptstyle \Kappa$};
		\node[left] at (1.7,0) {$\scriptstyle \Kappa'$};
		\node[left] at (1.7,0.5) {$\scriptstyle \varnothing,u_2$};
		\node[left] at (1.7,1) {$\scriptstyle \varnothing,u_1$};
		\node[below] at (0.8,0.45) {$\scriptstyle \varnothing$};
		\node[above] at (0.8,0.05) {$\scriptstyle \bbox$};
	\end{tikzpicture}
\end{array}+\begin{array}{c}
	\begin{tikzpicture}[scale=-1]
		\draw[thick] (0,0) to[out=0,in=180] (1,0.5) to[out=0,in=180] (1.7,1) (0,0.5) to[out=0,in=180] (1,0) to[out=0,in=180] (1.7,0.5) (0,1) -- (1,1) to[out=0,in=180] (1.7,0); 
		\node[right] at (0,0) {$\scriptstyle \varnothing,u_1$};
		\node[right] at (0,0.5) {$\scriptstyle \bbox,u_2$};
		\node[right] at (0,1) {$\scriptstyle \Kappa$};
		\node[left] at (1.7,0) {$\scriptstyle \Kappa'$};
		\node[left] at (1.7,0.5) {$\scriptstyle \varnothing,u_2$};
		\node[left] at (1.7,1) {$\scriptstyle \varnothing,u_1$};
		\node[below] at (0.8,0.45) {$\scriptstyle \bbox$};
		\node[above] at (0.8,0.05) {$\scriptstyle \varnothing$};
	\end{tikzpicture}
\end{array}\,.\\
\end{array}$}
\end{align}

Subtracting from the first equation the second one with a multiplier $P_5/P_6$ we derive the following equation:
\begin{align}
	\begin{split}
	&\CL_{\varnothing,\Box}(u_1)\CL_{\varnothing,\Box}(u_2)-\frac{P_4}{Q_4}\CL_{\varnothing,\Box}(u_2)\CL_{\varnothing,\Box}(u_1)\\
	&\underline{-\frac{P_5}{P_6}	\CL_{\varnothing,\varnothing}(u_1)\CL_{\varnothing,\bbox}(u_2)+\frac{P_4Q_5}{Q_4Q_6}	\CL_{\varnothing,\varnothing}(u_2)\CL_{\varnothing,\bbox}(u_1)}=0\,.
	\end{split}
\end{align}
Substituting expressions for the Lax operators \eqref{Lax}, substituting an expression \eqref{P4/Q4} for $P_4/Q_4$, dividing this equation by $\myt_{\ff_1}(u_1)\myt_{\ff_2}(u_2)$ from the left, and noticing that after switching to $\simeq$ equality underlined terms do not contribute, we derive the following relation between the $\tilde e^{(a)}(u)$-operators:
\begin{align}
	\tilde e^{(a)}(u_1)\,\tilde e^{(b)}(u_2)\;\simeq\;\varphi^{a\Leftarrow b}\left(u_{12}\right) \times\tilde e^{(b)}(u_2)\,\tilde e^{(a)}(u_1)\,.
\end{align}
This relation coincides with the one for the $e^{(a)}(u)$ operators of the quiver Yangian.

We can similarly derive the $ff$ relation.

\subsubsection{\texorpdfstring{$ef$ relations}{ef relations}}

To produce the $ef$ relations, the authors of \cite{Litvinov:2020zeq} exploit certain properties of the pole structure in the $ee$ and $ff$ relations.
Unfortunately, due to the no-go arguments of Section~\ref{sec:no-go}, this pole structure is lost for generic $\myY$.
Therefore we are unable to derive the $ef$ relation from the diagrammatic technique for the YBE.
For the cases not excluded by these no-go arguments, however, we expect that this derivation will work out (see also a derivation in \cite{Bao:2022fpk}). 
In those cases we confirm that the tilded operators $\tilde \myt_{\ff}(z)$, $\tilde e^{(a)}(z)$, $\tilde f^{(a)}(z)$ indeed generate the algebra $\myY$.

\section{Soliton corrections for \texorpdfstring{$\myY(\widehat{\fg\fl}_1)$}{Y{h1,h2}(gl1)}}\label{app:solit_corr_gl_1}

\subsection{Form-factors}
The form of the soliton corrections \eqref{S_k_form} suggests that the properties of $S_k$ can be encoded in a scalar rational form-factor function
$$
g_k(x_1,x_2,\ldots,x_k|y_1,y_2\ldots,y_k)
$$
of $2k$ variables in the following way: 
\begin{align}\label{S_k_form_expl}
    \begin{split}
        S_k|\Kappa_1\rangle_{u_1}\otimes|\Kappa_2\rangle_{u_2}=&\sum\lm_{\Box_1\in{\rm Add}\left(\Kappa_1\right)}\ldots\sum\lm_{\Box_k\in{\rm Add}\left(\Kappa_1-\sum\lm_{i=1}^{k-1}\Box_i\right)}
        \sum\lm_{\tilde\Box_1\in{\rm Add}\left(\Kappa_2\right)}\ldots\sum\lm_{\tilde\Box_k\in{\rm Add}\left(\Kappa_2+\sum\lm_{i=1}^{k-1}\tilde\Box_i\right)}\\
        &
        g_k\left(u_1+h_{\Box_1},\ldots,u_1+h_{\Box_k}|u_2+h_{\tilde\Box_1},\ldots,u_2+h_{\tilde\Box_k}\right)
        \\
        &\times\left[\Kappa_1\to\Kappa_1-\Box_1\to\ldots\to\Kappa_1-\sum\lm_{i=1}^{k-1}\Box_i\to\Kappa_1-\sum\lm_{i=1}^{k}\Box_i\right]\\
        &\times\left[\Kappa_2\to\Kappa_2+\tilde\Box_1\to\ldots\to\Kappa_2+\sum\lm_{i=1}^{k-1}\tilde\Box_i\to\Kappa_2+\sum\lm_{i=1}^{k}\tilde\Box_i\right]\\
        &\times\left|\Kappa_1-\sum\lm_{i=1}^k\Box_i\right\rangle_{u_1}\otimes\left|\Kappa_2+\sum\lm_{i=1}^k\tilde\Box_i\right\rangle_{u_2}\,.
    \end{split}
\end{align}

\subsection{Soliton corrections for \texorpdfstring{$\myY(\widehat{\fg\fl}_1)$}{Y(gl1)}}

The non-trivial coproduct is known \cite{Prochazka:2015deb} for $\myY(\widehat{\fg\fl}_1)$.
It is given in terms of the mode-expansions:\footnote{Note that the mode expansions \eqref{MEgl1app} used in this subsection is slightly \emph{different} from the one used in most of the text (defined in \eqref{shift_psi_exp}). (Since in this subsection we do not consider any shifts, the difference is $k$ vs.\ $k+1$ in the exponent of $z$.)
This is to match with the one ususally adopted in the literature on the affine Yangian of $\mathfrak{gl}_1$.}
\begin{align}\label{MEgl1app}
    e(u)=\sum\lm_{k=0}^{\infty}\frac{e_k}{u^{k+1}},\quad f(u)=\sum\lm_{k=0}^{\infty}\frac{f_k}{u^{k+1}},\quad \psi(u)=1+\sigma_3\sum\lm_{k=0}^{\infty}\frac{\psi_k}{u^{k+1}}\,,
\end{align}
where $\sigma_3\equiv\texttt{h}_1\texttt{h}_2\texttt{h}_3$, where $\texttt{h}_i$ are the equivariant weight parameters of the $\IC^3$ quiver:
\begin{align}
		\begin{array}{c}
			\begin{tikzpicture}
				\draw[thick,postaction={decorate},decoration={markings, 
					mark= at position 0.75 with {\arrow{stealth}}}] (0,0) to[out=60,in=0] (0,1) to[out=180,in=120] (0,0);
				\node[above] at (0,1) {$\scriptstyle (X_1,\texttt{h}_1)$};
				\begin{scope}[rotate=120]
					\draw[thick,postaction={decorate},decoration={markings, 
						mark= at position 0.75 with {\arrow{stealth}}}] (0,0) to[out=60,in=0] (0,1) to[out=180,in=120] (0,0);
					\node[left] at (0,1) {$\scriptstyle (X_2,\texttt{h}_2)$};
				\end{scope}
				\begin{scope}[rotate=240]
					\draw[thick,postaction={decorate},decoration={markings, 
						mark= at position 0.75 with {\arrow{stealth}}}] (0,0) to[out=60,in=0] (0,1) to[out=180,in=120] (0,0);
					\node[right] at (0,1) {$\scriptstyle (X_3,\texttt{h}_3)$};
				\end{scope}
			\draw[fill=\myblue] (0,0) circle (0.1);
			\end{tikzpicture}
		\end{array},\quad W=\Tr\left[X_1,X_2\right]X_3,\quad \texttt{h}_1+\texttt{h}_2+\texttt{h}_3=0\,.
	\end{align}

The coproducts on the (essential) low level modes are \cite{Prochazka:2015deb}:
\begin{align}
	\begin{split}
		\Delta e_0=&e_0\otimes 1+1\otimes e_0\,,\\
		\Delta f_0=&f_0\otimes 1+1\otimes f_0\,,\\
		\Delta \psi_3=&\psi_3\otimes 1+1\otimes \psi_3+\sigma_3\psi_2\otimes\psi_0+\sigma_3\psi_0\otimes\psi_2+\sigma_3\psi_1\otimes\psi_1\\
		&\quad -6\sigma_3\sum\lm_{m=1}^{\infty}\frac{m}{(m-1)!^2}\;{\rm ad}_{f_1}^{m-1}f_0\otimes{\rm ad}_{e_1}^{m-1}e_0\,.
	\end{split}
\end{align}
The expressions for the action of the coproduct on the remaining generators can be derived from the fact that these three modes ($e_0$, $f_0$ and $\psi_3$) can generate all the higher modes via mode relations, together with the fact that the coproduct is an algebra homomorphism.

Let us rewrite the coproduct expression for generator $\psi_3$ in the following form:
\begin{align}
	\Delta\psi_3=\Delta_1\psi_3+\sum\lm_{m=1}^{\infty}P_m,\quad {\rm deg}\,P_m=2m\,.
\end{align}
Substituting this relation into \eqref{iso}, we derive the following recurrence relation for $S_k$:
\begin{align}\label{S_k_gl_1_rec}
	\left[\Delta_1\psi_3,S_k\right]+\sum\lm_{m=1}^{k-1}P_mS_{k-m}+P_k=0,\quad \forall k\geq 1\,.
\end{align}
And substituting the ansatz \eqref{S_k_form_expl} leads to a recurrence relation for $g_k$:
\begin{align}\label{Y_gl_1_g_k_recurr}
	\begin{split}
		&\left(\sum\lm_{i=1}^kx_i-\sum\lm_{i=1}^ky_i\right)g_k(x_1,\ldots,x_k|y_1,\ldots,y_k)+\\
		&\hspace{2cm}\sigma_3\sum\lm_{m=1}^{k}m \,g_{k-m}(x_1,\ldots,x_{k-m}|y_1,\ldots,y_{k-m})\times\\
		&\hspace{4cm}\chi_m(x_{k-m+1},\ldots,x_k)\chi_m(y_{k-m+1},\ldots,y_k)=0\,,\\
	\end{split}
\end{align}
where we imply $g_0\equiv 1$, and $\chi_m$ are order $m-1$ polynomials defined in the following way:
\begin{align}
	\chi_m(x_1,\ldots,x_m)=\sum\lm_{j=1}^m\frac{(-1)^j}{(j-1)!(m-j)!}\,x_1\cdot x_2\cdot\ldots\cdot x_{j-1}\cdot 1\cdot x_{j+1}\cdot\ldots\cdot x_m\,.
\end{align}

The recurrence relation \eqref{Y_gl_1_g_k_recurr} has a unique solution.
In particular, for the first two form-factors we have:
\begin{align}
	\begin{split}
		g_1&=-\frac{\sigma_3}{x_1-y_1}\,,\\
		g_2&=-2\sigma_3\frac{-\frac{\sigma_3}{2(x_1-y_1)}+(x_1-x_2)(y_1-y_2)}{x_1+x_2-y_1-y_2}\,.\\
	\end{split}
\end{align}

\section{Coproduct for  \texorpdfstring{$\myY(\widehat{\fg\fl}_{m|n})$}{Yh1h2(gl(mn))} and soliton corrections}\label{app:solit_corr_gl_m|n}

\subsection{Coproduct}

In Section~\ref{sec:co-mul} we derived an expression for $S_1$ for a general toric quiver.
For higher soliton corrections starting with $S_2$,
the relevant equations turn out to be rather involved for a generic toric quiver.
We can nevertheless incorporate other indirect methods and derive actual coproducts, at least for a special family of quivers.

Resolutions of the so-called generalized conifold $xy=z^mw^n$ gives a large family of toric Calabi-Yau 3-fold  without compact four-cycles.
The corresponding BPS algebra is the affine Yangian $\myY_{\texttt{h}_1,\texttt{h}_2}\left(\widehat{\fg\fl}_{m|n}\right)$. In order to specify the algebra we need to choose a signature (cf.\ \cite{Nagao:2009rq,2019arXiv191208729B}):
\begin{align}
	\Sigma_{m,n}:\quad \{1,2,\ldots,m+n \}\longrightarrow \{+1,-1\},\;\mbox{so that}\;\#(+1)=m,\;\#(-1)=n \;,
\end{align}
where in the following we consider indices modulo $m+n$. 

This information could be reinterpreted in the form of a quiver, or a Dynkin diagram, and can be summarized in the following table:
\begin{align}\renewcommand{\arraystretch}{1.3}
	\begin{array}{c|c|c}
		\mbox{spin arrangement} & \sigma_i\sigma_{i+1}=1& \sigma_i\sigma_{i+1}=-1\\
		\hline
		\IZ_2\mbox{-parity} &\mbox{Even} & \mbox{Odd} \\
		\hline
		\mbox{Dynkin node} & \begin{array}{c}
			\begin{tikzpicture}
				\draw[ultra thick] (-0.5,0) -- (0.5,0);
				\draw[fill=white] (0,0) circle (0.15);
			\end{tikzpicture}
		\end{array}& \begin{array}{c}
			\begin{tikzpicture}
				\draw[ultra thick] (-0.5,0) -- (0.5,0);
				\draw[fill=white] (0,0) circle (0.15);
				\draw (-0.2,-0.2) -- (0.2,0.2) (0.2,-0.2) -- (-0.2,0.2);
			\end{tikzpicture}
		\end{array}\\
		\hline
		\mbox{quiver node} & \begin{array}{c}
			\begin{tikzpicture}
				\draw (0,0) circle (0.15);
				\draw[-stealth] (-1,0.05) -- (-0.141421,0.05) node[pos=0.2,above] {$A_i$};
				\draw[stealth-] (-1,-0.05) -- (-0.141421,-0.05) node[pos=0.2,below] {$B_i$};
				\draw[stealth-] (1,0.05) -- (0.141421,0.05) node[pos=0.2,above] {$A_{i+1}$};
				\draw[-stealth] (1,-0.05) -- (0.141421,-0.05) node[pos=0.2,below] {$B_{i+1}$};
				\draw[postaction={decorate},decoration={markings, 
					mark= at position 0.85 with {\arrow{stealth}}}] (-0.05,0.141421) to[out=120,in=180] node[pos=0.7,above left] {$C_i$} (0,0.7) to[out=0,in=60] (0.05,0.141421);
				\draw[white] (-0.1,1.1) -- (0.1,1.1);
			\end{tikzpicture}
		\end{array}& \begin{array}{c}
			\begin{tikzpicture}
				\draw (0,0) circle (0.15);
				\draw[-stealth] (-1,0.05) -- (-0.141421,0.05) node[pos=0.2,above] {$A_i$};
				\draw[stealth-] (-1,-0.05) -- (-0.141421,-0.05) node[pos=0.2,below] {$B_i$};
				\draw[stealth-] (1,0.05) -- (0.141421,0.05) node[pos=0.2,above] {$A_{i+1}$};
				\draw[-stealth] (1,-0.05) -- (0.141421,-0.05) node[pos=0.2,below] {$B_{i+1}$};
				\draw[white] (-0.1,1.1) -- (0.1,1.1);
			\end{tikzpicture}
		\end{array}\\
		\hline
		\mbox{superpotential }\delta W_i & \sigma_i\,\Tr\, C_i\left(B_{i+1}A_{i+1}-A_{i}B_{i}\right)& -\sigma_i\,\Tr\,B_{i+1}A_{i+1}A_iB_i\\
	\end{array}
\end{align}

The equivariant weights associated to the quiver arrows are:
\begin{align}
	\begin{split}
		h(A_i)=\sigma_i\left(-\texttt{h}_1-\texttt{h}_2\right)\;,\quad h(B_i)=\sigma_i\left(-\texttt{h}_1+\texttt{h}_2\right)\;,\quad h(C_i)=2\sigma_i\texttt{h}_1\;.
	\end{split}
\end{align}

Another way to encode the information contained in the Dynkin diagram is to construct two matrices---the Cartan matrix and the auxiliary matrix---associated with the signature choice $\Sigma$ following \cite{2019arXiv191208729B}:
\begin{align}
	\begin{split}
		A_{ij}&=\left(\sigma_i+\sigma_{i+1}\right)\delta_{i,j}-\sigma_i\delta_{i,j+1}-\sigma_j\delta_{i+1,j}\;,\\
		M_{ij}&=\sigma_i\delta_{i,j+1}-\sigma_j\delta_{i+1,j}\;.
	\end{split}
\end{align}

The admissible algebra $\myY_{\texttt{h}_1,\texttt{h}_2}\left(\widehat{\fg\fl}_{m|n}\right)$ (without negative shifts) is generated by a set of triplets $(e_n^{(a)},f_n^{(a)},\psi_n^{(a)})$, $n\in \IZ_{\geq 0}$ associated to nodes $a=1,\ldots, m+n$.
Raising/lowering generators have $\IZ_2$-grading corresponding to that of the node, whereas Cartan generators are always bosonic.
The generating functions satisfying defining relations \eqref{QiuvYangian} read (here we have adopted a canonical normalization of the $\psi^{(a)}(z)$ fields so that the first-order term in the expansion is 1):\footnote{Note that the mode expansions \eqref{EXglmnapp} used in this subsection is slightly different from the one used in most of the text (defined in \eqref{shift_psi_exp}). (Since in this subsection we do not consider any shifts, the difference is $n$ vs.\ $n+1$ in the exponent of $z$.)
This is to match with the one ususally adopted in the literature on the affine Yangian of $\mathfrak{gl}_{m|n}$.}
\begin{align}\label{EXglmnapp}
	e^{(a)}(z)=\sum\lm_{n=0}^{\infty}\frac{e_n^{(a)}}{z^{n+1}},\quad 	f^{(a)}(z)=\sum\lm_{n=0}^{\infty}\frac{f_n^{(a)}}{z^{n+1}},\quad 	\psi^{(a)}(z)=1+\sum\lm_{n=0}^{\infty}\frac{\psi_n^{(a)}}{z^{n+1}}\,.
\end{align}


Let us consider some relations for generator modes (in what follows we assume $m+n\geq 3$):
\begin{align}
	\begin{split}
		& \left[\psi_{n+1}^{(a)},e_k^{(b)}\right]-\left[\psi_n^{(a)},e_{k+1}^{(b)}\right]-A_{ab}\texttt{h}_1\left\{\psi_n^{(a)},e_k^{(b)}\right\}+M_{ab}\texttt{h}_2\left[\psi_n^{(a)},e_k^{(b)}\right]=0\,,\\
		& \left[\psi_{n+1}^{(a)},f_k^{(b)}\right]-\left[\psi_n^{(a)},f_{k+1}^{(b)}\right]+A_{ab}\texttt{h}_1\left\{\psi_n^{(a)},f_k^{(b)}\right\}+M_{ab}\texttt{h}_2\left[\psi_n^{(a)},f_k^{(b)}\right]=0\,,\\
		& \left[e_n^{(a)},f_k^{(b)}\right\}=-\nu_a\delta_{ab}\psi_{n+k}^{(a)}\,.
	\end{split}
\end{align}
where
\begin{align}
	\nu_a=\left\{\begin{array}{ll}
		-h(C_a)^{-1}, & \mbox{if }a\mbox{ is even;}\\
		1, & \mbox{if }a\mbox{ is odd.}\\
	\end{array}\right.
\end{align}

Clearly, we see that the subalgebra generated by the zero modes is simply $\widehat{\fg\fl}_{m|n}$:
\begin{align}
	\left[\psi_0^{(a)},e_0^{(b)}\right]=2A_{ab}\texttt{h}_1e_0^{(b)},\quad 
	\left[\psi_0^{(a)},f_0^{(b)}\right]=-2A_{ab}\texttt{h}_1f_0^{(b)},\quad \left[e_0^{(a)},f_0^{(b)}\right\}=-\nu_a\delta_{ab}\psi_0^{(a)}\,.
\end{align}

Higher modes could be ``derived" through the action of $\psi_1$:
\begin{align}\label{mode_shift}
	\left[\psi_1^{(a)},e_k^{(b)}\right]=2A_{ab}\texttt{h}_1e_{k+1}^{(b)}+A_{ab}\texttt{h}_1\left\{\psi_0^{(a)},e_k^{(b)}\right\}-M_{ab}\texttt{h}_2\left[\psi_0^{(a)},e_k^{(b)}\right]\,.
\end{align}

Since the zero modes form a $\widehat{\fg\fl}_{m|n}$ subalgebra, it is natural to expect that the coproduct for this part is trivial and becomes non-trivial starting with $\psi_1$:
\begin{align}
	\begin{split}
		&\Delta e_0^{(a)}=e_0^{(a)}\otimes 1+1\otimes e_0^{(a)}\,,\\
		&\Delta f_0^{(a)}=f_0^{(a)}\otimes 1+1\otimes f_0^{(a)}\,,\\
		&\Delta \psi_0^{(a)}=\psi_0^{(a)}\otimes 1+1\otimes \psi_0^{(a)}\,,\\
		&\Delta \psi_1^{(a)}=\psi_1^{(a)}\otimes 1+1\otimes \psi_1^{(a)}+\psi_0^{(a)}\otimes \psi_0^{(a)}+\CP^{(a)}\,.
	\end{split}
\end{align}
It is enough to know the corrections $\CP^{(a)}$ to extend the action of the coproduct to all generators via the algebra homomorphism and the relations \eqref{mode_shift}.

The coproduct in this form is known for affine Yangians $\myY(\widehat{\fg\fl}_{m|n})$ \cite{Bao:2022fpk,guay2018coproduct} (see also \cite{2019arXiv191106666U} for $\myY(\widehat{\fg\fl}_{m})$).

In what follows we propose a mechanism to construct corrections iteratively.
So, for instance, we derive:
\begin{align}
	2A_{ab}\texttt{h}_1\,\Delta e_1^{(b)}=2A_{ab}\texttt{h}_1\left(e_1^{(b)}\otimes 1+1\otimes e_1^{(b)}\right)+\left[\CP^{(a)},e_0^{(b)}\otimes 1+1\otimes e_0^{(b)}\right]\,.
\end{align}

Commuting further with $\Delta f_0^{(c)}$ we derive an equation for $\CP^{(a)}$:
\begin{align}\label{P_eq}
	2A_{ab}\texttt{h}_1\delta_{bc}\nu_b\left(\psi_0^{(b)}\otimes\psi_0^{(b)}+\CP^{(b)}\right)+\left[\left[\CP^{(a)},e_0^{(b)}\otimes 1+1\otimes e_0^{(b)}\right\},f_0^{(c)}\otimes 1+1\otimes f_0^{(c)}\right\}=0\,.
\end{align}

The term $\CP^{(a)}$ appears as a result of commuting $\Delta_0 \psi^{(a)}_1$ with $U$, so it also admits a degree decomposition:
\begin{align}
	\CP^{(a)}=\sum\lm_{k=1}^{\infty}\CP^{(a)}_k,\quad {\rm deg}\,\CP^{(a)}_k=2k\,.
\end{align}
This decomposition transforms \eqref{P_eq} into a recurrent set of equations to determine $\CP^{(a)}_k$:
\begin{align}
	\begin{split}
		&2A_{ab}\texttt{h}_1\delta_{bc}\nu_b\,\psi_0^{(b)}\otimes\psi_0^{(b)}+\left[\left[\CP^{(a)}_1,e_0^{(b)}\otimes 1\right\},1\otimes f_0^{(c)}\right\}=0\,,\\
		&2A_{ab}\texttt{h}_1\delta_{bc}\nu_b\,\CP^{(b)}_1+\left[\left[\CP^{(a)}_2,e_0^{(b)}\otimes 1\right\},1\otimes f_0^{(c)}\right\}+\\
		&+\left[\left[\CP^{(a)}_1,1\otimes e_0^{(b)}\right\},1\otimes f_0^{(c)}\right\}+\left[\left[\CP^{(a)}_1,e_0^{(b)}\otimes 1\right\},f_0^{(c)}\otimes 1\right\}=0\,,\\
		&2A_{ab}\texttt{h}_1\delta_{bc}\nu_b\,\CP^{(b)}_{k-1}+\left[\left[\CP^{(a)}_k,e_0^{(b)}\otimes 1\right\},1\otimes f_0^{(c)}\right\}+\\
		&+\left[\left[\CP^{(a)}_{k-1},1\otimes e_0^{(b)}\right\},1\otimes f_0^{(c)}\right\}+\left[\left[\CP^{(a)}_{k-1},e_0^{(b)}\otimes 1\right\},f_0^{(c)}\otimes 1\right\}+\\
		&+\left[\left[\CP^{(a)}_{k-2},1\otimes e_0^{(b)}\right\},f_0^{(c)}\otimes 1\right\}=0,\quad k\geq 3\,.
	\end{split}
\end{align}

The solutions for the first few levels read:
\begin{align}
	\begin{split}
		&\CP^{(a)}_1=2\texttt{h}_1\sum\lm_x\nu_x^{-1}A_{ax}\,f_0^{(x)}\otimes e_0^{(x)}\,,\\
		&\CP^{(a)}_2=-\sum\lm_{x,y:\,A_{xy}\neq 0}\nu_x^{-1}\nu_y^{-1}\frac{A_{ax}+A_{ay}}{2A_{xy}}\,\left[f_0^{(x)},f_0^{(y)}\right\}\otimes \left[e_0^{(x)},e_0^{(y)}\right\}\,.
	\end{split}
\end{align}

In particular, we derive:\footnote{Let us note that $$
\left[e_n^{(a)},e_k^{(b)}\right\}=\left[f_n^{(a)},f_k^{(b)}\right\}=0,\quad\mbox{if}\, A_{ab}=0\,.
$$}
\begin{align}
	\Delta e_1^{(a)}=e_1^{(a)}\otimes 1+1\otimes e_1^{(a)}+\psi_0^{(a)}\otimes e_0^{(a)}+\sum\lm_x\nu_x^{-1}\,f_0^{(x)}\otimes \left[e_0^{(a)},e_0^{(x)}\right\}\quad{\rm mod}\;A_5^{\otimes 2}\,.
\end{align}

\subsection{Soliton corrections}

The soliton corrections can be defined from an equation analogous to \eqref{S_k_gl_1_rec}:
\begin{align}
	\left[\Delta_1\psi_1^{(a)},S_k\right]+\sum\lm_{m=1}^{k-1}\CP^{(a)}_mS_{k-m}+\CP^{(a)}_k=0,\quad \forall k\geq 1\,.
\end{align}

First of all let us note the following relation:
\begin{align}
	\left[\Delta_1\psi^{(a)}_1,f_x^{(b)}\otimes e_y^{(b)}\right]=-2A_{ab}\texttt{h}_1\left(\hat x-\hat y\right)\cdot f_x^{(b)}\otimes e_y^{(b)}\,,
\end{align}
where operators $\hat x$ and $\hat y$ raise the mode number by 1 in the first and the second factor, respectively.

Thus we derive for the first order soliton correction (cf. \eqref{1-solit}):
\begin{align}
	\begin{split}
		&S_1|\Kappa_1\rangle_{u_1}\otimes|\Kappa_2\rangle_{u_2}=\sum\lm_{a\in Q_0}\sum\lm_{\sqbox{$a$}_1\in {\rm Rem}(\Kappa_1)}\sum\lm_{\sqbox{$a$}_2\in {\rm Add}(\Kappa_2)}\\
		&\nu_a^{-1}\frac{\left[\Kappa_1\to\Kappa_1-\sqbox{$a$}_1\right]\left[\Kappa_2\to\Kappa_2+\sqbox{$a$}_2\right]}{u_{12}+h_{\sqbox{$a$}_1}-h_{\sqbox{$a$}_2}}|\Kappa_1-\sqbox{$a$}_1\rangle_{u_1}\otimes|\Kappa_2+\sqbox{$a$}_2\rangle_{u_2}
	\end{split}
\end{align}

The higher order correction can also be organized as a form-factor $G_2$:
\begin{align}
	\begin{split}
		&S_2|\Kappa_1\rangle_{u_1}\otimes|\Kappa_2\rangle_{u_2}=\sum\lm_{a,b\in Q_0}\sum\lm_{\sqbox{$a$}_1,\sqbox{$b$}_2\in {\rm Rem}(\Kappa_1)}\sum\lm_{\sqbox{$a$}_3,\sqbox{$b$}_4\in {\rm Add}(\Kappa_2)}\\
		&G_2^{(ab)}(u_1+h_{\sqbox{$a$}_1},u_1+h_{\sqbox{$b$}_2}|u_2+h_{\sqbox{$a$}_3},u_2+h_{\sqbox{$b$}_4})\times\\
		&\left[\Kappa_1\to \Kappa_1-\sqbox{$a$}_1\to\Kappa_1-\sqbox{$a$}_1-\sqbox{$b$}_2\right]\left[\Kappa_2\to \Kappa_2-\sqbox{$a$}_3\to\Kappa_2-\sqbox{$a$}_3-\sqbox{$b$}_4\right]\times\\
		&|\Kappa_1-\sqbox{$a$}_1-\sqbox{$b$}_2\rangle_{u_1}\otimes|\Kappa_2+\sqbox{$a$}_3+\sqbox{$b$}_4\rangle_{u_2}\,.
	\end{split}
\end{align} 

For $G_2$ we have:
\begin{align}
	\begin{split}
	&G_2^{(ab)}(x_1,x_2|y_1,y_2)=\frac{\nu_a^{-1}\nu_b^{-1}}{\left(x_1-x_2+A_{ab}\texttt{h}_1+M_{ab}\texttt{h}_2\right)\left(y_1-y_2-A_{ab}\texttt{h}_1+M_{ab}\texttt{h}_2\right)}\times\\
	&\left(1+\left(M_{ab}\texttt{h}_2+A_{ab}\texttt{h}_1\right)\left(\frac{1}{x_1-y_1}+\frac{1}{x_2-y_2}\right)+\frac{M_{ab}\texttt{h}_2^2-A_{ab}^2\texttt{h}_1^2}{\left(x_1-y_1\right)\left(x_1-y_1\right)}\right)\,.
	\end{split}
\end{align}

\section{Flavor Janus interface}\label{app:flavor}

For simplicity we consider a model of the 2D $\CN=(2,2)$ chiral field charged with respect to the flavor symmetry with fugacity $u$.
This model can be easily derived from the standard $U(1)$ gauged linear sigma model \cite{D-book_1} by freezing the gauge field and assigning a vacuum expectation value $u$ to the complex scalar in the gauge multiplet.
The resulting action reads:
\begin{align}\label{action}
	\begin{split}
		S_0
		=\int dx^0dx^1\,\Big[&|\p_0\phi|^2-|\p_1\phi|^2+\\
		&+\I \bar\psi_-\left( \overset{\leftrightarrow}{\p}_0+\overset{\leftrightarrow}{\p}_1 \right)\psi_-+\I \bar\psi_+\left( \overset{\leftrightarrow}{\p}_0-\overset{\leftrightarrow}{\p}_1 \right)\psi_++\\
		&+|{\bf F}|^2
		-|u\phi|^2-\bar\psi_-u\psi_+-\bar\psi_+\bar u\psi_-\Big]\,.
	\end{split}
\end{align}

This action is invariant with respect to the following SUSY transformations up to boundary terms under the assumption that $u$ is a constant parameter:
\begin{align}\label{SUSY_chir}
\begin{split}
	&\delta \phi=\epsilon_+\psi_--\epsilon_-\psi_+\,,\\
	&\delta \psi_+=\I\bar\epsilon_-(\p_0+\p_1)\phi+\epsilon_+ {\bf F}-\bar\epsilon_+\bar u \phi\,,\\
	&\delta \psi_-=-\I\bar\epsilon_+(\p_0-\p_1)\phi+\epsilon_- {\bf F}+\bar\epsilon_- u \phi\,,\\
	&\delta {\bf F}=-\I\bar\epsilon_+(\p_0-\p_1)\psi_+-\I\bar\epsilon_-(\p_0+\p_1)\psi_-+(\bar\epsilon_+\bar u\psi_-+\bar\epsilon_- u\psi_+)\,.
\end{split}
\end{align}

If $u$ is a function of the spacial coordinate $x^1$, the variation of the action \eqref{action} with respect to \eqref{SUSY_chir} produces a bulk term:
\begin{align}\label{uncompensated}
	\delta S_0=\int dx^0dx^1\,\left(-\I\bar \phi\left(\p_1\bar u\left(\epsilon_-\psi_-\right)+\p_1 u\left(\epsilon_+\psi_+\right)\right)\right)+{\rm c.c.}\,.
\end{align}

A background expectation value of the field $u$ varying with the spacial coordinate $x^1$ breaks the invariance of the action \eqref{action} with respect to spacial translations.
Since the translation generators---momenta---are elements of the superalgebra, not all the initial supercharges can be preserved in the presence of an interface.
We could choose a B-type supersymmetry that is preserved by the interface; a family of such choices is parameterized by a complex phase $e^{\I\vartheta}$ (which is a natural choice for interfaces in 2D theories, see e.g. \cite{Gaiotto:2010be}):
\begin{align}
	\epsilon_-=e^{\frac{\I\vartheta}{2}}\epsilon,\quad \epsilon_+=e^{\frac{\I\vartheta}{2}}\epsilon\,.
\end{align}

The subsequent computation will be simplified if we assume that along the interface only ${\rm Im}\,e^{-\I\vartheta}u$ varies whereas ${\rm Re}\,e^{-\I\vartheta}u$ remains constant.
In this case, the uncompensated term \eqref{uncompensated} is the variation of the following term:
\begin{align}\label{Janus}
	-\delta S_0=\delta_{\vartheta}S_{\rm Janus}:=\delta_{\vartheta}\int dx^0dx^1\,{\rm Im}\left(e^{-\I\vartheta}\p_1u\right)|\phi|^2\,.
\end{align}
Therefore a modified action:
\begin{align}
	S_0+S_{\rm Janus}
\end{align}
preserves $\vartheta$-supersymmetry of the initial $\CN=(2,2)$ supersymmetry.

Also varying the Euclidean action with respect to ${\rm Im}\,e^{-\I\vartheta}u$ we observe:
\begin{align}\label{rem}
	\frac{\delta\left(S_0^{\rm Eucl}+S_{\rm Janus}^{\rm Eucl}\right)}{\delta\left({\rm Im}\,e^{-\I\vartheta}u\right)}=\mathsf{Q}\Lambda+\left(e^{\I\vartheta}\bar\phi {\bf F}+e^{-\I\vartheta}\phi \bar {\bf F}\right)\,,
\end{align}
where the last term disappears due to the compatibility of the flavor symmetry with the supersymmetry and the superpotential (if there is one, otherwise simply ${\bf F}=0$), and the first term is $\mathsf{Q}$-exact for the supecharge generated by $\vartheta$-supersymmetry, where
\begin{align}
	\Lambda=\bar\phi\left(e^{\frac{\I\vartheta}{2}}\psi_++e^{\frac{-\I\vartheta}{2}}\psi_-\right)+{\rm (c.c.)}\,.
\end{align}

Importantly for us, \eqref{rem} indicates that small deformations of the interface path $u(x^1)$ produce $\mathsf{Q}$-exact terms canceled during localization, therefore the interface partition function is insensitive to those deformations and delivers a flat parallel transport description.

We have not yet considered a variation of ${\rm Re}\,e^{-\I\vartheta}u$ along the interface.
In principle, there is no need to do this since the $R$-matrix interface permutes the ordering we have defined by the ordering function ${\rm Re}\,u$ (see \eqref{ordering}).
The Janus interface for $\vartheta=\pi/2$ will capture the corresponding modifications for the ordering of the tensor factors.

The corresponding computations will be more involved if an explicit construction for both ${\rm Re}\,e^{-\I\vartheta}u$ and ${\rm Im}\,e^{-\I\vartheta}u$ varying along the interface is in question.
The reason is that the complex counterpart for the operator $|\phi|^2$ we inserted in \eqref{Janus} is a non-local defect operator dual to the momentum of the phase of $\phi$.
The complete complex twisted chiral field:
\begin{align}
    \tilde \phi(x)=|\phi(x)|^2+\I\int^x dx'\,\frac{\I\delta}{\delta({\rm arg}\,\phi(x'))}
\end{align}
is a defect operator describing an insertion of a 2D vortex with a Dirac string attached to it.
The field $\tilde\phi$ is neutral with respect to the flavor symmetry.

In the mirror dual picture an effective superpotential generated for the field $\tilde \phi$ reads \cite{Hori:2000kt}:
\begin{align}
    W_{\rm eff}(\tilde\phi)=u\, \tilde \phi+e^{-\tilde \phi}\,.
\end{align}
It is easy to give a complete description of this model if we switch the description of the twisted chiral field to a completely analogous description of a chiral field  swapping A- and B-twists simultaneously.
In this case the action and supersymmetry transforms for the field $\tilde \phi$ are given by \eqref{action} and \eqref{SUSY_chir} for $u=0$, respectively; in addition there is a superpotential term:
\begin{align}
S_0^{\rm dual}(\tilde \phi)=S_0(\tilde \phi,u=0)+\int dx^0dx^1\,\left(\tilde{\bf F}\p_{\tilde \phi}W_{\rm eff}-\p_{\tilde \phi}^2W_{\rm eff}\tilde\psi_+\tilde\psi_-+{\rm c.c.}\right)\,.
\end{align}

The uncompensated term for varying $u$ in this case reads:
\begin{align}
    \delta S_0^{\rm dual}=-\I\int dx^0dx^1\,\left(\bar\epsilon_+\tilde\psi_+-\bar\epsilon_-\tilde\psi_-\right)\p_u\p_{\tilde \phi}W_{\rm eff}\,\p_1 u+{\rm c.c.}\,.
\end{align}
The corresponding Janus interface action reads:
\begin{align}
    S_{\rm Janus}^{\rm dual}=\I \int dx^0dx^1\,\p_uW_{\rm eff}\,\p_1 u+{\rm c.c.}\,,
\end{align}
so that the resulting total action $S_0^{\rm dual}+S_{\rm Janus}^{\rm dual}$ is invariant under the A-type $\vartheta$-symmetry:
\begin{align}
    \epsilon_+=e^{\I\vartheta}\epsilon,\quad \epsilon_-=e^{-\I\vartheta}\bar\epsilon\,.
\end{align}


\section{Comparison between the rational and trigonometric \texorpdfstring{$\bDelta_0$}{D0} and \texorpdfstring{$\Delta_1$}{D1}
}\label{sec:splitting}

One potential source of confusion is the fact that the rational $\bDelta_{0}$, $\Delta_1$ 
and the trigonometric $\ddot\bDelta_{0}$, $\ddot\Delta_1$ 
have rather similar structures but differ in important properties. 
In particular, while the trigonometric $\ddot\bDelta_0$ factorizes and $\ddot\Delta_1$ is a valid coproduct (although diagonal), the rational 
$\bDelta_0$ does not factorize whereas $\Delta_1$ is not an algebra homomorphism, see the table at the end of section \ref{sec:co-mul}.
In this section, we compare the rational $\bDelta_{0}$ and $\Delta_1$ 
with the trigonometric $\ddot\bDelta_{0}$ and $\ddot\Delta_1$, and try to pinpoint how their differences arise at a technical level. 

\subsection{Rational vs. trigonometric \texorpdfstring{${\bDelta}_0$}{D0} }
\label{sec:app:Delta0}

In this section we comment more on the ability/inability of the naive tensor product \eqref{chain_naiv_co} to factorize \eqref{split}.

Let us show that the naive representation \eqref{chain_naiv_co} indeed does not factorize.
For this we transform it into a nearly-factorize form.
Consider a naive representation of the raising operator acting on a two-site crystal chain:\footnote{To reduce clutter, in this section we have dropped the color $a$ dependence of the generators and the atoms since here we never have to consider the interaction between two colors; it is easy to restore this dependence.}
\begin{align}\label{spl0}
	\begin{split}
	&\bDelta_0e(z)\,|\Kappa_1\rangle_{u_1}\otimes|\Kappa_2\rangle_{u_2}=(e(z)\otimes 1)\, |\Kappa_1\rangle_{u_1}\otimes|\Kappa_2\rangle_{u_2}+\\&+\sum\lm_{\Box\in{\rm Add}(\Kappa_2)}\Psi_{\Kappa_1}(h_{\Box}+u_2-u_1)\frac{[\Kappa_2\to\Kappa_2+\Box]}{z-\left(h_{\Box}+u_2\right)}\, |\Kappa_1\rangle_{u_1}\otimes|\Kappa_2+\Box\rangle_{u_2}\,,
	\end{split}
\end{align}
where we have represented the first term as an element of $\myY\otimes\myY$ to stress that it is the second term that is troublesome.

To separate the contribution from both crystals to the $\Psi$-term we apply the usual contour integral trick:
\begin{align}
	\Psi_{\Kappa_1}\left(h_{\Box}+u_2-u_1\right)=\oint_{\CC}\frac{dw}{w-\left(h_{\Box}+u_2\right)}\Psi_{\Kappa_1}(w-u)\,,
\end{align}
where the integration cycle $\CC$ encircles all the poles of the form $h_{\Box}+u_2$ for the atoms $\Box\in{\rm Add}(\Kappa_2)$, which are included in the summation in \eqref{spl0}.
For the two denominator factors appearing in the result we can apply a simple algebraic manipulation:
\begin{align}
	\frac{1}{w-\left(h_{\Box}+u_2\right)}\cdot \frac{1}{z-\left(h_{\Box}+u_2\right)}=\frac{1}{z-w}\left(\frac{1}{w-\left(h_{\Box}+u_2\right)}-\frac{1}{z-\left(h_{\Box}+u_2\right)}\right)\,.
\end{align}
Gathering all the elements under the cycle integration we derive the following relation:
\begin{align}
	\begin{split}
	&\left(\bDelta_0e(z)-e(z)\otimes 1\right)|\Kappa_1\rangle_{u_1}\otimes|\Kappa_2\rangle_{u_2}=\\
	&=-\oint_{\CC}dw\,\psi(w)\otimes\frac{e(z)-e(w)}{z-w}|\Kappa_1\rangle_{u_1}\otimes|\Kappa_2\rangle_{u_2}\,.
	\end{split}
\end{align}

The next naive step would be to simply strip off the representation factors $|\Kappa_1\rangle_{u_1}\otimes|\Kappa_2\rangle_{u_2}$ from both sides and check if the operators defined this way, namely
\begin{align}\label{trial_split}
	\tilde \Delta e(z)\mathop{=}\lm^{???}e(z)\otimes 1-\oint_{\CC}dw\,\psi(w)\otimes\frac{e(z)-e(w)}{z-w}
\end{align}
gives rise to a valid coproduct corresponding to $\bDelta_0$ in the representation.
However, one can see that $\tilde \Delta $ defined in \eqref{trial_split}  also does not factorize since the integration cycle $\CC$ only encircles the poles of the $e(z)$ generator (that correspond to the atoms in ${\rm Add}(\Kappa_2)$), whereas the eigenvalue of the $\psi(z)$ generator   also have poles corresponding to the atoms in ${\rm Add}(\Kappa_1)\cup {\rm Rem}(\Kappa_1)$, and there is no way to separate those poles from those of $e(z)$ acting on $\Kappa_2$.\footnote{One might try to deform the contour $\CC$ to encircle $\infty$ so that all the poles from both $\psi(z)$ and $e(z)$ can  contribute, but then the expression \eqref{trial_split} would not reproduce the action  \eqref{spl0} that we started with.
}

Let us now examine why the trigonometric $\ddot\Delta_1$ does not have this problem with factorization. 
The trigonometric counterpart of the $e$-actions on a two-site crystal chain \eqref{spl0} is
\begin{align}\label{spl0trig}
	\begin{split}
	&\ddot{\bDelta}_0e(z)\,|\Kappa_1\rangle_{u_1}\otimes|\Kappa_2\rangle_{u_2}=(e(z)\otimes 1)\, |\Kappa_1\rangle_{u_1}\otimes|\Kappa_2\rangle_{u_2}+\\&+\sum\lm_{\Box\in{\rm Add}(\Kappa_2)}\Psi^{-}_{\Kappa_1}(h_{\Box}+u_2-u_1)\,[\Kappa_2\to\Kappa_2+\Box]\,\mathfrak{p}(z-\left(h_{\Box}+u_2\right))\, |\Kappa_1\rangle_{u_1}\otimes|\Kappa_2+\Box\rangle_{u_2}\,,
	\end{split}
\end{align}
where the propagator  $\mathfrak{p}(z)$ is the formal delta function that is defined differently for the rational/trigonometric/elliptic cases:\footnote{In Section~\ref{ssec:spec_param}, $\mathfrak{p}(z)$ is called $\delta^{\textrm{rat.}}$ and $\delta^{\textrm{trig.}}$ for the rational and trigonometric cases, respectively.}
\begin{align}
    \mathfrak{p}(z)=\begin{cases}
    \frac{1}{z} \qquad \qquad & \textrm{rational}\\
    \delta(Z)=\sum_{n\in\mathbb{Z}} Z^n \qquad \qquad &\textrm{trig./elliptic}
    \end{cases}
\end{align}
Then using
\begin{align}\label{trigdelta}
    \Psi^{-}_{\Kappa_1}(h_{\Box}+u_2-u_1)\,\mathfrak{p}(z-\left(h_{\Box}+u_2\right)) =\Psi^{-}_{\Kappa_1}(z-u_1)\,\mathfrak{p}(z-\left(h_{\Box}+u_2\right))\,,
\end{align}
\eqref{spl0trig} can be simplified into 
\begin{align}
	\begin{split}
	&\ddot{\bDelta}_0e(z)\,|\Kappa_1\rangle_{u_1}\otimes|\Kappa_2\rangle_{u_2}=(e(z)\otimes 1 +\psi^{-}(z)\otimes e(z))\, |\Kappa_1\rangle_{u_1}\otimes|\Kappa_2\rangle_{u_2}
	\end{split}
\end{align}
Namely, the trigonometric $\ddot\bDelta_0$ does factorize. 
From this comparison, one can also see that the relation \eqref{trigdelta}, which is necessary for the simplification and hence for the factorization, does not have a counterpart in the rational case. 

\subsection{Rational vs. trigonometric \texorpdfstring{$\Delta_1$}{D1}}
\label{sec:app:Delta1}

Now we explain why $\Delta_1$ is not an algebra homomorphism of $\myY$ whereas $\ddot\Delta_1$ is an algebra homomorphism of $\ddot\myY$, although their definitions have the same form. 

The first thing to emphasize is that the algebraic relations of the quiver BPS algebra are usually written in terms of the so-called fields $(e^{(a)}(z),\psi^{(a)}_{\pm}(z),f^{(a)}(z) )$, and one needs to plug in their mode expansions to obtain the corresponding relations in terms of their modes $(e^{(a)}_j,\psi^{(a)}_{\pm, j},f^{(a)}_j)$. 
Namely, the mode relations are the true defining relations of these algebras, whereas the field relations are convenient ways to package this information.\footnote{We emphasize that the difference between the field relations and the mode relations is not just a technicality. For example, in the gluing basis (as opposed to the crystal basis used here) of \cite{Gaberdiel:2018nbs,Li:2019nna,Li:2019lgd},  the numerator and the denominator of the bond factor might share a common factor, which should not be canceled in order to produce the correct mode relations, although whether or not to cancel these common factors makes no difference to the field relations.} 

When checking whether $\Delta_1$ (resp.\ $\ddot\Delta_1$) is an algebra homomorphism of $\myY$ (resp.\ $\ddot{\myY}$), if we had used the field relations, we would have concluded (mistakenly) that they are both algebra homomorphisms. This is not too surprising since the field relations of $\myY$ and $\ddot{\myY}$ take very similar forms. 
However, the $\Delta_1$  (resp.\ $\ddot\Delta_1$) preserving the field relations of $\myY$ (resp.\ $\ddot{\myY}$) is only a necessary condition for it to be an algebra homomorphism: one needs to check whether it preserves the mode relations.\footnote{For the affine Yangian of $\mathfrak{gl}_1$, it was first emphasized in \cite{Prochazka:2015deb} that $\Delta_1$ does not preserve the mode relations of the algebra.} It is here that the rational $\Delta_1$ and the trigonometric $\ddot\Delta_1$ differ in their behaviors. 

It is enough to illustrate this with the $\psi-e$ relation. 
What makes the difference is that in the rational case, the coproduct of the $e/f$ generators in terms of the $e_k/f_k$ modes involve a summation whose range depends on the mode number $k$:
\begin{align}
    &\Delta_1(e_k)=e_k\otimes 1 +\sum^{k}_{j=0}\psi_{j}\otimes e_{k-j}\,,
\end{align}
where we have used $\psi_{0}=1$.
One can check that the $\psi-e$ relation in the rational case, in terms of the modes (see eq.\ (4.20) of \cite{Li:2020rij}), is not preserved by $\Delta_1$, precisely due to this dependence on $k$ in the summation.
In constrast, in the trigonometric case, this dependence is absent:
\begin{align}
    & \ddot\Delta_1(\ddot e_k)=\ddot e_k\otimes 1+ \sum_{j\leq 0}\ddot\psi^{-}_{j}\otimes \ddot e_{k-j}\,,
\end{align}
and one can check that the mode relations of the  $\psi^{\pm}-e$ relation for the trigonometric case are preserved by $\ddot\Delta_1$.

\section{Miki automorphism vs.\ soliton corrections}\label{app:Miki}

A trigonometric generalization of $\myY(\widehat{\fg\fl}_1)$ is known as a quantum toroidal algebra \cite{feigin2017finite} (see also \cite{MR1324698,Ding:1996mq,Miki2007,MR2793271,Feigin:2013fga,MR2566895,Bezerra:2019dmp}), sometimes denoted as $U_{q_1,q_2}(\widehat{\widehat{\fg\fl}}_1)$ \cite{Negut:2020npc,Garbali:2021qko}.
In this section let us  denote $\myY(\widehat{\fg\fl}_1)$ as $\myY$, and the quantum toroidal algebra of $\fg\fl_1$ as $\ddot{\myY}$.
It is parameterized by the exponentiated equivariant weights $q_i=e^{\beta\texttt{h}_i}$, $i=1,2,3$ subjected to the relation:
\begin{align}
	q_1q_2q_3=1\,.
\end{align}

In this section we will work only with $\ddot\myY$, therefore we use simplified notations.
The algebra is generated by a triplet of generator modes $(e_n,f_n,h_n)$, $n\in\IZ$ and has two central elements $C$ and $C^{\perp}$.
Modes can be organized in generating functions:
\begin{align}
	\begin{split}
		e(z)=\sum\lm_{n\in\IZ}e_n Z^{-n},\quad f(z)=\sum\lm_{n\in\IZ}f_n Z^{-n},\quad
	\psi^{\pm}(z)=\left(C^{\perp}\right)^{\mp 1}\exp\left(\sum\lm_{r=1}^{\infty}\kappa_r h_{\pm r}Z^{\mp r}\right)\,,
	\end{split}
\end{align}
where we adopt the convention
\begin{align}
    Z\equiv e^{\beta z}\,, \quad W\equiv e^{\beta w}\,, \quad \textrm{etc.}
\end{align}
and
\begin{align}
    \kappa_r=(1-q_1^r)(1-q_2^r)(1-q_3^r)\,.
\end{align}

Introducing a notation for a ``half" of the trigonometric bond factor:
$$
\eta(Z,W)=(Z-q_1 W)(Z-q_2 W)(Z-q_3 W)\,,
$$
we can write the algebraic relations in the following form (cf.~\eqref{QiuvYangian}):
\begin{align}
	\begin{split}
		&\left[h_r,h_s\right]=\delta_{r+s,0}\frac{1}{r}\frac{C^r-C^{-r}}{\kappa_r}\,,\\
		&\left[h_r,e_n\right]=-\frac{1}{r}e_{n+r}C^{\left(-r-|r|\right)/2},\quad 		\left[h_r,f_n\right]=\frac{1}{r}f_{n+r}C^{\left(-r+|r|\right)/2}\,,\\
		&\left[e(Z),f(W)\right]=\frac{1}{\kappa_1}\left(\delta\left(\frac{CW}{Z}\right)\psi^+(W)-\delta\left(\frac{CZ}{W}\right)\psi^-(Z)\right)\,,\\
		&\eta(Z,W)e(Z)e(W)+\eta(W,Z)e(W)e(Z)=0\,,\\
		&\eta(W,Z)f(Z)f(W)+\eta(Z,W)f(W)f(Z)=0\,.
	\end{split}
\end{align}

The coproduct for the generators reads:\footnote{This coproduct is the one denoted as $\ddot\Delta_1$ in the main text, which arises from the naive crystal chain representation in the trigonometric case $\ddot\bDelta_0=\ddot{\rm Rep}\circ \ddot\Delta_1$; for the difference between the rational and trigonometric cases see Appendix~\ref{sec:app:Delta0}.}
\begin{align}
	\begin{split}
	\Delta e_n&=e_n\otimes 1+\sum\lm_{j\geq 0}\psi_j^+C^n\otimes e_{n-j}\,,\\
	\Delta f_n&=\sum\lm_{j\geq 0} f_{n+j}\otimes \psi_{-j}^-C^n+1\otimes f_n\,,\\
	\Delta h_r&=h_r\otimes C^{-r}+\otimes h_r\,,\\
	\Delta h_{-r}&=h_{-r}\otimes 1+C^r\otimes h_{-r}\,,
	\end{split}
\end{align}
where 
\begin{align}
    \psi^{\pm}(Z)=\sum\lm_{\pm n\geq 0}\psi_n^{\pm}Z^{-n}\,.
\end{align}

The Miki automorphism $\vartheta$ is defined as:
\begin{align}
	e_0\mapsto h_{-1},\quad h_{-1}\mapsto f_0,\quad f_0\mapsto h_{1},\quad h_1\mapsto e_0,\quad C^{\perp}\mapsto C,\quad C\mapsto \left(C^{\perp}\right)^{-1}\,.
\end{align}

We define a perpendicular coproduct as\footnote{The coproduct $\Delta^{\perp}$ is the ``true'' coproduct that corresponds to a non-diagonal $R$-matrix on lowest weight crystal representations of $\ddot\myY$; and we have denoted it as $\ddot\Delta$ in the main text. 
If one considers lowest weight Fock modules in $\myY^{\perp}$ instead (see e.g. \cite{2021CMaPh.384.1971G}) the roles of $\Delta$ and $\Delta^{\perp}$ will be interchanged.}
\begin{align}
	\Delta^{\perp}=\vartheta^{-1}\circ\Delta\circ\vartheta: \quad \ddot A\longrightarrow \ddot A\otimes\ddot A\,,
\end{align}
where $\vartheta$ on $\ddot A\otimes\ddot A$ acts simply as $\vartheta\otimes\vartheta$.

For the lowest few generators, we define the following explicit expressions:
\begin{align}
	\begin{array}{lcl}
	\Delta e_0=\sum\lm_{j\geq 0} e_0\otimes 1+\psi_j^+\otimes e_{-j}\,,&&\Delta^{\perp}e_0=e_0\otimes 1+ \left(C^{\perp}\right)^{-1}\otimes e_0\,,\\
	\Delta f_0=f_j\otimes\sum\lm_{j\geq 0}\psi_{-j}^-+1\otimes f_0\,,&&\Delta^{\perp}f_0= f_0\otimes C^{\perp}+1\otimes f_0\,,\\
	\Delta h_1=h_1\otimes C^{-1}+1\otimes h_1\,, &&\Delta^{\perp} h_1=h_1\otimes 1+C\otimes h_1+\CJ_1\,, \\
	\Delta h_{-1}=h_{-1}\otimes 1+C\otimes h_{-1}\,,&&\Delta^{\perp} h_{-1}=h_{-1}\otimes C^{-1}+1\otimes h_{-1}+\CJ_{-1}\,,
	\end{array}
\end{align}
where
\begin{align}
	\begin{split}
	\CJ_1&=\kappa_1\sum\lm_{j=1}^{\infty}(-1)^{j}\, \left[{\rm ad}_{f_0}^jh_1,h_{-1}\right]\otimes {\rm ad}_{e_0}^jh_1\,,\\
	\CJ_{-1}&=\kappa_1\sum\lm_{j=1}^{\infty}(-1)^{j+1}\,{\rm ad}_{f_0}^jh_{-1}\otimes \left[{\rm ad}_{e_0}^jh_1,h_{-1}\right]\,.
	\end{split}
\end{align}

In complete analogy with \eqref{iso} we define an operator $\ddot U$ intertwining the coproduct and the perpendicular coproduct:
\begin{align}
	\bDelta^{\perp}\cdot \ddot U=\ddot U\cdot \bDelta\,.
\end{align}

When $C=1$, the operators $\psi^{\pm}$ become Cartan and diagonal in the crystal basis.
In this case we could search for $\ddot U$ using the ansatz \eqref{S_k_form_expl}, where the variables $x_i$ and $y_i$ should replaced by $X_i=e^{\beta x_i}$ and $Y_i=e^{\beta y_i}$, namely each equivariant weight $h$ should be replaced by its corresponding exponentiated weight $H=e^{\beta h}$.
As the result we find:
\begin{align}\label{tor_ff}
	g_1(X|Y)=\kappa_1 \, \frac{1}{1-Y/X}\,.
\end{align}

So we expect that the dimensional reduction of the transformation $\ddot U$ induced by the Miki automorphism corresponds to the soliton correction matrix $U$:
\begin{align}
	\lim\lm_{\beta\to 0}\ddot U=U\,.
\end{align}


\bibliographystyle{utphys} 
\bibliography{biblio}


\end{document}